\documentclass[journal]{IEEEtran}
\usepackage{amsmath}
\usepackage{amsthm}
\newtheorem{Definition}{Definition}
\newtheorem{Lemma}{Lemma}
\newtheorem{Theorem}{Theorem}

\usepackage{subfigure}
\usepackage{amssymb}
\usepackage{booktabs}
\usepackage{mathtools}
\begin{document}  
\title{Evolutionary Dynamics with Self-Interaction Learning in Networked Systems}
\author{Ziyan~Zeng, 
Minyu~Feng,~\IEEEmembership{Senior Member,~IEEE,}
and Attila~Szolnoki
\thanks{This work was supported by the National Natural Science Foundation of China (NSFC) (Grant No. 62206230),  the Natural Science Foundation of Chongqing (Grant No. CSTB2023NSCQ-MSX0064), and the National Research, Development and Innovation Office (NKFIH) (Grant No. K142948). \textit{(Corresponding author: Minyu Feng)}}   
\thanks{Ziyan~Zeng and Minyu~Feng are with the College of Artificial Intelligence, Southwest University, Chongqing, China. (email: myfeng@swu.edu.cn)}
\thanks{Attila~Szolnoki is with the institute of Technical Physics and Materials Science, Centre for Energy Research, P.O. Box 49, H-1525 Budapest, Hungary.}
}
\markboth{IEEE TRANSACTIONS ON NETWORK SCIENCE AND ENGINEERING, VOL.~, NO.~, 2025
}{Zeng \MakeLowercase{\textit{et al.}}: Evolutionary Dynamics with Self-Interaction Learning}
\maketitle
\begin{abstract}
The evolution of cooperation in networked systems helps to understand the dynamics in social networks, multi-agent systems, and biological species. The self-persistence of individual strategies is common in real-world decision making. The self-replacement of strategies in evolutionary dynamics forms a selection amplifier, allows an agent to insist on its autologous strategy, and helps the networked system to avoid full defection. In this paper, we study the self-interaction learning in the networked evolutionary dynamics. We propose a self-interaction landscape to capture the strength of an agent's self-loop to reproduce the strategy based on local topology. We find that proper self-interaction can reduce the condition for cooperation and help cooperators to prevail in the system. For a system that favors the evolution of spite, the self-interaction can save cooperative agents from being harmed. Our results on random networks further suggest that an appropriate self-interaction landscape can significantly reduce the critical condition for advantageous mutants, especially for large-degree networks. 
\end{abstract}
\begin{IEEEkeywords}
Evolutionary Games, Evolutionary Dynamics, Networked Systems, Self-interaction Learning
\end{IEEEkeywords}
\IEEEpeerreviewmaketitle
\section{Introduction}
\small
The evolution of species and behaviors has attracted significant research attention since Darwin laid the main principle~\cite{darwin2023origin}. In biological systems, successful species with high fecundity can survive and dominate the whole system~\cite{shakarian2012review,ohtsuki2007evolutionary,allen2023symmetry}. In human populations, profitable strategies that bring agents high payoff or fitness can be propagated and fixed in society~\cite{nowak2004evolutionary,nowak2006five}, the most important enigma is the evolution of cooperation. The evolutionary game theory provides a powerful framework to address it~\cite{axelrod1981evolution}, where fruitful results are obtained in well-mixed systems. Additionally, with the development of complex network theory~\cite{newman_10}, structured systems that are described by graphs with vertices and edges are found to be a useful tool to model realistic environments~\cite{nowak1992evolutionary,santos2005scale}. The most representative social dilemma situations are captured via the celebrated prisoner's dilemma~\cite{rapoport_70,wang2023imitation} or public goods games~\cite{szolnoki2010reward}.

The evolution of cooperation is a stochastic process with two absorbing barriers~\cite{lieberman2005evolutionary, fu2009evolutionary}, including pure cooperation and pure defection, if we do not consider strategy mutation. If the fixation probability of cooperation is higher than the neutral drift in the Moran process~\cite{pattni2015evolutionary,diaz2021survey}, then cooperation can be favored in this system. A well-known condition for cooperation is that the benefit-to-cost ratio should be higher than the mean degree~\cite{ohtsuki2006simple} or the mean degree of nearest neighbors~\cite{konno2011condition} if the system is large enough. However, this condition may be too demanding, especially in large-scale networks, where an agent may connect to more than hundreds of neighbors. The question is, what network structure and dynamic mechanism can facilitate or promote the evolution of cooperation? To answer this question, researchers have proposed numerous models~\cite{zhu2022networked,10272996}. For example, the game transition model in evolutionary dynamics shows a behavior-based changing environment can result in a favorable outcome for cooperation~\cite{su2019evolutionary,feng2023evolutionary}. Another simple and powerful framework to promote cooperation is the reputation mechanism~\cite{fehr_n04,ren2023reputation,li2016changing,hu2021adaptive}, which may have arisen in early human culture. 

This article emphasizes the strategy update potential from self-replacement angle. Naturally, in real-world scenarios, the offspring can take the place of its parent, and an agent may also insist on its original strategy when compared to the neighbor information. Recent research shows that a graph with self-loops can be a strong amplification of selection for advantageous mutants~\cite{pavlogiannis2018construction,adlam2015amplifiers}. This suggests that the fixation of cooperation may be enhanced with self-interaction. Recent studies~\cite{wang2023greediness,wang2023conflict} confirmed that such self-interaction can somehow promote cooperation in isothermal graphs via the identity-by-descent method~\cite{allen2014games} and pair approximation~\cite{gardiner1985handbook}. If mutation is considered in the evolutionary dynamics, self-interaction is detrimental for the advantageous mutants~\cite{sharma2023self}. However, these works mainly focus on networks with regular topology. Additionally, the self-interaction weights are often the same for all vertices in a network to simplify the problem. To our best knowledge it is still unclear what kind of self-interaction is beneficial for the fixation of cooperation beyond the isothermal graph and homogeneous self-interaction strength. To fill this gap, in this paper, we study the effect of self-interaction learning in networked systems including particular graph families, synthetic networks, and real-world network data sets. The contribution of this article can be highlighted as follows: 
\begin{enumerate}
    \item We propose an evolutionary game model in networked system with self-interaction landscapes, where each agent has a certain probability to maintain its strategy based on independent self-loop in learning graph. 
    \item We present the critical condition for cooperation with self-interaction strength. We identify the condition of self-interaction that can always allow the system to favor cooperation in several graph families. 
    \item We analyze the effect of self-interaction in several synthetic and real networks. We show that proper self-interaction can significantly reduce the condition for cooperation, especially if the network degree is high. 
\end{enumerate}

The organization of the remaining part is as follows: In Sec.~\ref{sec: literature review}, we provide a literature review for evolutionary dynamics in networks. In Sec.~\ref{sec: III}, we propose the self-interaction landscape and introduce the evolutionary donation game in networked systems. In Sec.~\ref{sec: theorems}, we study the condition for cooperation in specific graph families. In Sec.~\ref{sec: simulation}, we analyze the results for specific graph families and perform further simulations in random graphs and real-world network data sets. In Sec.~\ref{sec: conclusion}, we conclude our work, discuss some open issues, and outline some potential future research directions. 

\section{Literature Review}\label{sec: literature review}
\small
In this section, we review the current literature on evolutionary dynamics in complex networks that establishes the research context. 

To reveal the emergence of mutual cooperation from games with defective Nash equilibrium in social systems, spatial structure was first introduced into prisoner's dilemmas in~\cite{nowak1992evolutionary}, demonstrating chaotically changing spatial patterns in which cooperators and defectors both persist. Since then, spatial structures have been widely regarded as a framework to overcome social dilemmas such as the prisoner's dilemma~\cite{szabo1998evolutionary} and the public goods game~\cite{szabo2002phase}. However,~\cite{hauert2004spatial} revealed that spatial structure does not always enhance cooperation; in snowdrift game, the cooperation level can even fall below that in the mixed equilibrium of well-mixed populations. Studies on scale-free networks suggested that heterogeneity offers a unifying explanation for the promotion of cooperation~\cite{santos2005scale, santos2006evolutionary}. With the development of network theory, researchers have expanded their focus to multilayer networks~\cite{wang2015evolutionary}, time-varying networks~\cite{10844908,zeng2022spatial}, and higher-order networks~\cite{majhi2022dynamics,alvarez2021evolutionary}. 

In addition to the effects of network topology on cooperation, scholars have explored various behavioral and social mechanisms. Punishment~\cite{helbing2010punish,szolnoki2017second} and reward~\cite{szolnoki2012evolutionary} have been shown to promote cooperation under specific conditions. Reputation-based mechanisms~\cite{fu2008reputation,xia2023reputation}, which update the accumulated behavioral history of individuals, also play a significant role in strategy selection. With the rise of large language models, reinforcement learning has emerged as an important topic in this area~\cite{kaisers2010frequency,shi2022analysis}.~\cite{pi2024memory} proposed a memory-based evolutionary game involving dynamic interactions between learners and profiteers, demonstrating that memory mechanisms can promote cooperative behavior among profiteers and that higher learning rates lead to more cooperators. Similarly,~\cite{xie2025reputation} introduced a theoretically grounded double Q-learning algorithm, which significantly enhances the level of cooperation.

Obtaining exact solutions for evolutionary games on finite networks remains challenging. Fortunately, several mathematical frameworks have been developed to analyze cooperation and interpret phenomena observed in Monte Carlo simulations. A key contribution was made in~\cite{ohtsuki2006simple}, which derived the condition $b/c>k$ for sufficiently large regular networks using pairwise approximation. This was extended to heterogeneous networks in~\cite{konno2011condition}, where the condition becomes $b/c>\langle k_{nn}\rangle$. A more rigorous result was later provided in~\cite{taylor2007evolution}, yielding $b/c>(N-2)/(N/k-2)$, valid for any population size $N>2$. The general voter model perturbation method~\cite{chen2013sharp} introduced algebraically explicit first-order approximations for fixation probabilities based on coalescing random walks (CRW).~\cite{allen2014games} formalized the identity-by-descent (IBD) approach to derive cooperation conditions in vertex-transitive graphs.

More recently, with advancements in mathematical methodology, researchers have investigated the role of various social mechanisms. For example,~\cite{allen2012mutation} studied cooperation under mutation, finding that high mutation rates reduce clustering among cooperators and hinder cooperative success. By extending the stochastic game model~\cite{hilbe2018evolution} to structured populations,~\cite{su2019evolutionary} showed that the cooperation level in games of different values is closely related to node degrees. Further studies on high-order and heterogeneous networks include~\cite{sheng2024strategy}, which demonstrated that higher-order interactions favor cooperation in hypergraphs with multiple communities, and~\cite{su2023strategy}, which found that spatial heterogeneity tends to suppress cooperation, while temporal heterogeneity tends to promote it. To derive closed-form results, several studies have focused on regular networks. For instance,~\cite{wang2023optimization} examined incentive strategies in structured populations and found that optimal incentive protocols, both positive and negative, are time-invariant. Meanwhile,~\cite{wang2024evolutionary} derived the replicator equation for any $n$-strategy multiplayer game under weak selection, and proposed the “Satisfied-Cooperate, Unsatisfied-Defect” update rule based on aspiration, which outperforms other aspiration-based rules in promoting cooperation in the prisoner's dilemma game.

There are also several research papers that investigate the effects of real-world social mechanisms on individual cooperative behavior. ~\cite{zhang2014evolution} described an economic experiment, the Mutual Aid game, which tested theoretical predictions and found that pool punishment emerges only when second-order free riders are punished, while peer punishment is more stable than previously expected. ~\cite{leibbrandt2013rise} compared competitiveness across traditional fishing societies, where local environmental factors determine whether fishermen operate individually or within collectives. ~\cite{traulsen2010human} employed a setting in which individuals were virtually arranged on a spatial lattice and observed that spontaneous strategy changes, representing mutations or exploratory behavior, occur more frequently than typically assumed in theoretical models. ~\cite{jiang2013if} conducted a series of economic experiments to explore whether severe punishment is more effective than mild punishment, and found that the effectiveness of punishment stems not only from imposed fines but also from underlying psychological factors. ~\cite{grujic2010social} designed an experiment to study the emergence of cooperation when humans play the prisoner's dilemma on a network of a size comparable to that used in simulations, revealing that the cooperation level gradually declines to an asymptotic state with low but nonzero cooperation. Overall, these empirical studies highlight the complexity of human behavior and underscore the importance of investigating realistic social mechanisms. 

\section{Evolutionary Dynamics with Self-interaction Landscapes}\label{sec: III}
\small
In this section, we propose the evolutionary dynamics model based on self-interaction landscapes. We first present the network structure and our new definition of self-interaction landscapes. Then, we introduce the evolutionary dynamic model on the actual network considering the non-mutation case. 

\subsection{The Network Structures and Self-interaction Learning}
We consider an undirected network framework $\mathcal{G}=(\mathcal{V}, \mathcal{E}, \mathcal{W})$ composed of $N=\vert\mathcal{V}\vert$ vertices, where $\mathcal{V}$ is the vertex set, $\mathcal{E}\in\mathcal{V}\times\mathcal{V}$ denotes the edge set, and $\mathcal{W}$ presents the weight set of all edges. The unweighted network class can be presented by setting $w_{ij}=1$ for all edges $(i,j)$. For simplicity, we do not consider multiple edges in our model. 

To study the effect of self-interaction among agents, here we add weighted self-loops for all vertices based on each agent's property to capture the strength of self-interaction. Formally, we assume that the weights of self-loops are based on the following self-interaction landscapes. 
\begin{Definition}\label{def: 1}
\textit{Self-interaction landscape} is a vector $\mathbf{\mathcal{L}}=[\ell_1, \ell_2, \cdots, \ell_N]$ that describes the intensity of each agent's self-loop. For $i\in\mathcal{V}$, the weight of self-loop $w_{ii}$ is defined as a non-negative function $\ell_i(\mathcal{T}_i): \mathcal{R}\rightarrow\mathcal{R^+}$, where $\mathcal{T}_i$ is a symbol that describes a topology property of the agent $i$. 
\end{Definition}

According to Def.~\ref{def: 1}, we describe the self-interaction strength as the self-loop weight, depending on the local topology property of each agent. Under this self-interaction mechanism, each agent tends to insist on its own strategy based on the local structure. In this paper, we mainly discuss the case that $\mathcal{T}_i=k_i$, i.e., the self-loop weight is proportional to the degree of agent $i$. The vertex degree in a network is often bounded, therefore the self-interaction strength is finite, i.e., $\ell(k_i)<+\infty$. 

\subsection{Evolutionary Dynamics with Self-interaction Landscapes}
We consider a two-person two-strategy donation game consisting of cooperation ($C$) and defection ($D$) in the network structure. Each agent takes one position of a vertex in the network and can choose to be a cooperator or a defector. In the whole network, cooperators pay a cost ($c$) to obtain the benefit ($b$), while defectors pay no cost to get the same benefit. Here, the value $b/c$ is denoted as the benefit-to-cost ratio. Therefore, for two agents playing such a game, the mutual cooperation brings each player the payoff $b-c$, and mutual defection leads to the payoff $0$ for each player. For one-way cooperation, the cooperator receives no benefit for a $-c$ payoff, while the defector earns the full benefit $b$. In the reasonable $b>c>0$ parameter range, this game is well known as the prisoner's dilemma game~\cite{axelrod1980effective}, where the defector obtains the highest expected payoff, and mutual defection becomes the unique Nash equilibrium point in a single round game. 

We define the strategy vector of the network as $\mathbf{\mathcal{S}}=(s_{i})_{i\in\mathcal{V}}\in\{0,1\}^{\mathcal{V}}$, where 0 and 1 denote the defection and cooperation, respectively. The total payoff of an agent is the sum of payoffs from interacting with all neighbors and itself, denoted as $f_i(\mathbf{\mathcal{S}})$. We translate this payoff into fitness as the reproductive rate of the vertex $i$ in the exponential form $F_i(\mathbf{\mathcal{S}})=\exp[\delta f_i(\mathbf{\mathcal{S}})]$, where $\delta\in[0, 1]$ denotes the strength of selection. It is worth noting that we are mainly interested in the weak selection ($\delta\ll1$) limit where games have only a small effect on microscopic dynamics~\cite{fu2009evolutionary}. Regarding the strategy update, we consider a death-birth (DB) update process in the continuous time axis. The reason for considering DB is that this kind of replacement dynamics often shows fundamental advantages for the fixation of cooperation. The strategy update events occur through a Poisson process at a rate of $1$. During an elementary step an agent is chosen randomly to die leaving an empty site. Then, its neighbors compete for the empty site with the probability proportional to their fitness. 

In the absence of mutation the above-introduced evolution process in a networked system gradually evolves into a pure cooperation or pure defection state. Therefore, we are interested in the fixation probabilities of the cooperation and defection, denoted by $\rho_C$ and $\rho_D$ respectively. Accordingly, $\rho_C$ ($\rho_D$) presents the probability that a single cooperator (defector) invades the pure defective (cooperative) networked system and finally prevails the whole system. 

\section{Condition for Cooperation in Particular Graph Families}\label{sec: theorems}
\small
In this section, we study the condition for $\rho_C>\rho_D$ with self-interaction in some graph families. In particular, we consider regular graphs, star graphs, and two modified stars. We first consider the following lemma that helps us to reveal the potential consequence of self-interactions~\cite{allen2017evolutionary}. 

\begin{Lemma}\label{Lemma: 1}
The fixation probability of cooperation in a network with $N$ vertices is
\begin{small}
\begin{equation}
\rho_C=\frac{1}{N}+\frac{\delta}{2N}\left[-c\eta^{(2)}+b(\eta^{(3)}-\eta^{(1)})\right]+O(\delta^2),  
\end{equation}
\end{small}
where $\eta^{(n)}$ is the expected coalescence time of all $n$-step random walks over the network. 
\end{Lemma}

The proof can be found in Appendix~\ref{Appendix: A} where we introduce some technical terms which are used in the following analysis. For the prisoner's dilemma, if the fixation probability of cooperation is higher than neutral drift, i.e., $\rho_C>1/N$, we say that the selection favors cooperation. In this case, we also have $\rho_D<1/N$, and therefore this condition is equivalent to $\rho_C-\rho_D>0$. We are interested in the critical point of the benefit-to-cost ratio $(b/c)^*_{\mathbf{\mathcal{L}}}$ for the networked system with self-interaction to favor cooperation. As Corollary 1 in~\cite{allen2017evolutionary} suggests, we have $\vert (b/c)^*\vert>1$ for any population structure with $N\geq 3$. For $1<(b/c)^*_{\mathbf{\mathcal{L}}}<+\infty$, there is a higher probability that the networked system will reach cooperation against defection if the benefit-to-cost ratio exceeds $(b/c)^*_{\mathbf{\mathcal{L}}}$. If $-\infty<(b/c)^*_{\mathbf{\mathcal{L}}}<-1$ (representing not a prisoner's dilemma), the dynamic process favors the evolution of spite instead of cooperation, where agents tend to pay a cost to harm others. The self-interaction landscape reduces the cooperation condition if $1<(b/c)^*_{\mathbf{\mathcal{L}}}<(b/c)^*$, i.e., the positive threshold with self-interaction is smaller than without self-interaction. 

We first consider the family of regular graphs, where each vertex has the same number of neighbors. 
Additionally, it is a natural assumption that each agent has the same self-interaction weight. The following theorem shows the condition for cooperation and the advantages of self-interaction. 
\begin{Theorem}\label{Theorem: 1}
For regular graphs with $N$ vertices and degree $k$, where self-interaction landscape is $\mathbf{\mathcal{L}}\triangleq [\ell(k)]^{\mathcal{V}}$, we have

(a) the critical benefit-to-cost ratio for cooperation is
\begin{small}
\begin{equation}\label{Eq: bcr regular}
(\frac{b}{c})^*_{\mathbf{\mathcal{L}}}=\frac{N\left[k^2+3k\ell(k)+2\ell^2(k)\right]-2\left[k+\ell(k)\right]^2}{N\left[k\ell(k)+k+2\ell^2(k)\right]-2\left[k+\ell(k)\right]^2},
\end{equation}
\end{small}

(b) if $N>2k$ and $k>2$, any $\mathbf{\mathcal{L}}$ can reduce the condition for cooperation, 

(c) if $N<2k$, $\mathbf{\mathcal{L}}$ can reduce the condition for cooperation with
\begin{small}
\begin{equation}\label{eq: regular spite to cooperation}
\ell(k)>\frac{N(4-k)+\sqrt{N^2k^2+8kN+16(k-N)(kN-N-k)}}{4(N-1)}, 
\end{equation}
\end{small}
otherwise, the selection favors the evolution of spite. 
\end{Theorem}

The precise proof is given in Appx.~\ref{Appendix: B}. Thm.~\ref{Theorem: 1} shows that cooperation can spread in a system distributed on a regular graph. Additionally, for a sufficiently sparse regular graph that satisfies $N>2k$ and $k>2$, the self-interaction can always reduce the critical condition for the favor of cooperation. For sparse regular graphs with $k\leq 2$, the self-interaction does not always reduce the cooperation condition. For a dense regular graph with $N<2k$, there is a positive phase transition point of $\ell(k)$ that saves the system from the evolution of spite to cooperation. 

Next we present the results obtained for star topology. Some recent works have shown that the stars with self-loops can be an amplifier of selection. Here, we quantify the effect of self-interaction in stars by coalescence random walk theory. A star with $N$ vertices is a complete bipartite graph with only one hub agent (H) and $N-1$ leaves (L). Therefore, at most one agent has a degree higher than $1$. For simplicity, we assume that the self-interaction strength is identical for each leaf agent, denoted as $\ell(1)$. For the hub agent, we have $\ell(N-1)$. In the following analysis, we simplify $p_{ij}^{(1)}$ as $p_{ij}$ and define $\alpha=\ell(1)$, $\beta=\ell(N-1)$, $\gamma=\ell(N)$, and $\epsilon=\ell(2)$ because the degrees in the following graph families can only take some of these values. 
\begin{Theorem}\label{Theorem: 2}
For a star graph with $N$ vertices and the self-interaction landscape $\mathbf{\mathcal{L}}\triangleq [\ell(N-1), \ell(1),\cdots,\ell(1)]$, define $\alpha=\ell(1)$ and $\beta=\ell(N-1)$, then we have

(a) the critical benefit-to-cost ratio for cooperation is
\begin{small}
\begin{equation}\label{eq: thm2 star}
(\frac{b}{c})^*_{\mathbf{\mathcal{L}}}=\frac{Num_{Star}}{Den_{Star}}, 
\end{equation}
\end{small}
where 
\begin{small}
\begin{equation}
\begin{aligned}
&Num_{Star}=3 (1 + \alpha) (-1 + N + \beta) \left\{ -\frac{8}{3} + N^3 \left( \frac{4}{3} + \frac{11 \alpha}{3}\right.\right.\\
&\left.+ \alpha^2 \right) + \frac{14 \beta}{3} - \frac{4 \beta^2}{3} + \alpha^2 \left( -2 + \frac{8 \beta}{3} \right) + \alpha \left( -6 + 10 \beta - \frac{8 \beta^2}{3} \right)\\
&+ N^2 \left[-\frac{16}{3} + 3 \beta + \alpha^2 \left( -4 + \frac{4 \beta}{3} \right) + \alpha \left( -\frac{40}{3} + 7 \beta \right) \right] +\\
&\left. N \left[ \frac{20}{3} + \alpha^2 (5 - 4 \beta) - \frac{23 \beta}{3} + \frac{4 \beta^2}{3} + \alpha \left( \frac{47}{3} - 17 \beta + \frac{8 \beta^2}{3} \right) \right] \right\},
\end{aligned}
\end{equation}
\end{small}
and
\begin{small}
\begin{equation}
\begin{aligned}
&Den_{Star}=2 \left\{ \beta \left[ -N^3 + N \left( -1 - \frac{\beta}{2} \right) + N^2 \left( 2 - \frac{\beta}{2} \right)\right.\right.\\
&\left. + \beta \right] + \alpha^3 \left[ 2 + N^4 - 5 \beta+ 4 \beta^2 + N^3 \left( -5 + \frac{5 \beta}{2} \right) + N \left( -7 +\right.\right.\\
&\left.\left.\frac{25 \beta}{2} - 6 \beta^2 \right) + N^2 \left( 9 - 10 \beta + 2 \beta^2 \right) \right] + \alpha \left[ 4 + N^4 - 11 \beta + 12 \beta^2\right.\\
&- 2 \beta^3 + N^3 \left( -7 + \frac{5 \beta}{2} \right) + N^2 \left( 15 - 16 \beta + 5 \beta^2 \right) + N \left( -13 + \right.\\
&\left.\left.\frac{49 \beta}{2} - 17 \beta^2 + 2 \beta^3 \right) \right] + \alpha^2 \left[ 8 + 4 N^4 - 22 \beta + 21 \beta^2 - 4 \beta^3 + \right.\\
&\left.N^3 \left( -20 + 12 \beta \right) + N^2 \left( 36 - 46 \beta + \frac{27 \beta^2}{2} \right) +\right.\\
&\left.\left.N \left( -28 + 56 \beta - \frac{69 \beta^2}{2} + 4 \beta^3 \right) \right] \right\}.
\end{aligned}
\end{equation}
\end{small}

(b) if self-interaction only applies to the hub agent (i.e., $\alpha=0$), the evolution always favors the spite instead of the cooperation if $N>2$, and $(\frac{b}{c})^*_{\mathbf{\mathcal{L}\vert\alpha=0}}\rightarrow -\infty$ with $N\rightarrow+\infty$, 

(c) if the self-interaction only applies to leaf vertices (i.e., $\beta=0$), there exists a positive critical threshold for the system to favor cooperation if $N>3$, and with $N\rightarrow\infty$,
\begin{small}
\begin{equation}\label{eq: star limit beta0}
(\frac{b}{c})^*_{\mathbf{\mathcal{L}\vert\beta=0}}\rightarrow\frac{3\alpha^3+14\alpha^2+15\alpha+4}{2\alpha^3+8\alpha^2+2\alpha}. 
\end{equation}
\end{small}
\end{Theorem}

We present the proof for Thm.~\ref{Theorem: 2} in Appx.~\ref{Appendix: C}. For the exceptional case of Thm.~\ref{Theorem: 2} (c) ($N=3$), we can also find a phase transition point $\alpha>-2+\sqrt{(3N-4)/(N-2)}=\sqrt{5}-2\approx0.236$ to save the system from the evolution of spite. 

In previous studies, star topology has been shown to be disadvantageous to the emergence of cooperation. Accordingly, the critical threshold for the fixation of cooperation is $\infty$ if we do not consider self-loop~\cite{allen2017evolutionary}. This is because of the equivalence between one-step and three-step expected coalescence time. Our proposed self-interaction landscape can buffer the random walk, as well as the replacement of strategy. If three-step expected coalescence time is greater than one-step, i.e., $(N-1)\pi_L\eta_L(p_{LL}+p_{LL}^{(2)})+\pi_H\eta_H(p_{HH}+p_{HH}^{(2)})-2>0$, the condition for cooperation is positive. Thm.~\ref{Theorem: 2} highlights that a proper self-interaction landscape can enhance the fixation of cooperation in stars. But the self-interaction of the hub agent is very harmful because the strategy replacement of the leader itself is likely to induce the defection of the entire system.  However, the self-interaction of the leaves is crucial and helpful for the system to overcome the defection. 

Additionally, researchers often consider the graph surgery to promote cooperation in stars. Here, we study the effect of self-interaction in two modified stars. The first one is a graph containing two stars joined by their hubs. The second one is a ceiling fan where each leaf has one edge to another leaf. In the following, we consider a joined star topology composed of two connected star graphs via the hub vertices. 
\begin{Theorem}\label{Theorem: 3}
For a graph that two stars with $N$ vertices are joined by hubs ($2N$ vertices in total), with $\mathbf{\mathcal{L}}\triangleq [\ell(N), \ell(N), \ell(1),\cdots,\ell(1)]$, and denote $\alpha=\ell(1)$ and $\gamma=\ell(N)$, then we have 

(a) the critical benefit-to-cost ratio for cooperation is
\begin{small}
\begin{equation}\label{eq: thm3 starhh}
(\frac{b}{c})^*_{\mathbf{\mathcal{L}}}=\frac{Num_{StarHH}}{Den_{StarHH}}, 
\end{equation}
\end{small}
where
\begin{small}
\begin{equation}
\begin{aligned}
&Num_{StarHH}=3(1+\alpha)(N+\gamma)\left\{N^4(\alpha^3+\frac{16}{3}\alpha^2+\frac{25}{3}\alpha+\frac{10}{3})\right.\\
&+N^3\left[\alpha^3(\frac{4}{3}\gamma-2)+\alpha^2(10\gamma-\frac{25}{3})+p(20\gamma-\frac{34}{3})+\frac{29}{3}\gamma-\frac{13}{3}\right]\\
&+N^2\left[\alpha^3(-\frac{7}{3}\gamma+\frac{1}{3})+\alpha^2(\frac{11}{3}\gamma^2-\frac{38}{3}\gamma+1)+\alpha(\frac{40}{3}\gamma^2-\frac{64}{3}\gamma\right.\\
&+1)\left.+\frac{26}{3}\gamma^2-9\gamma+\frac{1}{3}\right]+N\left[\alpha^3(-\frac{1}{3}\gamma+\frac{2}{3})+\alpha^2(-\frac{10}{3}\gamma^2-\gamma\right.\\
&+2)\left.+\alpha(\frac{8}{3}\gamma^3-\frac{32}{3}\gamma^2-\gamma+2)+3\gamma^3-\frac{16}{3}\gamma^2-\frac{1}{3}\gamma+\frac{2}{3}\right]+\\
&\left.\gamma\left[\frac{4}{3}\alpha^3+4\alpha^2+\alpha(-\frac{5}{3}\gamma^2+4)+\frac{1}{3}\gamma^3-\gamma^2+\frac{4}{3}\right]\right\}\\
\end{aligned}
\end{equation}
\end{small}
and
\begin{small}
\begin{equation}
\begin{aligned}
&Den_{StarHH}=2\left\{N^5(\alpha^4+\frac{11}{2}\alpha^3+9\alpha^2+5\alpha+2)+N^4\left[\alpha^4(\frac{5}{2}\gamma-\right.\right.\\
&\left.\frac{5}{2})+\alpha^3(\frac{33}{2}\gamma-\frac{23}{2})+\alpha^2(\frac{61}{2}\gamma-19)+\alpha(\frac{33}{2}\gamma-\frac{25}{2})+6\gamma-4\right]\\
&+N^3\left[\alpha^4(2\gamma^2-\frac{11}{2}\gamma+2)+\alpha^3(\frac{37}{2}\gamma^2-29\gamma+\frac{15}{2})+\alpha^2(44\gamma^2\right.\\
&\left.-\frac{107}{2}\alpha+11)+\alpha(29\gamma^2-36\gamma+7)+\frac{21}{2}\gamma^2-12\gamma+\frac{3}{2}\right]+\\
&N^2\left[\alpha^4(-\frac{7}{2}\gamma^2+2\gamma-\frac{1}{2})+\alpha^3\gamma(\frac{11}{2}\gamma^2-\frac{49}{2}\gamma+\frac{15}{2})+\alpha^2(\frac{47}{2}\gamma^3\right.\\
&\left.-\frac{111}{2}\gamma^2+\frac{23}{2}\gamma+\frac{7}{2})+\alpha(21\gamma^3-\frac{79}{2}\gamma^2+7\gamma+5)+9\gamma^3-14\gamma^2\right.\\
&\left.+\gamma+2\right]+N\left[\alpha^4(-\frac{1}{2}\gamma^2+\gamma-1)+\alpha^3(-5\gamma^3-2\gamma^2+6\gamma-\frac{11}{2})\right.\\
&\left.+\alpha^2(4\gamma^4-\frac{41}{2}\gamma^3-\frac{5}{2}\gamma^2+13\gamma-\frac{21}{2})+\alpha(6\gamma^4-16\gamma^3-\frac{5}{2}\gamma^2\right.\\
&\left.+12\gamma-\frac{17}{2})+\frac{7}{2}\gamma^4-\frac{13}{2}\gamma^3-\frac{3}{2}\gamma^2+4\gamma-\frac{5}{2}\right]+\alpha^4(2\gamma^2+1)+\\
&\left.\alpha^3(8\gamma^2-\frac{3}{2}\gamma+4)+\alpha^2(-\frac{5}{2}\gamma^4+\frac{25}{2}\gamma^2-\frac{9}{2}\gamma+6)+\alpha(\frac{1}{2}\gamma^5-2\gamma^4\right.\\
&\left.-\frac{1}{2}\gamma^3+9\gamma^2-\frac{9}{2}\gamma+4)+\frac{1}{2}\gamma^5-\gamma^4-\frac{1}{2}\gamma^3+\frac{5}{2}\gamma^2-\frac{3}{2}\gamma+1\right\}. 
\end{aligned}
\end{equation}
\end{small}

(b) if $N>2$, we have $(\frac{b}{c})^*_{\mathbf{\mathcal{L}}}>0$ that there exists a positive benefit-to-cost ratio for the system to favor cooperation regardless of $\alpha$ and $\gamma$. The system never favors the evolution of spite. 

(c) if $N\rightarrow+\infty$, the limit condition for cooperation is independent of $\gamma$ and presented as
\begin{small}
\begin{equation}
(\frac{b}{c})^*_{\mathbf{\mathcal{L}\vert N\rightarrow+\infty}}=\frac{3\alpha^4+19\alpha^3+41\alpha^2+35\alpha+10}{2\alpha^4+11\alpha^3+18\alpha^2+10\alpha+4}. 
\end{equation}
\end{small}
\end{Theorem}

The proof can be found in Appx.~\ref{Appendix: D}. Thm.~\ref{Theorem: 3} shows that the hub-joined star system can favor cooperation if $N>2$ with self-interaction. There is a positive critical threshold for the dominance of cooperation, which is not a strict condition. Therefore, a graph with two stars connected by hubs is an excellent structure for the evolutionary dynamics considering self-interaction. Additionally, the role of self-interaction in leaf vertices is much more prominent than that of the hub vertices with the increase of $N$. 

Next, we consider the so-called ceiling fan (CF) where each leaf of a star is only connected to one another by an edge. In this case, the system size $N$ should be an odd number. 
\begin{Theorem}\label{Theorem: 4}
For a ceiling fan with $N$ vertices and $\mathbf{\mathcal{L}}\triangleq [\ell(N-1), \ell(1),\cdots,\ell(1)]$, and denote $\epsilon=\ell(2)$ and $\beta=\ell(N-1)$, then we have 

(a) the critical benefit-to-cost ratio for cooperation is
\begin{small}
\begin{equation}\label{eq: thm4 cf}
(\frac{b}{c})^*_{\mathbf{\mathcal{L}}}=\frac{Num_{CF}}{Den_{CF}}, 
\end{equation}
\end{small}
\begin{small}
\begin{equation}
\begin{aligned}
&Num_{CF}=(2 + \epsilon) (-1 + N + \beta) \left\{ 78 + 4 N^2 (9 + 11 \epsilon + 2 \epsilon^2) + \right.\\
&\left.\epsilon^2 (17 - 22 \beta) - 122 \beta + 24 \beta^2 + \epsilon (83 - 125 \beta + 22 \beta^2) + \right.\\
&\left.N \left[-114 + 68 \beta + 5 \epsilon^2 (-5 + 2 \beta) + \epsilon (-127 + 74 \beta)\right] \right\},
\end{aligned}
\end{equation}
\end{small}
\begin{small}
\begin{equation}
\begin{aligned}
&Den_{CF}=-57 + N^3 (9 + 25 \epsilon + 34 \epsilon^2 + 6 \epsilon^3) + 157 \beta - 155 \beta^2 + \\
& 15 \beta^3 +\epsilon^3 (-12 + 29 \beta - 22 \beta^2) + \epsilon^2 (-76 + 201 \beta - 177 \beta^2 + 22 \beta^3)\\
& + \epsilon (-109 + 294 \beta - 273 \beta^2 + 28 \beta^3) + N^2 \left[ 25 (-3 + \beta) + 48 \epsilon^2 (-3\right.\\
&\left. + 2 \beta)+ 2 \epsilon^3 (-12 + 7 \beta) + 3 \epsilon (-53 + 26 \beta) \right] + N \left[ 123 - 182 \beta + \right.\\
&\left.47 \beta^2 + \epsilon^3 (30 - 43 \beta + 10 \beta^2) + 3 \epsilon^2 (62 - 99 \beta + 32 \beta^2) \right.\\
&\left.+ 3 \epsilon (81 - 124 \beta + 35 \beta^2) \right].
\end{aligned}
\end{equation}
\end{small}

(b) if $N\geq7$, there exists a positive benefit-to-cost threshold for the system to favor cooperation regardless of $\epsilon$ and $\beta$, and the system never favors spite, 

(c) with $N\rightarrow+\infty$, the limit condition for cooperation is independent of $\beta$ as follows
\begin{small}
\begin{equation}
(\frac{b}{c})^*_{\mathbf{\mathcal{L}\vert N\rightarrow+\infty}}=\frac{8\epsilon^3+60\epsilon^2+124\epsilon+72}{6\epsilon^3+34\epsilon^2+25\epsilon+9}. 
\end{equation}
\end{small}
\end{Theorem}

We present the detailed discussion of the proof in Appx.~\ref{Appendix: E}. There are two system sizes not included in Thm.~\ref{Theorem: 4} (b), which are $N=3$ and $N=5$. In these two cases, the evolutionary dynamics can favor the evolution of spite instead of cooperation, depending on the ranges of $\epsilon$ and $\beta$. Based on Eq.~\ref{eq: thm4 cf}, we can directly analyze these two cases. For $N=3$, the boundary point for $\epsilon$ (denoted as $\epsilon^*_{N=3}$) is the third root of $3\epsilon^3+13\epsilon^2-17\epsilon-15=0$ ($\epsilon^*_{N=3}\approx1.53$), if we arrange the roots from small to large. If $0\leq\epsilon<\epsilon^*_{N=3}$, the system can favor cooperation if $\beta>\beta^*_{N=3}$, where $\beta^*_{N=3}$ is the third root of $(22\epsilon^2+28\epsilon+15)\beta^3+(8\epsilon^3+111\epsilon^2+42\epsilon-14)\beta^2+(26\epsilon^3+174\epsilon^2-120\epsilon-164)\beta+24\epsilon^3+104\epsilon^2-136\epsilon-120=0$. If $\epsilon=\epsilon^*_{N=3}$, $\beta>0$ can make the system favor the fixation of cooperation. If $\epsilon>\epsilon^*_{N=3}$, the system is possible to favor cooperation if $\beta\geq0$. For $N=5$, the boundary point for $\epsilon$ (denoted as $\epsilon^*_{N=5}$) is the third root of $9\epsilon^3+47\epsilon^2+8\epsilon-6=0$ ($\epsilon^*_{N=3}\approx0.277$). If $0\leq\epsilon<\epsilon^*_{N=5}$, $\beta$ has to be greater than the third root of $(22\epsilon^2+28\epsilon+15)\beta^3+(28\epsilon^3+303\epsilon^2+252\epsilon+80)\beta^2+(164\epsilon^3+1116\epsilon^2+384\epsilon-128)\beta+288\epsilon^3+1504\epsilon^2+256\epsilon-192=0$. If $\epsilon=\epsilon^*_{N=5}$ and $\beta>0$, or $\epsilon>\epsilon^*_{N=5}$ and $\beta\geq0$, the system can favor cooperation instead of defection. 

This ends our analysis of the condition for cooperation in particular graph families. In Thm.~\ref{Theorem: 1}, we show that the self-interaction learning strategy can reduce the condition for cooperation in sparse regular graphs. In dense regular graphs, the self-interaction can save the system from the evolution of spite. In Thms. \ref{Theorem: 2}-\ref{Theorem: 4}, we show that in the star topology and the modified stars, as the increase of system sizes, the self-interaction of leaf vertices becomes crucial in the fixation of cooperation. 
\section{Simulations and Analysis}\label{sec: simulation}
\small
In this section, we perform simulations of the evolutionary dynamics to verify and further analyze the effects of self-loop learning in networked systems. 
\subsection{Methods}
\small
We perform our simulation in regular networks, stars, hub-hub joined stars, ceiling fans, several types of random networks, and real networks. We design the self-interaction landscape $\mathbf{\mathcal{L}}$ in two aspects. For regular networks, random networks, and real networks, we design continuous functions that are related to the network degree, including $(k+1)^{-1}$, $e^{-k}$, $\ln{k}$, and $1-k^{-1}$, e.g., $\ell_i(k_i)=\ln{k_i}$ is the self-interaction strength of the agent $i$ if we consider the third function. For stars, hub-hub joined stars, and ceiling fans, we directly assign real numbers as the self-interaction strength of each agent. For the game parameters, we fix the cost $c=1$. For the simulations of fixation probabilities, we show $N\times(\rho_C-\rho_D)$ instead of $\rho_C$ for more convincing results, as $\rho_C>\rho_D$ indicates the advantage of cooperation. The fitting lines of fixation probabilities are obtained by linear regression. All following simulations are carried out by Python 3.9 based on the \textit{networkx} package. 
\subsection{Regular Networks}
\begin{figure}
    \centering
    \subfigure[Sparse Regular Graphs]{\includegraphics[width=0.5\linewidth]{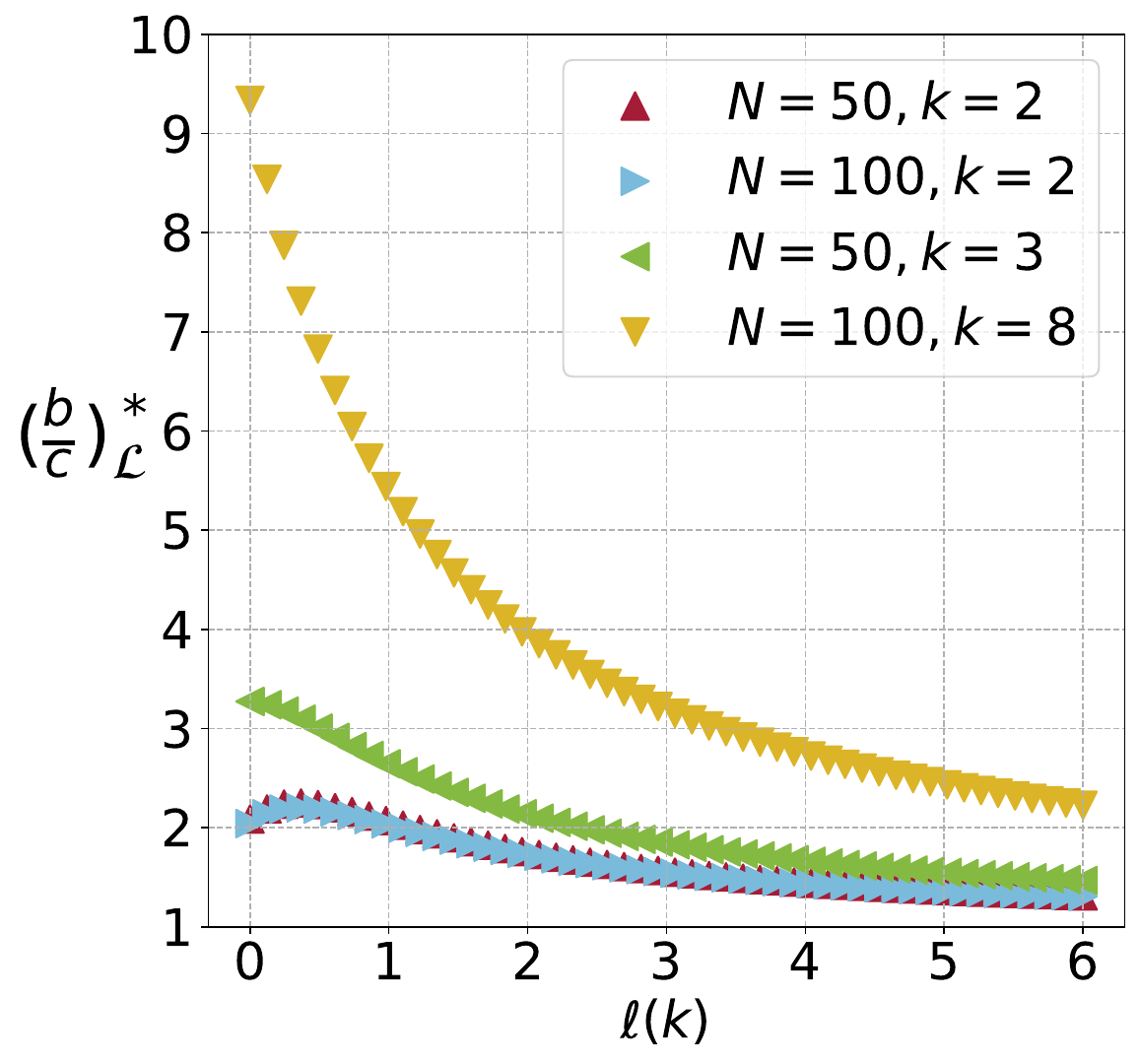}}
    \hspace{-3mm}
    \subfigure[Dense Regular Graphs]{\includegraphics[width=0.49\linewidth]{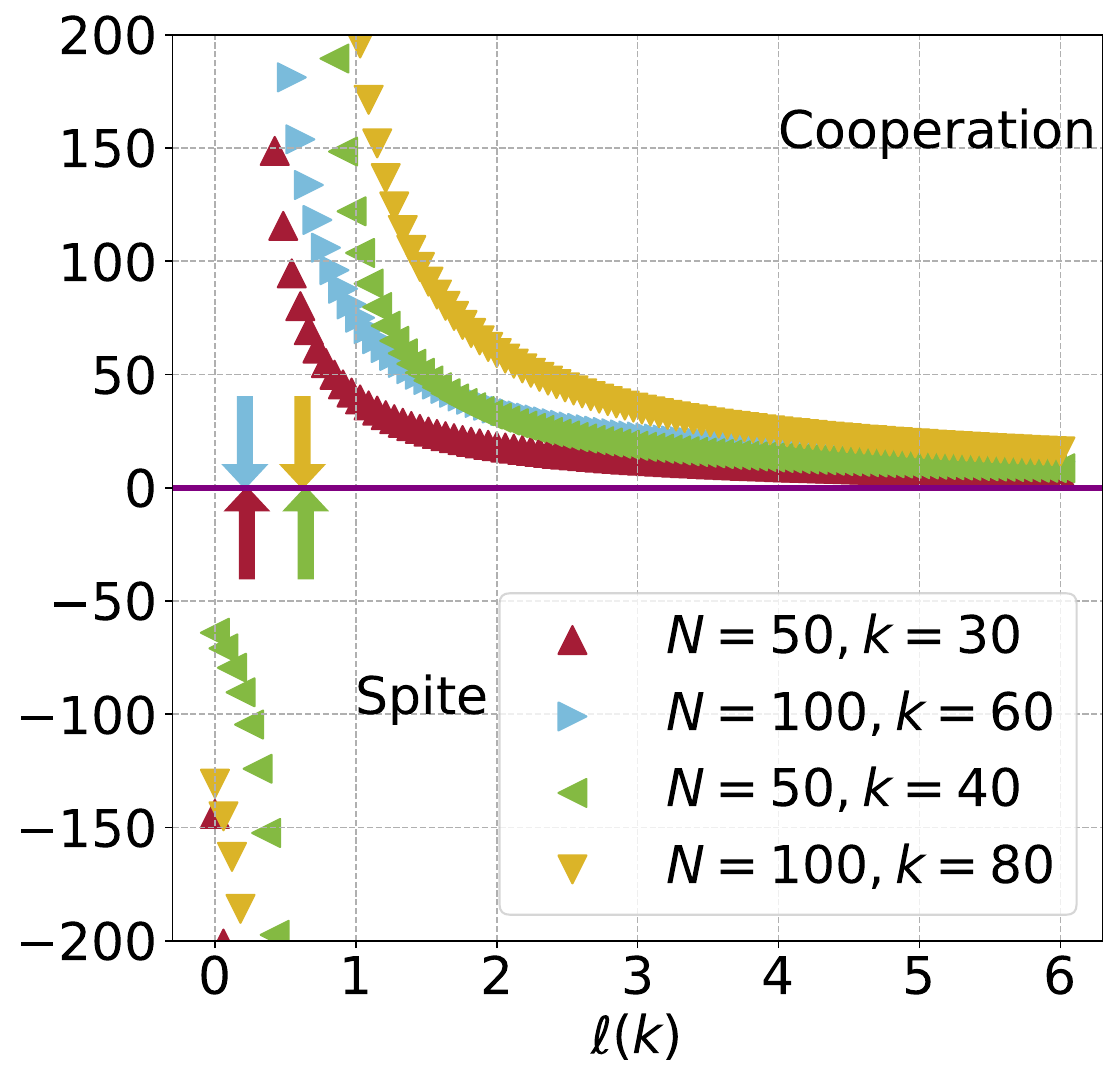}}
    \vspace{-3mm}

    \subfigure[$N=50$]{\includegraphics[width=0.49\linewidth]{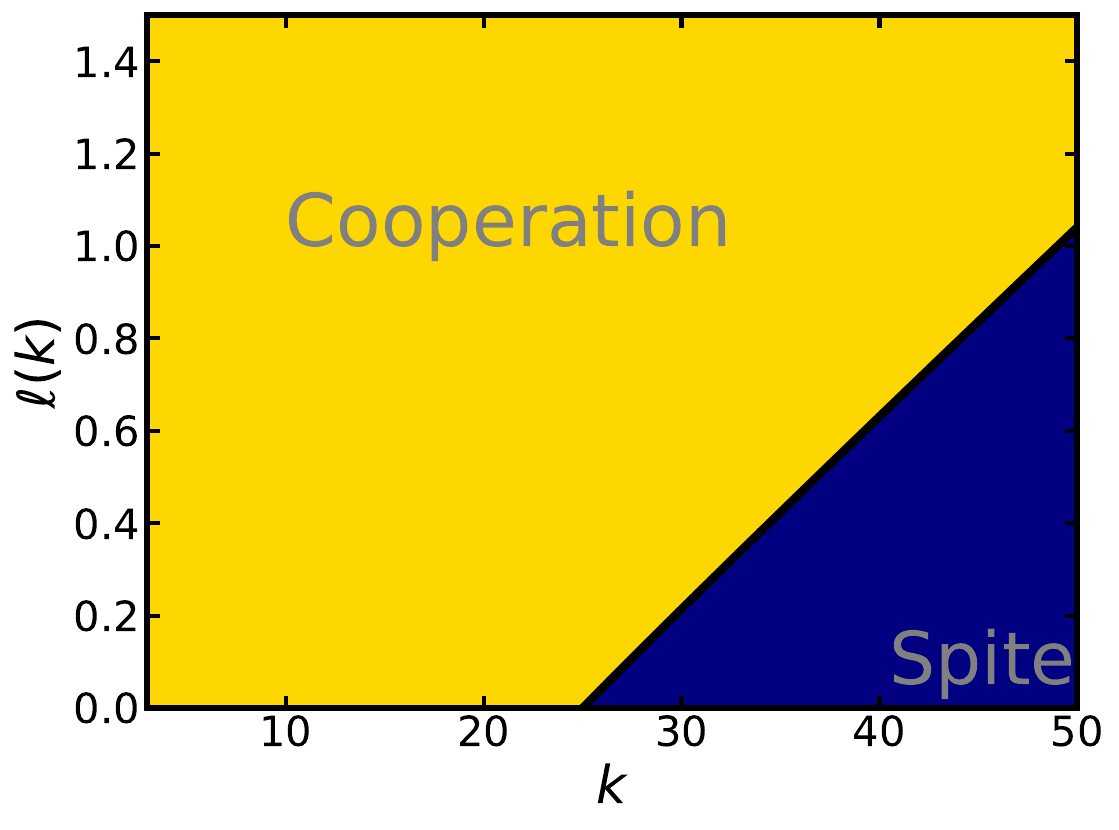}}
    \hspace{-3mm}
    \subfigure[$N=70$]{\includegraphics[width=0.49\linewidth]{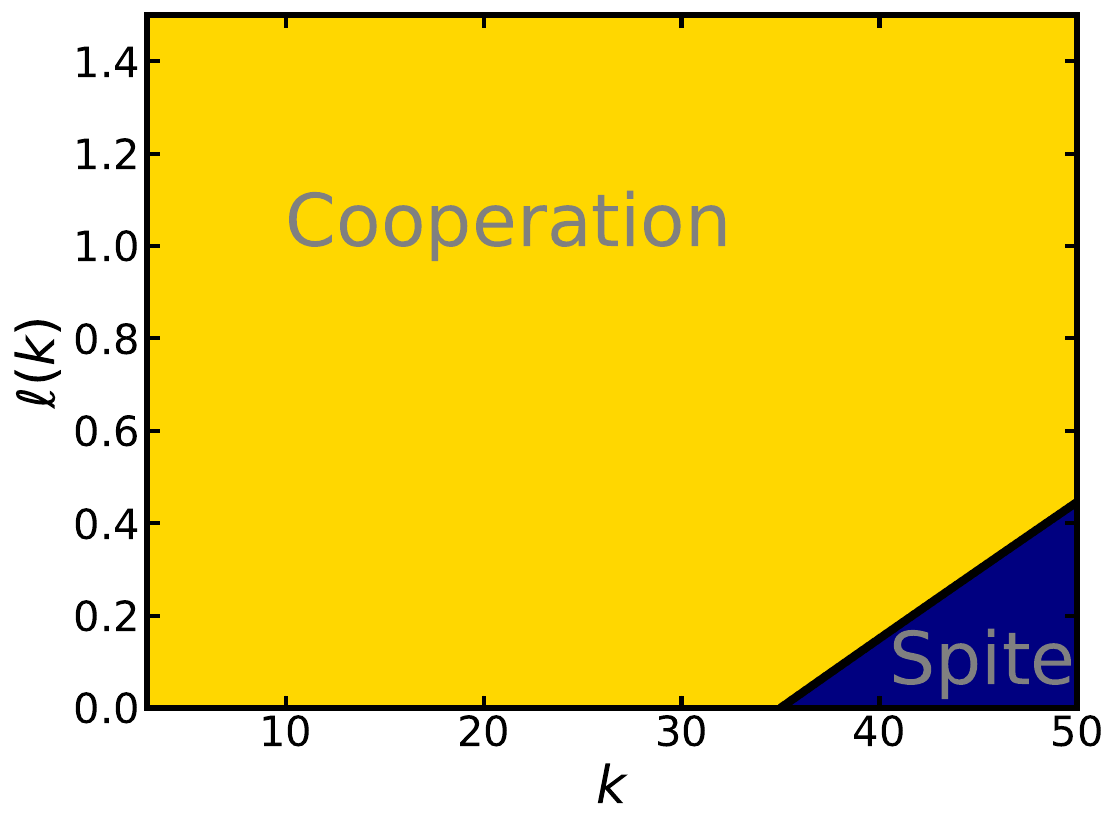}}
    \vspace{-3mm}
    \caption{\textbf{Critical thresholds for cooperation in sparse and dense regular graphs with self-interaction. }(a) For sparse regular graphs with $N>2k$ cooperation can always be favored. The network parameters are $(N,k)\in\left\{(50,2),(100,2),(50,3),(100,8)\right\}$. (b) For sense regular graphs with $N<2k$ spite is always favored to win over cooperation. The network parameters are $(N,k)\in\left[(50,30),(100,60),(50,40),(100,80)\right]$. Each transition point of $\ell(k)$ from spite to cooperation is marked with arrows by the corresponding color of the legend. (c) and (d) Phase diagrams of cooperation and spite with $N=50$ and $N=70$, respectively. The yellow and blue areas indicate cooperation and spite respectively. (color online)}
    \label{fig: regular equation}
\end{figure}
\begin{figure}
    \centering
    \subfigure[Examples of Lattices]{\includegraphics[width=0.45\linewidth]{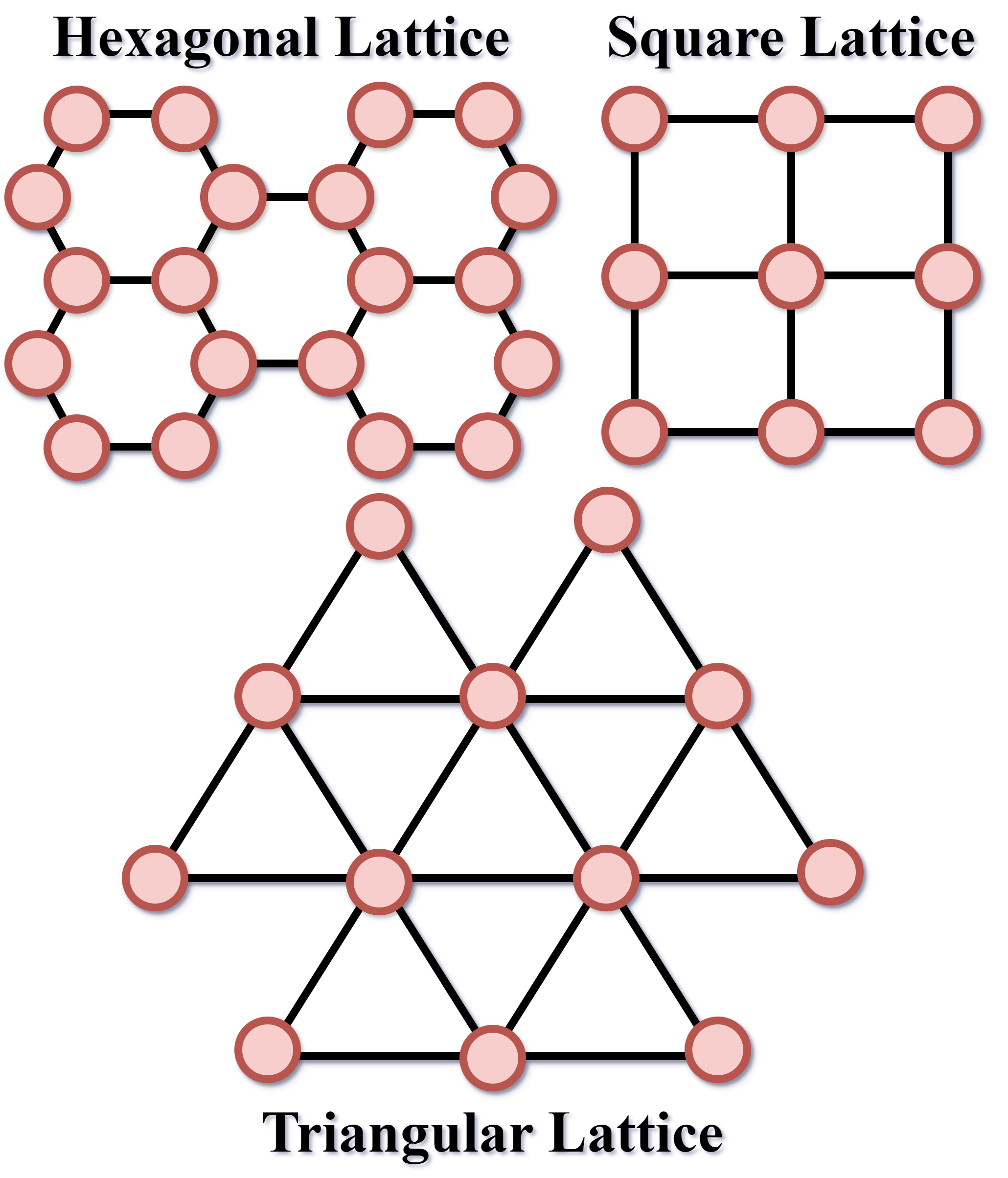}}
    \subfigure[Hexagonal Lattice, $k=3$]{\includegraphics[width=0.53\linewidth]{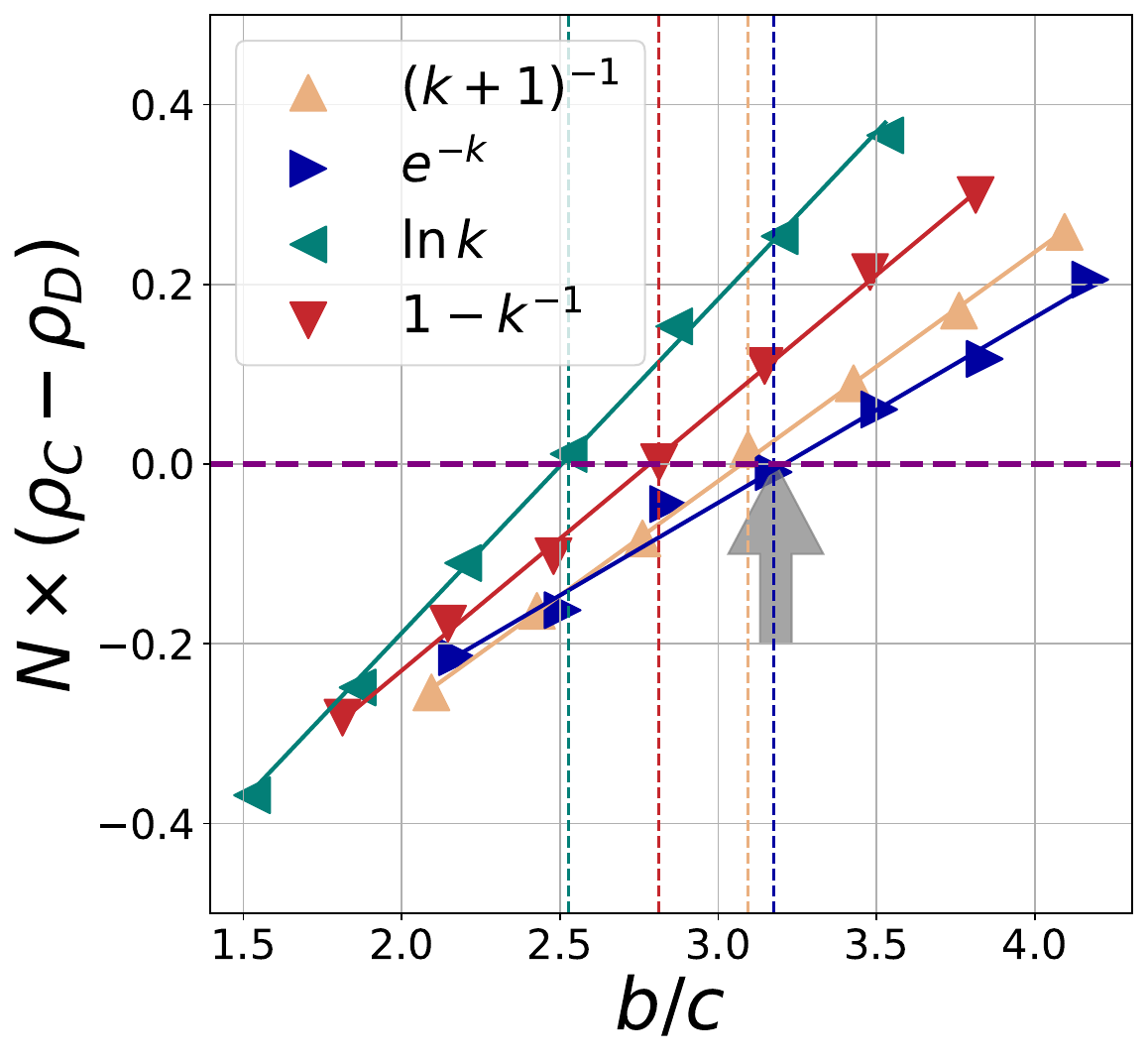}}
    
    \subfigure[Square Lattice, $k=4$]
    {\includegraphics[width=0.51\linewidth]{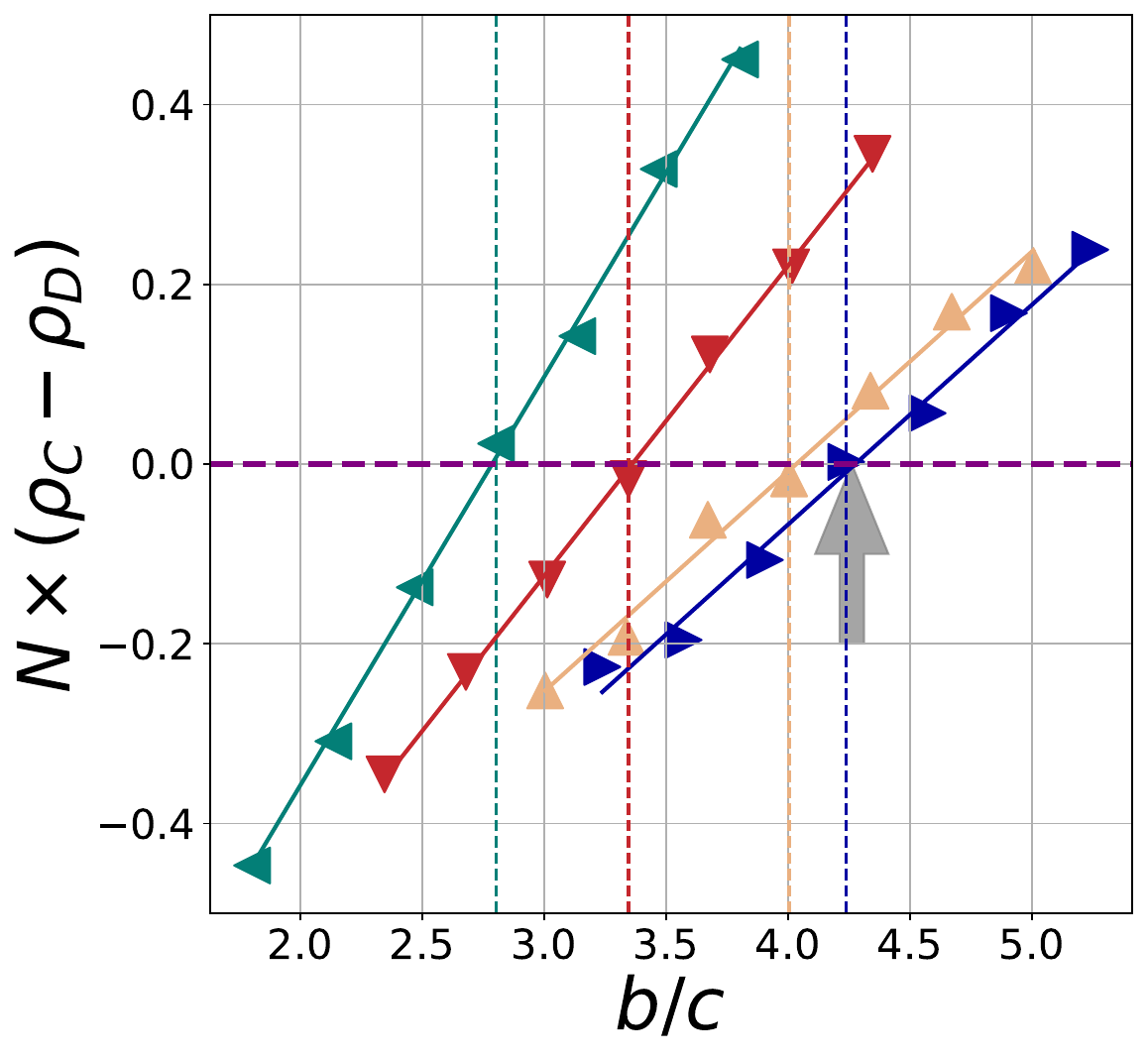}}
    \subfigure[Triangular Lattice, $k=6$]{\includegraphics[width=0.47\linewidth]{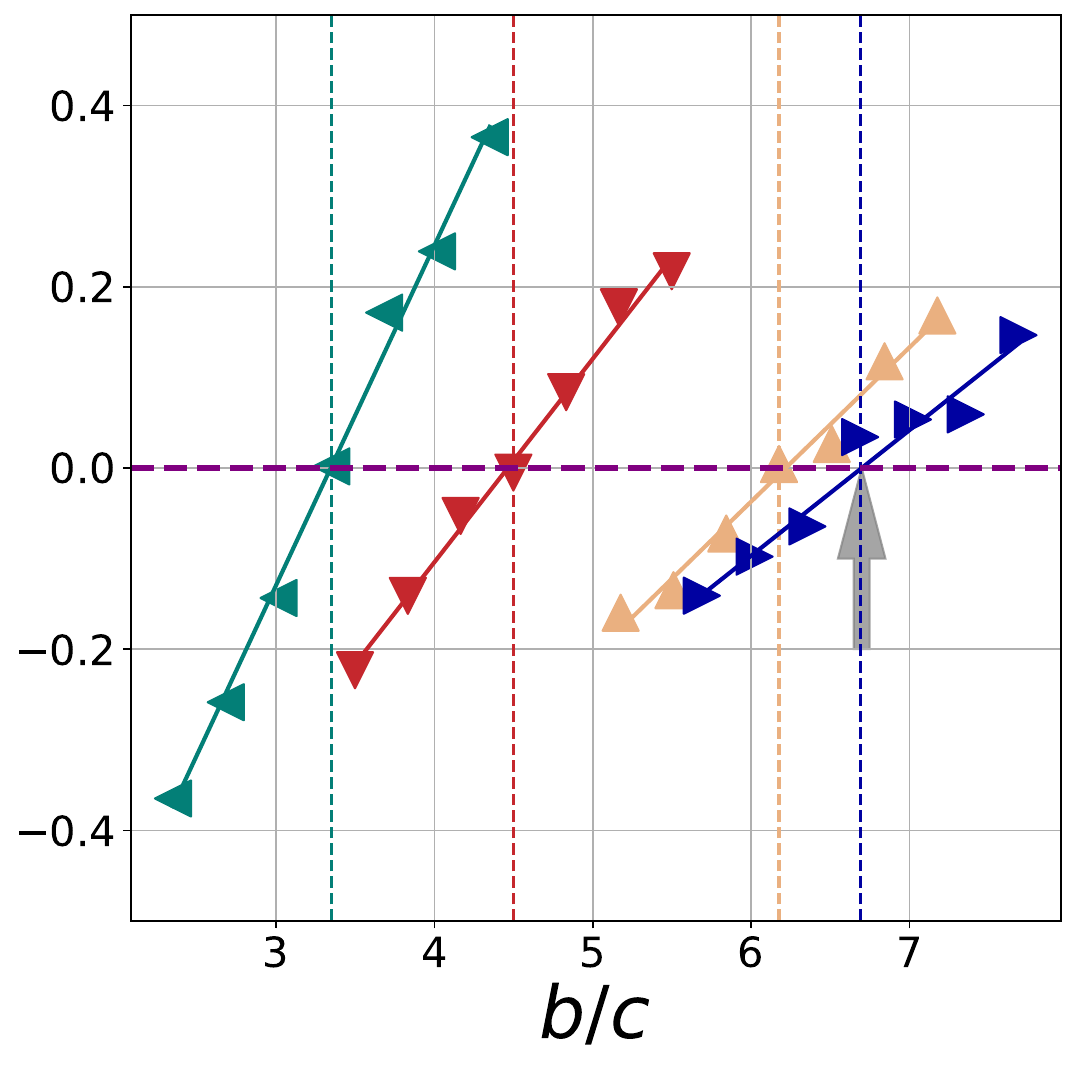}}
    \caption{\textbf{Fixation probabilities and conditions for cooperation in lattices with specific self-interaction function. }(a) Topology structures of the hexagonal lattice ($k=3$), the square lattice ($k=4$), and the triangular lattice ($k=6$). The periodic boundaries are not shown here but are considered in the simulations. (b)-(d) Fixation probabilities $N\times(\rho_C-\rho_D)$ as the increase of $b/c$ and the variation of self-loop strength functions. Here we expand $\rho_C-\rho_D$ by $N$ times to unify the scale of the vertical axis. The system sizes for (b), (c), and (d) are $72$, $100$, and $98$ respectively. Each data point marks the ratio of cooperation fixation over $2\times10^6$ independent runs, fitted by linear regression. The vertical lines indicate the theoretical results presented as Eq.~\ref{Eq: bcr regular}. The grey arrow in each panel is the condition for cooperation without self-interaction, which is $(N-2)/(N/k-2)$. (color online)}
    \label{fig: regular fixation probability}
\end{figure}
We first consider the regular network family, where each agent has the same degree $k$. The condition for cooperation without self-interaction is $(N-2)/(N/k-2)$ in unweighted networks. The boundary point for the sparse and dense structure is $N=2k$. Therefore, in Fig.~\ref{fig: regular equation}, we show the critical condition $(\frac{b}{c})^*_{\mathbf{\mathcal{L}}}$ that we have mentioned in Thm.~\ref{Theorem: 1} in sparse and dense regular networks as the increase of $\ell(k)$. In Fig.~\ref{fig: regular equation}(a), we present the critical threshold in sparse regular networks with $N>2k$, where the evolutionary dynamics always favor cooperation. In Thm.~\ref{Theorem: 1}(b), we show that the critical condition for cooperation can be reduced by self-interaction if $k>2$. In our examples, we can see that when $k=3$ or $k=8$, any self-interaction strength can reduce the condition for cooperation compared to the baseline ($\ell(k)=0$). However, if $k=2$, the cooperation condition first increases and then decreases as the growth of $\ell(k)$, i.e., regular graphs with $k=2$ may not benefit from the self-interaction. 

In Fig.~\ref{fig: regular equation}(b), we further discuss the dense regular graph with $N<2k$. The evolutionary dynamics can favor cooperation with sufficiently large $\ell(k)$, while the evolution of spite is favored if $\ell(k)$ is small. Therefore, in dense regular graphs, appropriately large self-interaction strength can save the system from spite and lead to cooperation. The quantitative condition for the transition is given in Thm.~\ref{Theorem: 1}(c) as Eq.~\ref{eq: regular spite to cooperation}. In Fig.~\ref{fig: regular equation}(b), we also mark this condition with arrows with the same colors as the curves. 

In Figs.~\ref{fig: regular equation}(c)-(d), we further show two phase diagrams with $N=50$ and $N=70$ to give examples of the phase transition from spite to cooperation. Evidently, the curves separating cooperation and spite areas are linear, as demonstrated in Thm.~\ref{Theorem: 1}(c) and Eq.~\ref{eq: regular spite to cooperation}. With an increase in the degree of the network, we find that the self-interaction strength should also be enlarged to guarantee the favor of cooperation. However, if the network degree is smaller than $N/2$, cooperation can be favored regardless of the strength of self-interaction. 

We present three regular graphs to show the fixation probability, including the hexagonal lattice with $k=3$, the square lattice with $k=4$, and the triangular lattice with $k=6$, each of which has periodic boundary conditions. In Fig.~\ref{fig: regular fixation probability}(a), we illustrate the structures of these three lattices. 
In Figs.~\ref{fig: regular fixation probability}(b)-(c), we show the results of $N(\rho_C-\rho_D)$ with several self-interaction functions. The vertical colored lines indicate the theoretical results in Eq.~\ref{Eq: bcr regular}, and the grey arrows present the threshold condition without self-interaction. Apparently, the critical condition for cooperation is reduced for all four self-interaction strength functions, among which $\ell(k)=\ln{k}$ is the most helpful for the fixation of cooperation. Additionally, $\ell(k)=e^{-k}$ results in the weakest positive consequence due to self-interaction. In this case, a vertex holds its origin strategy with a very small probability, hence the reduction of cooperation condition is the smallest. 
\subsection{Stars}
\begin{figure}
    \centering
    \subfigure[Favor of Spite, $\alpha=0$]{\includegraphics[width=0.54\linewidth]{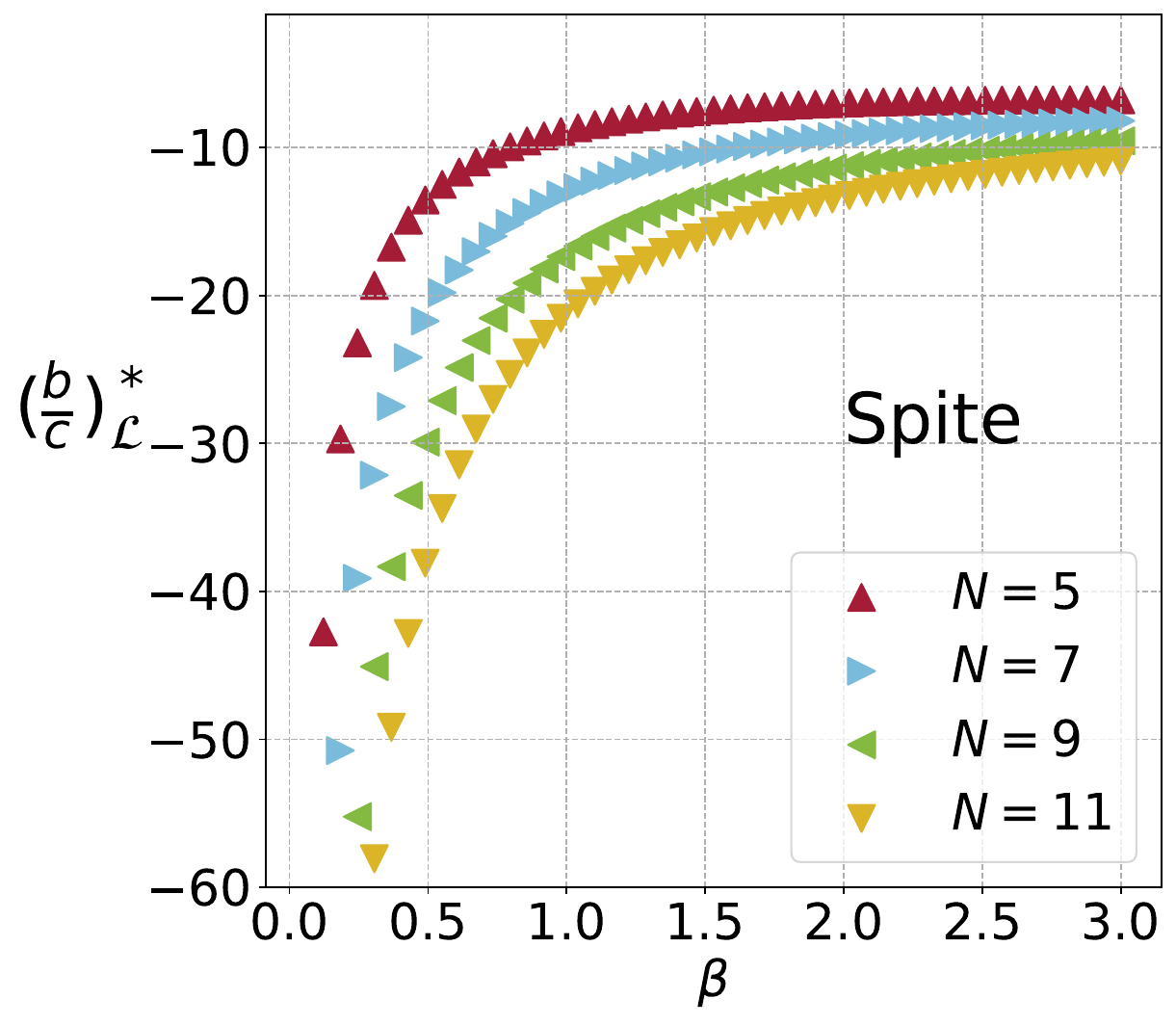}}
    \hspace{-3mm}
    \subfigure[Favor of Cooperation, $\beta=0$]{\includegraphics[width=0.46\linewidth]{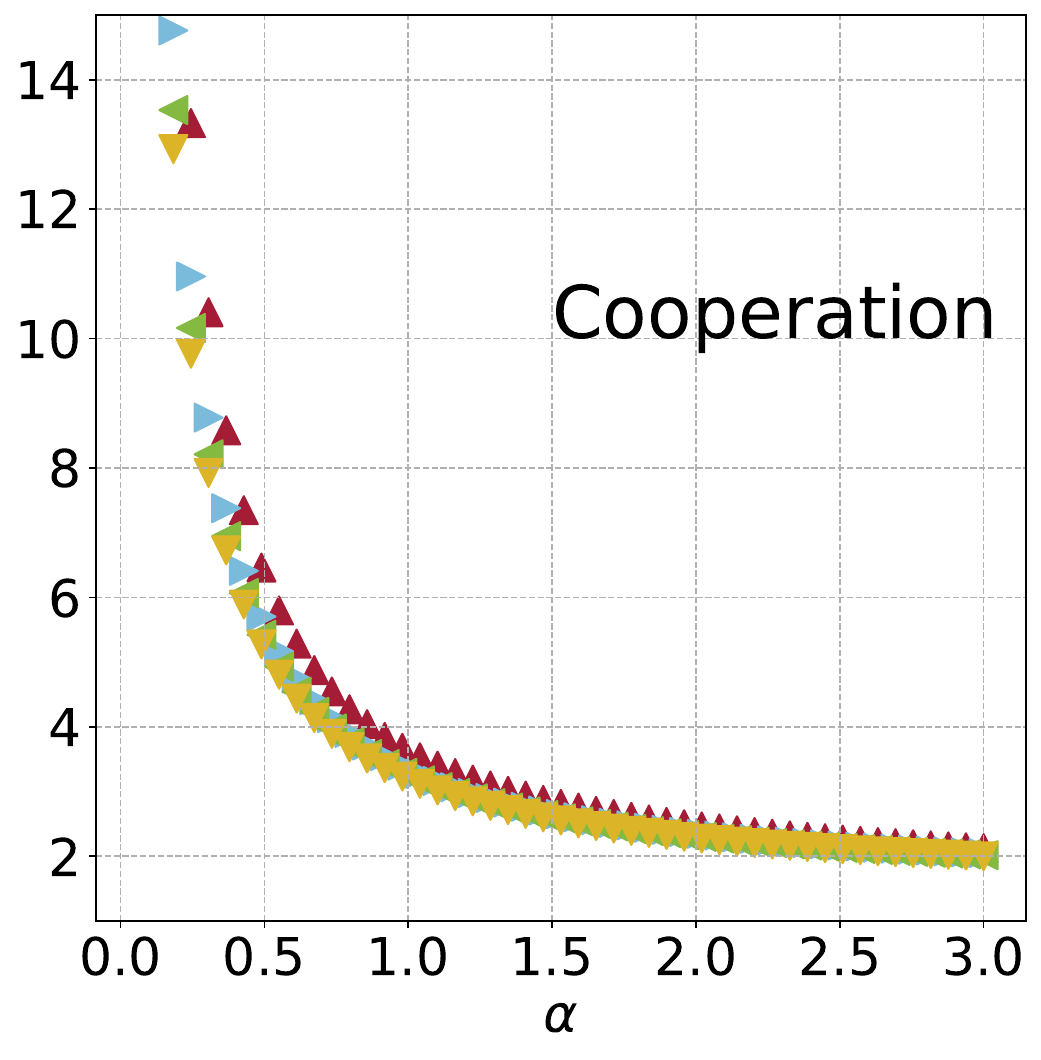}}
    \vspace{-3mm}

    \subfigure[$N=15$]{\includegraphics[width=0.49\linewidth]{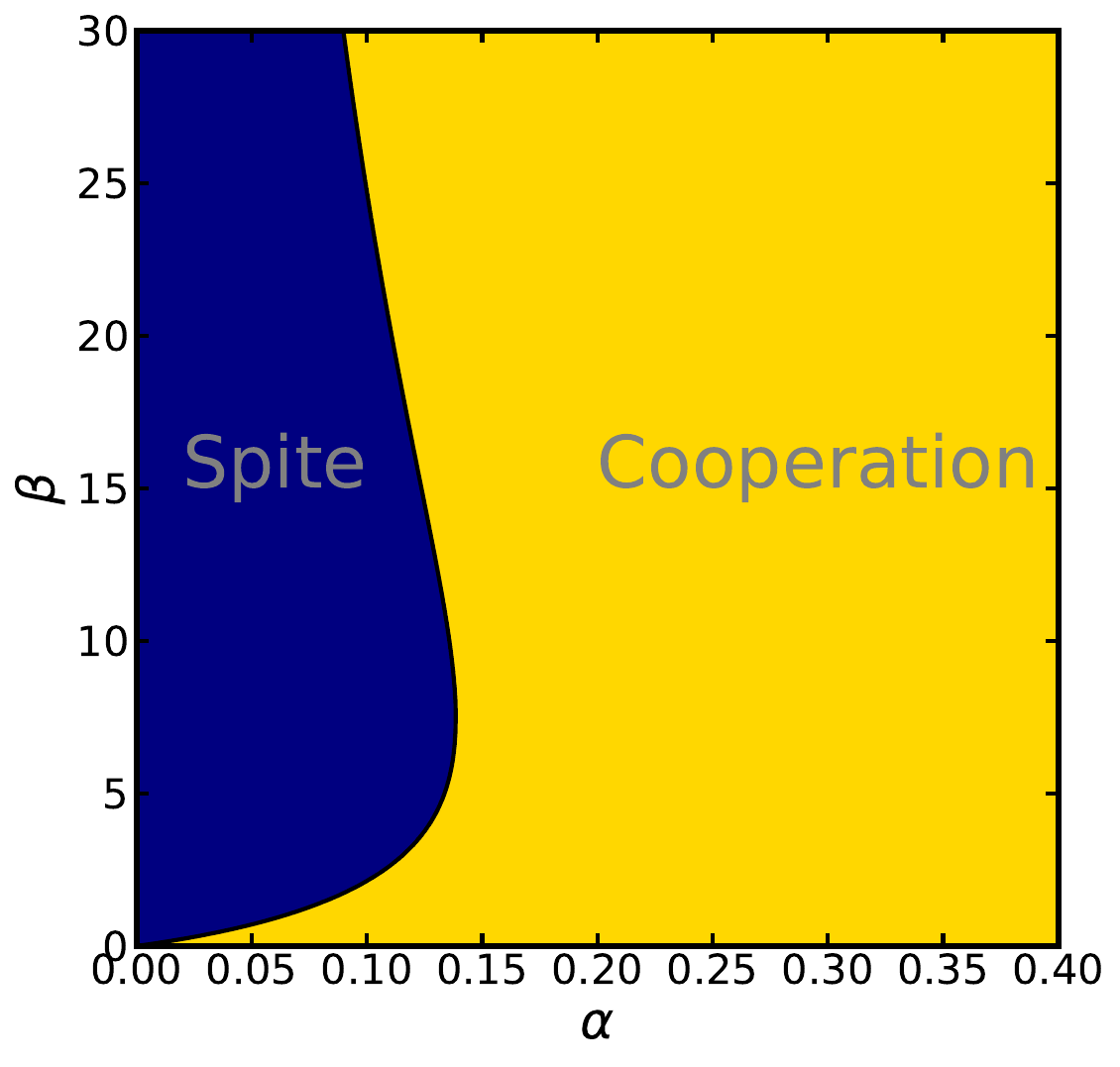}}
    \hspace{-3mm}
    \subfigure[$N=25$]{\includegraphics[width=0.49\linewidth]{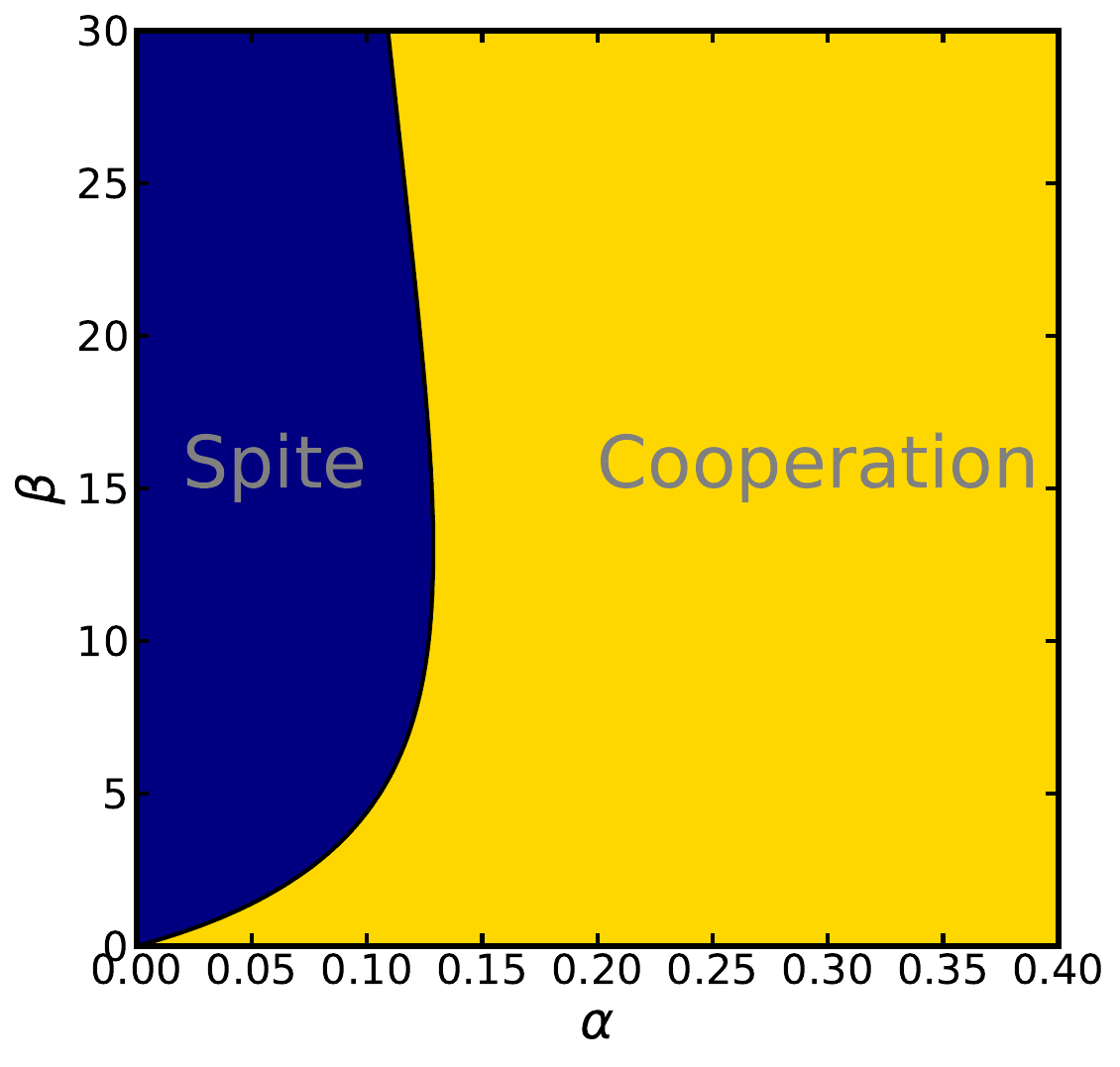}}
    \vspace{-3mm}
    \caption{\textbf{Evolution of cooperation and spite in stars with self-interaction.} $\alpha$ and $\beta$ denotes the self-interaction strength of hub and leaves, respectively. (a) Self-interaction only applies to the hub vertex, i.e., $\alpha=0$. The evolution of spite is always favored in this case, which is harmful to the system. (b) Self-interaction only applies to the leaf vertices, i.e., $\beta=0$. The system can always favor the evolution of cooperation. The sizes of stars are $N\in\{5,7,9,11\}$ as indicated in the legend. (c) and (d) Phase diagrams of the cooperation (yellow) and spite (blue) with $N=15$ and $N=25$, respectively. (color online)}
    \label{fig: star equation}
\end{figure}
In the following we check the case of star topology. If we do not consider self-loops, the star system cannot favor cooperation for all benefit-to-cost ratios~\cite{allen2017evolutionary}. Thm.~\ref{Theorem: 2} shows that proper self-interaction strength in stars can promote the fixation of cooperation. In Fig.~\ref{fig: star equation}, we discuss the effects of self-loops of stars in two extreme cases, including $\alpha=0$ and $\beta=0$. Fig.~\ref{fig: star equation}(a) shows the case that $\alpha=0$, i.e., there is no self-interaction among leaves. As a result, the evolution selection always favors spite, and cooperation cannot be favored, according to Thm.~\ref{Theorem: 2}(b). The critical condition decreases with the increase of system size $N$, indicating that an agent only needs to pay a small cost to bring tremendous damage to its neighbors. However, as shown in Fig.~\ref{fig: star equation}(b), the self-loops of leaves can promote cooperation. 
A slight self-interaction strength among leaves can reduce the cooperation condition to a single-digit number, as shown in Thm.~\ref{Theorem: 2}(c). 

In Figs.~\ref{fig: star equation}(c)-(d), we further present the phase diagrams for $N=15$ and $N=25$ in stars by setting coordinate axes as $\alpha$ and $\beta$. These phase diagrams also show that the self-interaction of the hub vertex is not beneficial for the fixation of cooperation. However, a star only needs a small self-interaction strength of leaves to maintain the possibility of cooperation, even if the hub vertex's self-interaction strength is large. Summing up, self-interaction of the leader (hub agent) is detrimental, while the same used for the majority (for leaves) could be helpful. 

\begin{figure}[h!]
    \centering
    \subfigure[Star Topology]{\centering\includegraphics[width=0.47\linewidth]{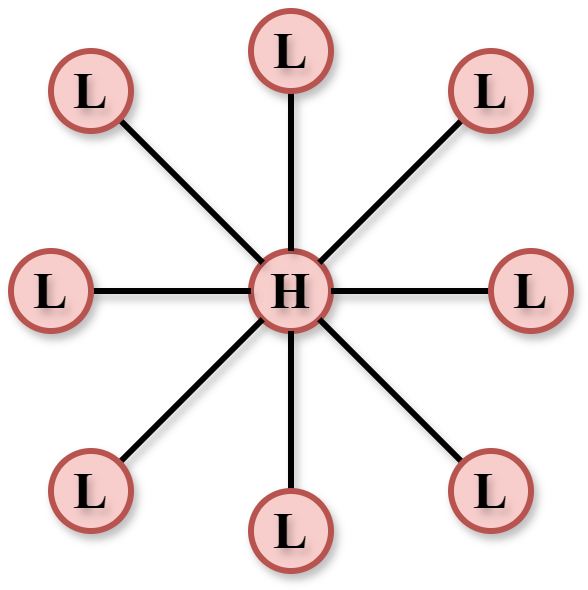}}
    \hspace{-3mm}
    \subfigure[Fixation Probability]{\includegraphics[width=0.53\linewidth]{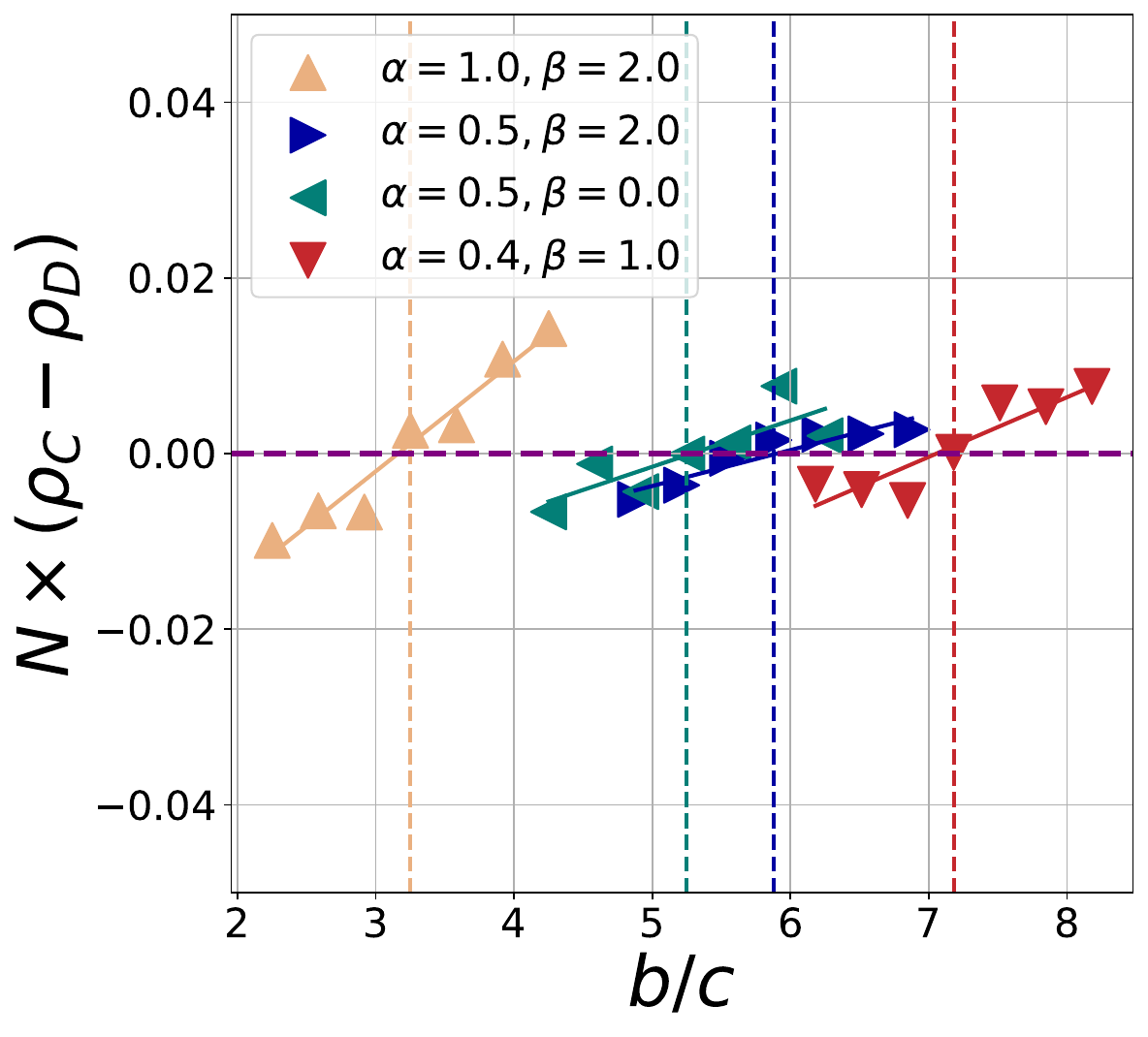}}
    \vspace{-3mm}
    \caption{\textbf{Fixation probabilities and conditions for cooperation in stars. }(a) An example of a star. The hub vertex has the greatest degree $N-1$, while each leaf has the degree $1$. (b) The fixation probability for different self-interaction strength of hub and leaves $(\alpha,\beta)\in\{(1.0,2.0),(0.5,2.0),(0.5,0.0), (0.4, 1.0)\}$. The system size is $N=10$. The vertical lines present the theoretical results in Eq.~\ref{eq: thm2 star}. The condition for cooperation without self-interaction is $\infty$. Each data point is the ratio of cooperation fixation in $2\times10^6$ independent runs.  (color online)} 
    \label{fig: star fixation probability}
\end{figure}

We further present the fixation probabilities in stars with self-interaction. In Fig.~\ref{fig: star fixation probability}(a), we illustrate the typical star topology where leaves are connected only to a central hub. In Fig.~\ref{fig: star fixation probability}(b), we show the fixation probabilities for four groups of parameters, including a special case that $\beta=0$, i.e., the self-interaction only applies to the leaves. We can see that the increase in self-interaction among leaves often reduces the condition for cooperation and magnifies the fixation probability of cooperation. We also show that self-interaction for hub could also be useful if it is attached with the self-interaction of leaves. 

\subsection{Hub-hub Joined Stars}
\begin{figure}
    \centering
    \subfigure[$\alpha=0$]{\includegraphics[width=0.53\linewidth]{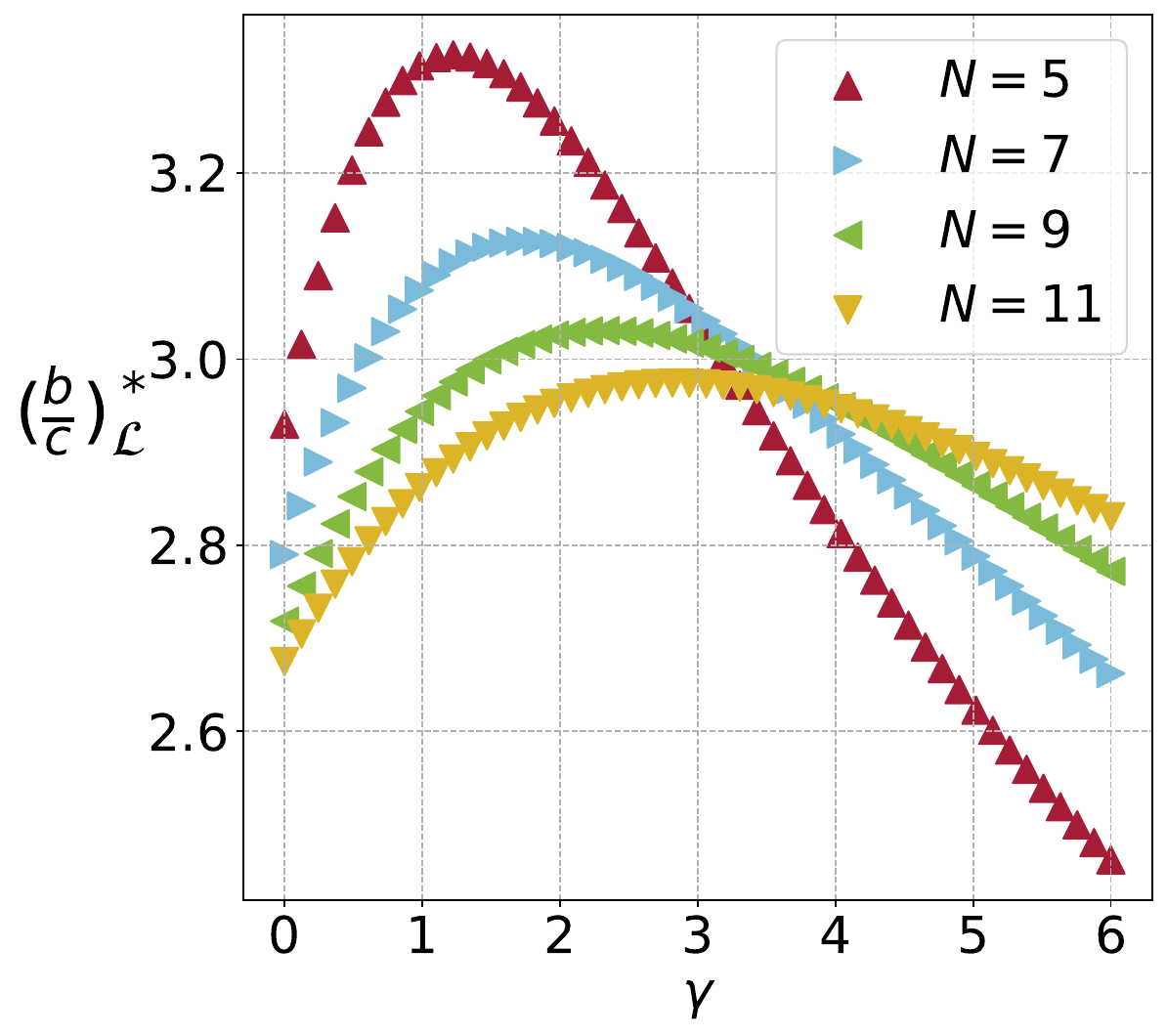}}
    \hspace{-3mm}
    \subfigure[$\gamma=0$]{\includegraphics[width=0.47\linewidth]{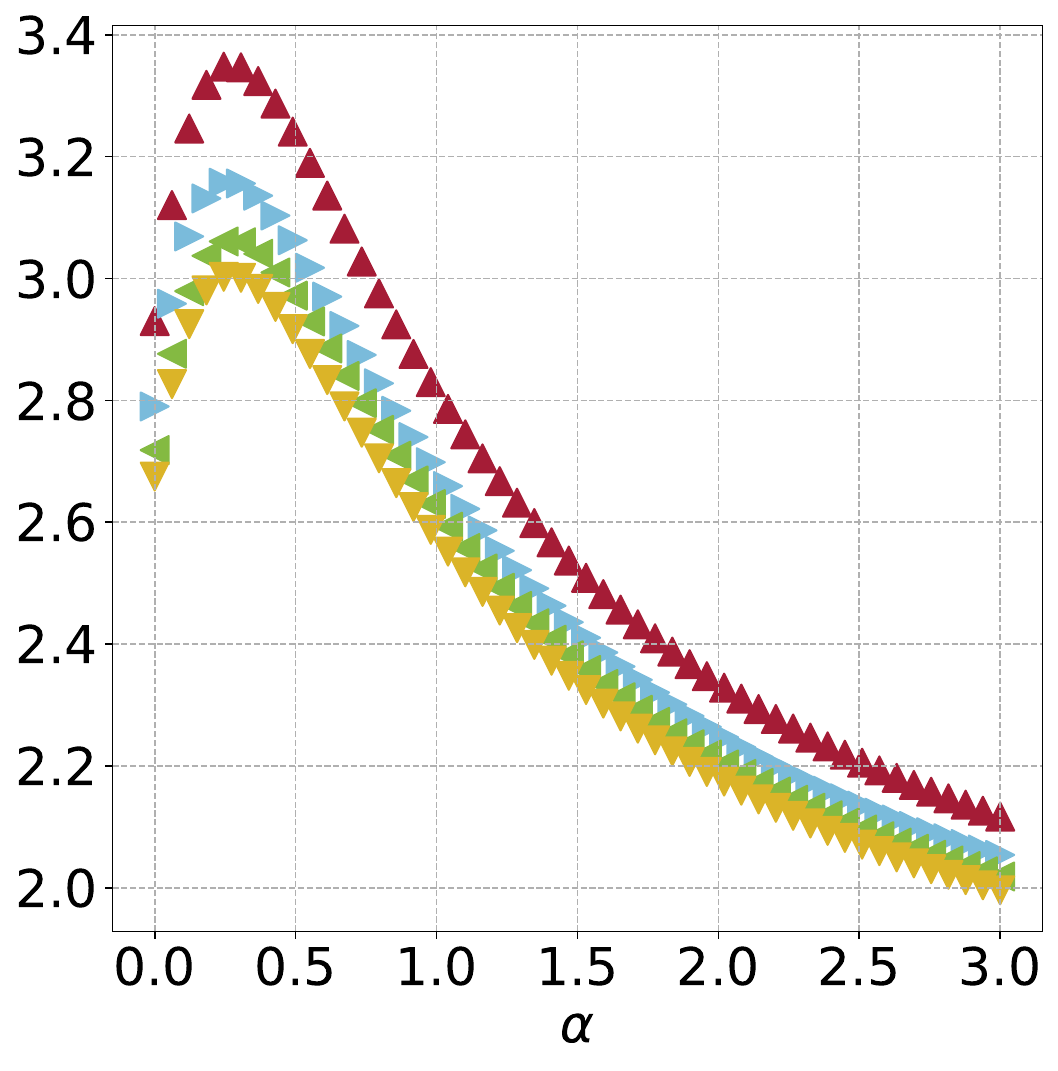}}
    \vspace{-3mm}

    \subfigure[$N=15$]{\includegraphics[width=0.49\linewidth]{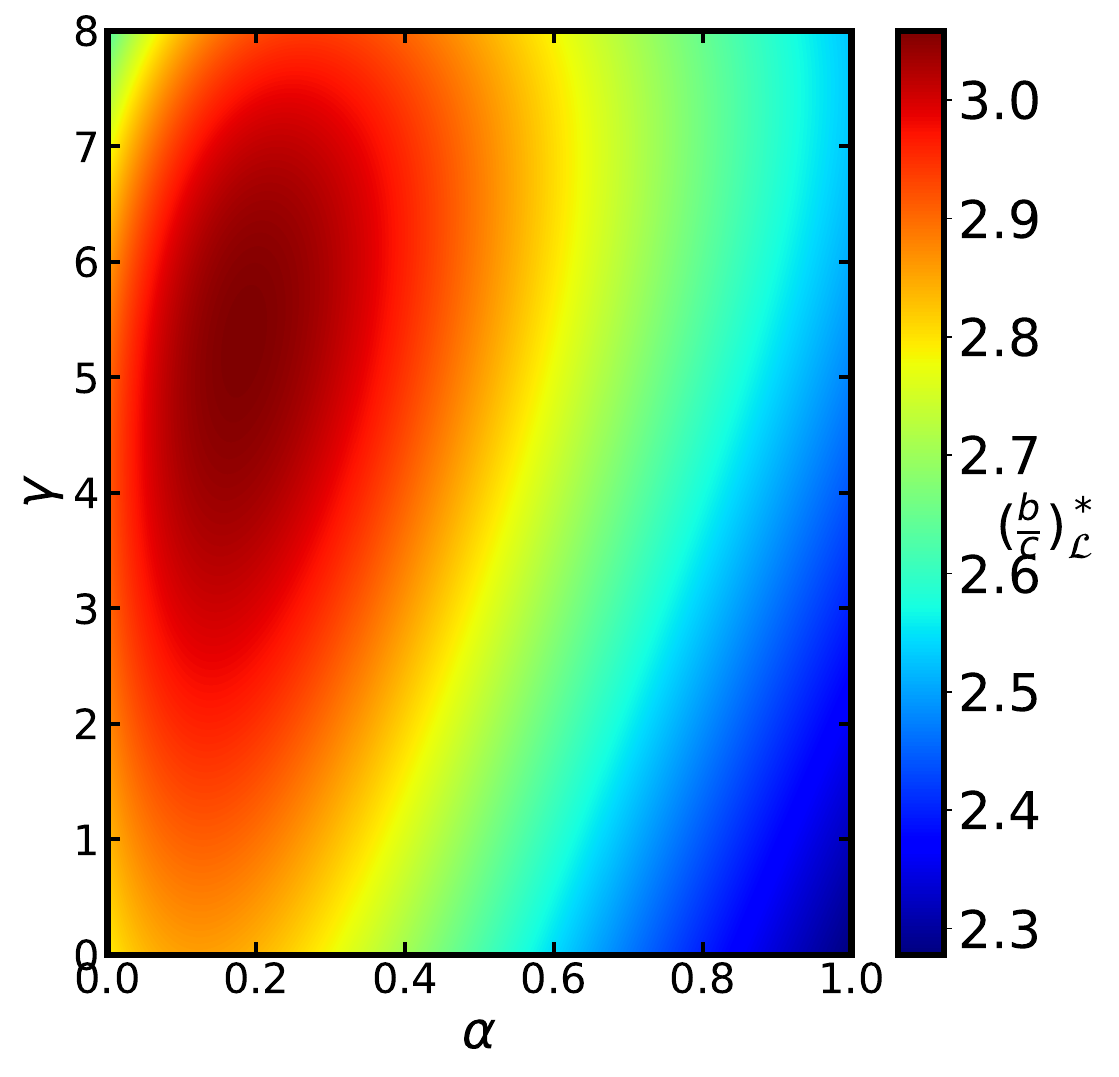}}
    \hspace{-3mm}
    \subfigure[$N=25$]{\includegraphics[width=0.49\linewidth]{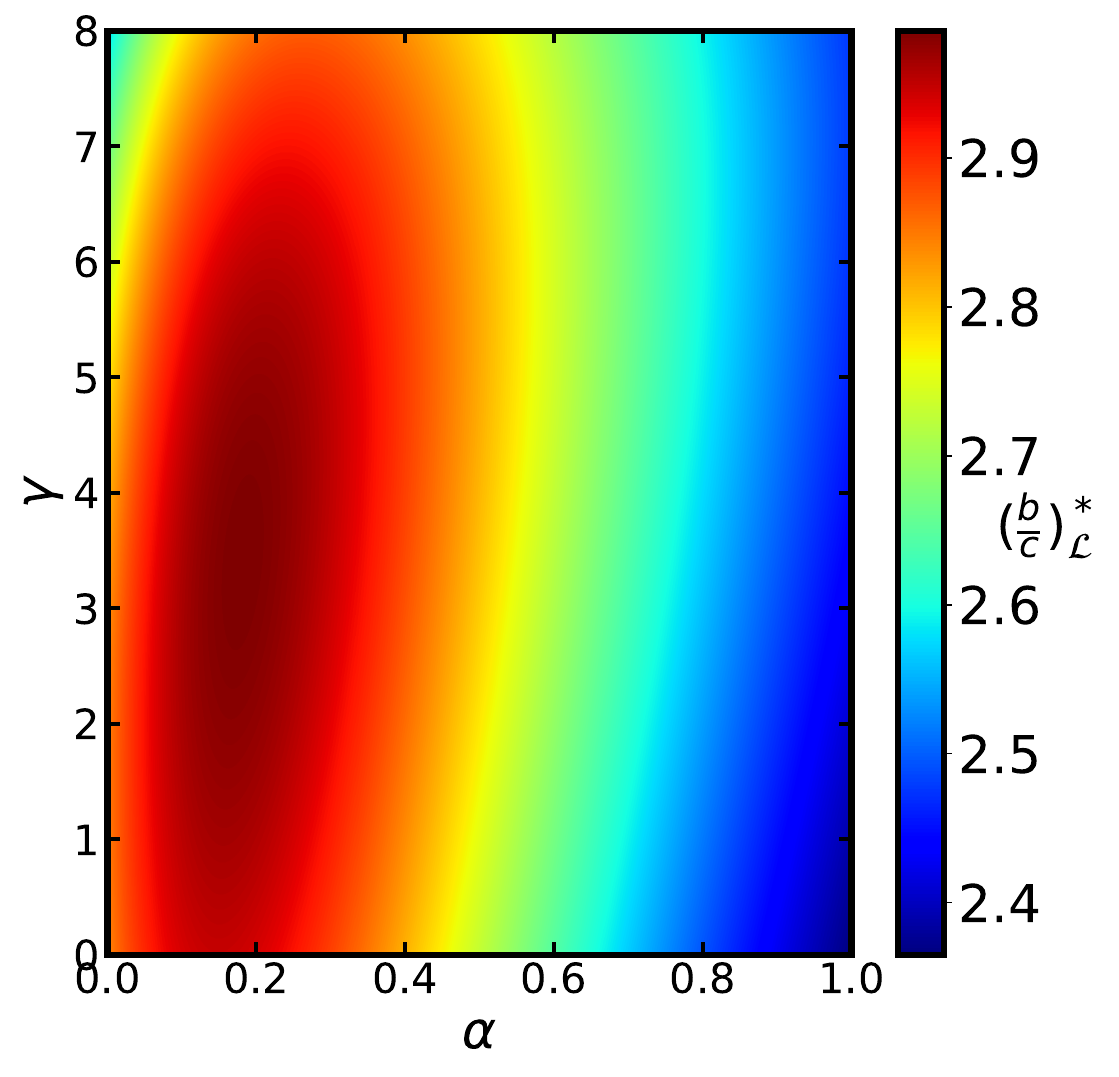}}
    \vspace{-3mm}
    \caption{\textbf{Evolution of cooperation in hub-hub joined stars with self-interaction. } $\alpha$ and $\gamma$ indicate the self-interaction strength for leaves and hubs, respectively. (a) The self-interaction only applies to two hub vertices with $\alpha=0$. (b) The self-interaction only applies to the leaf vertices with $\gamma=0$. The system sizes are $N\in\left\{5,7,9,11\right\}$. The hub-hub joined stars always favor cooperation and never favor spite. (c) and (d) Heatmaps of cooperation conditions with $N=15$ and $N=25$, respectively. The red area indicates that the requested condition of cooperation is higher. (color online)} 
    \label{fig: star hh equation}
\end{figure}
Next, we focus on the stars that are joined by hubs. When self-interaction is not considered, a graph of two stars joined by their hubs is an advantageous structure for the fixation of cooperation. In Figs.~\ref{fig: star hh equation}(a) and (b), we show the critical condition for cooperation with $\alpha=0$ and $\gamma=0$ respectively. Though the cooperation condition is always positive as Thm.~\ref{Theorem: 3}(b) shows, the self-interaction strength does not always promote the fixation of cooperation like for regular graphs. If the self-interaction only applies to both hub vertices, the cooperation condition first increases and then decreases as we grow $\gamma$. For a large system size $N$, we note that the maximal critical threshold is smaller as the increase of $\gamma$, but it also takes a larger $\gamma$ to make the condition smaller than the baseline without self-interaction. The reverse applies for a smaller $N$, that the maximal condition is much higher, but it takes a smaller $\gamma$ to reduce the condition for cooperation. The observed crossover in Fig.~\ref{fig: star hh equation}(a) suggests that larger population sizes tend to exhibit reduced sensitivity to variations in the self-interaction strength of hub vertices. Intuitively, this is because, as the number of vertices increases, the relative influence or weight of the hub strategies diminishes, i.e., the growing number of leaf nodes plays a more dominant role in shaping the strategic evolution within the network. From a theoretical perspective, this crossover arises from a combination of factors: specifically, a reduction in the two-step expected coalescence time and an increase in the three-step expected coalescence time. When a random walker starts at a hub vertex, a larger network offers more potential paths for the next step, even under the same level of self-interaction strength at the hub.

In Figs.~\ref{fig: star hh equation}(c)-(d), we further show the heatmaps of the cooperation condition given $N=15$ and $N=25$, respectively. Evidently, the cooperation condition is more sensitive to the change of $\alpha$. With the growth of the self-interaction strength for leaves $\alpha$, there is a significant reduction of the cooperation condition, resulting in a more easily achievable cooperation. 

\begin{figure}[h!]
    \centering
    \subfigure[The Hub-hub Joined Star]{\centering\includegraphics[width=0.53\linewidth]{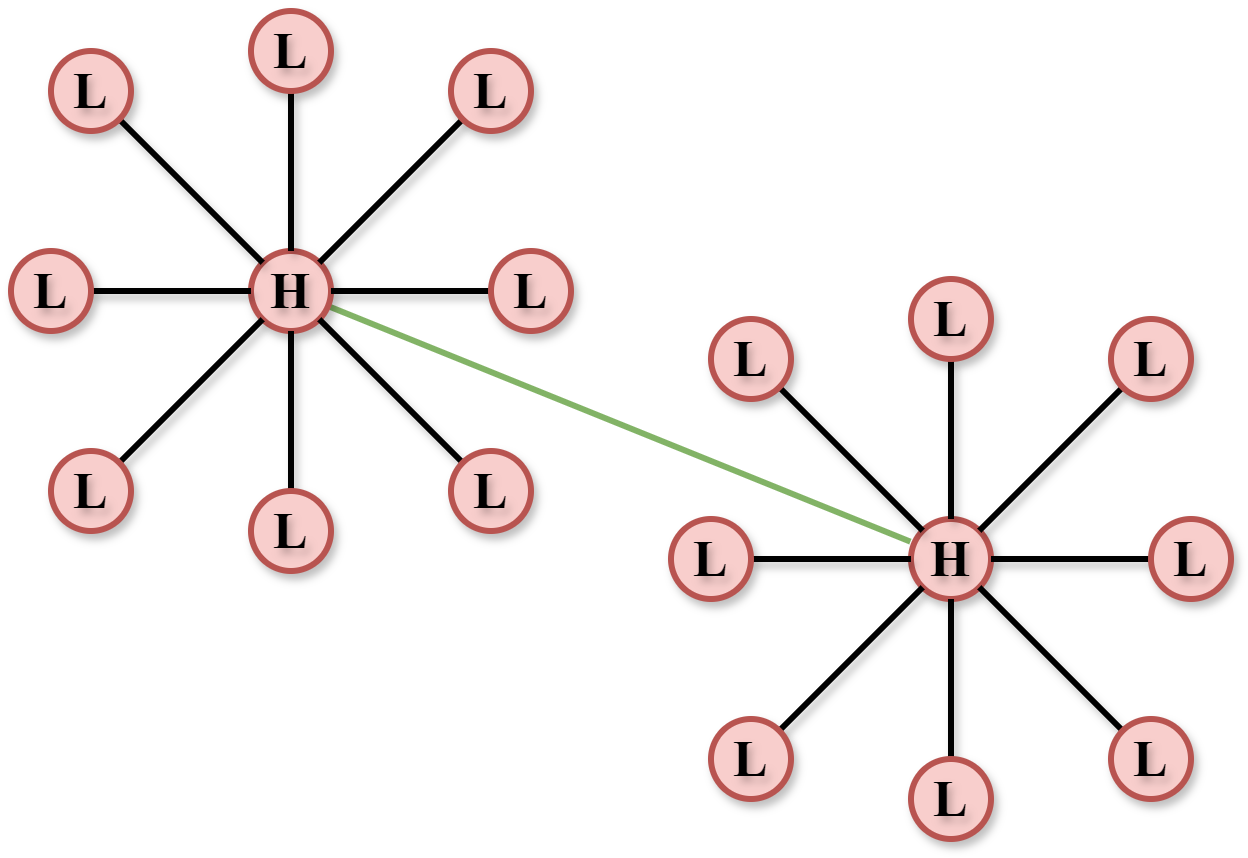}}
    \hspace{-3mm}
    \subfigure[Fixation Probability]{\includegraphics[width=0.47\linewidth]{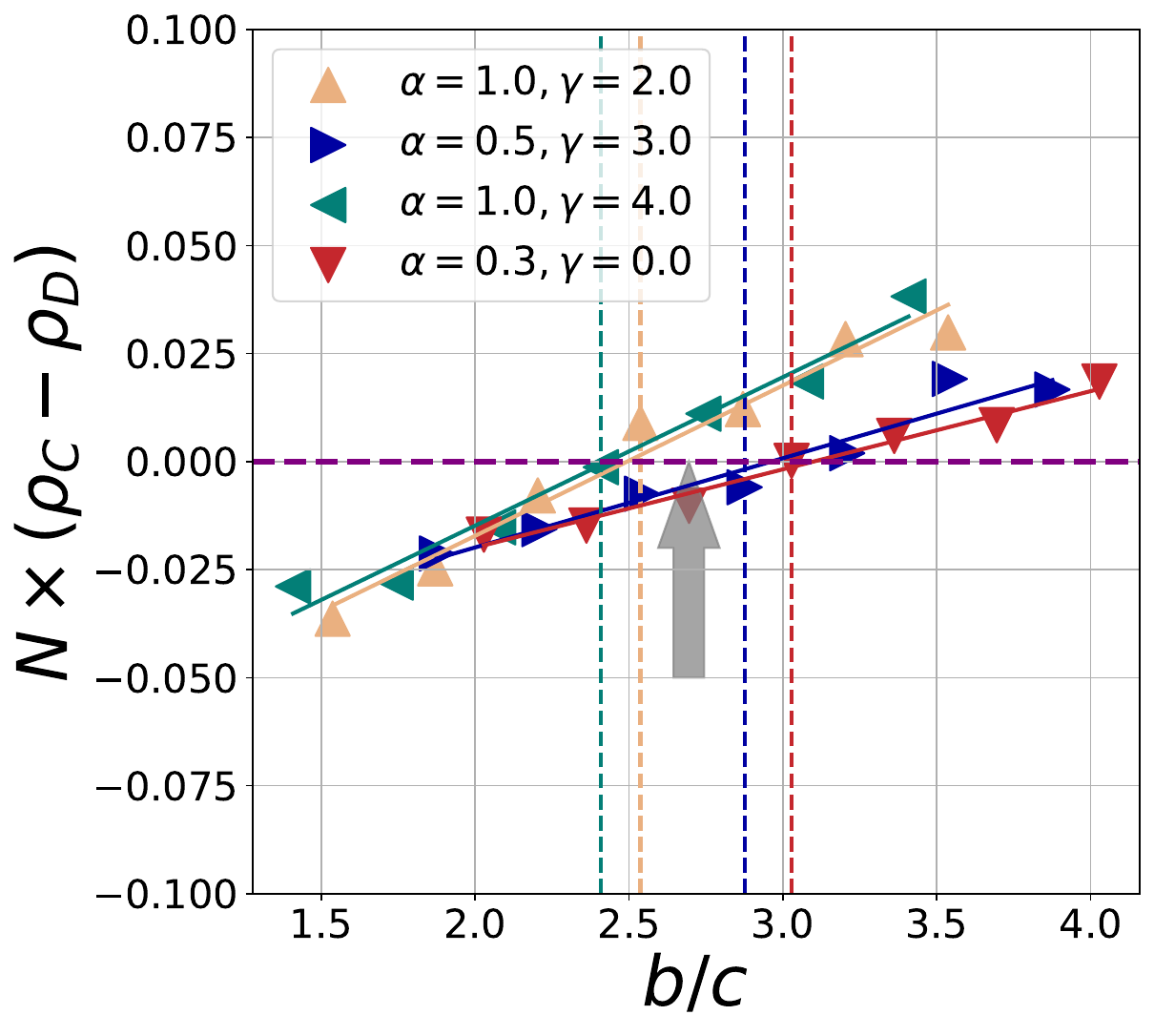}}
    \vspace{-3mm}
    \caption{\textbf{Fixation probabilities and conditions for cooperation in hub-hub joined stars. }(a) An example of hub-hub joined stars. The hub vertex of each star panel is connected via a link, shown by green solid line. (b) The fixation probability for different self-interaction parameters $(\alpha,\gamma)\in\{(1.0,2.0),(0.5,3.0),(1.0,4.0), (0.3, 0.0)\}$. The system size is $2N=20$, where each star panel has $N=10$ vertices. The vertical lines present the theoretical results of Eq.~\ref{eq: thm3 starhh}. The grey arrow indicates the cooperation condition without self-interaction. Each data point is averaged over $2\times10^6$ independent runs.  (color online)} 
    \label{fig: star hh fixation probability}
\end{figure}

In Fig.~\ref{fig: star hh fixation probability}, we present an additional simulation on the fixation probability of the hub-hub joined star with self-interaction. Fig.~\ref{fig: star hh fixation probability}(a) shows an example of a hub-hub joined star, where an edge is added between the hubs of separate stars. 
In Fig.~\ref{fig: star hh fixation probability}(b), we present the exact evolutionary fixation probabilities. The promotion of cooperation is guaranteed by the joint effect of $\alpha$ and $\gamma$. However, unilateral changes of leaves or hubs makes difficult to promote cooperation. 

\subsection{Ceiling Fans}
\begin{figure}
    \centering
    \subfigure[$\epsilon=0$]{\includegraphics[width=0.53\linewidth]{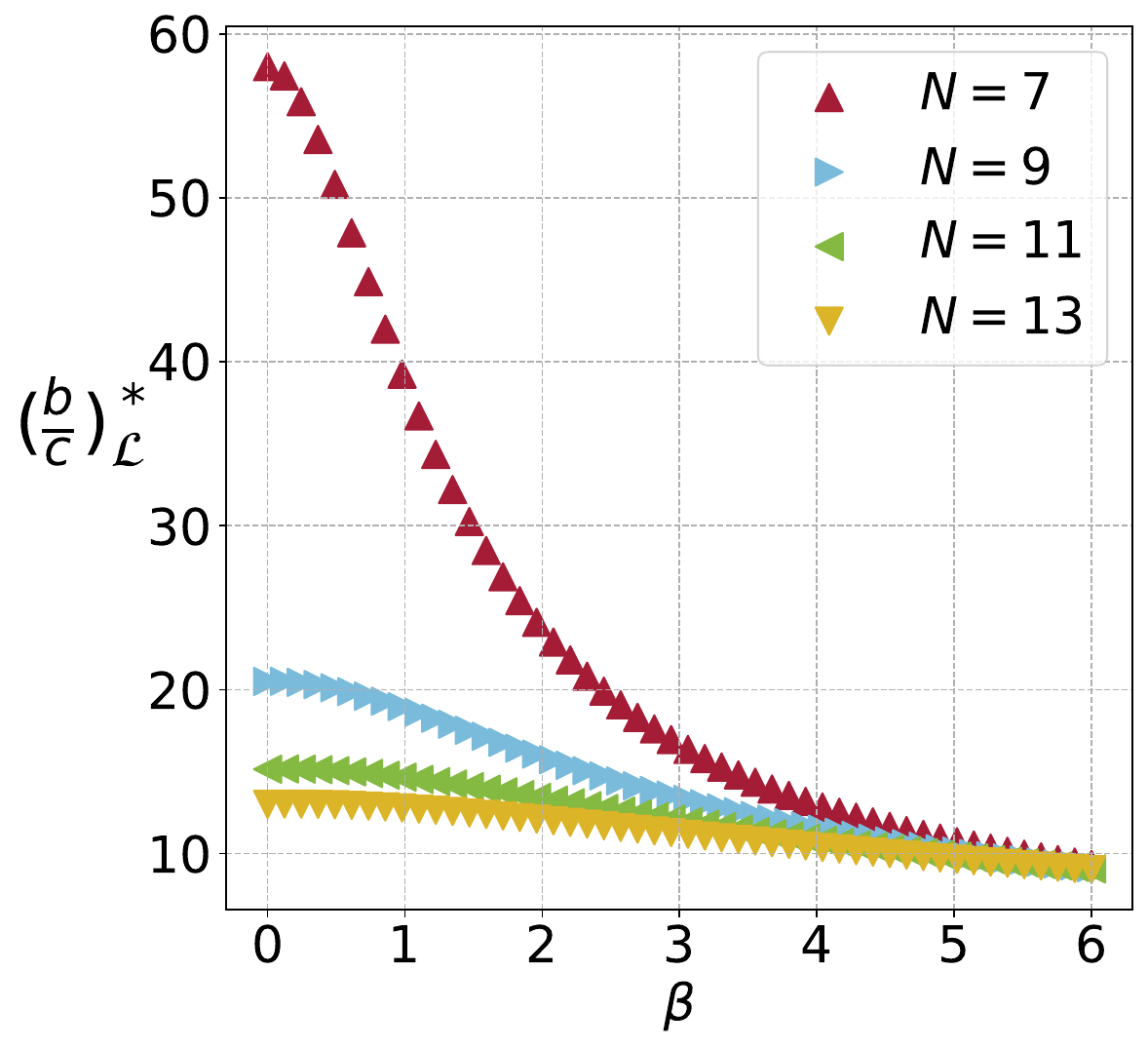}}
    \hspace{-3mm}
    \subfigure[$\beta=0$]{\includegraphics[width=0.47\linewidth]{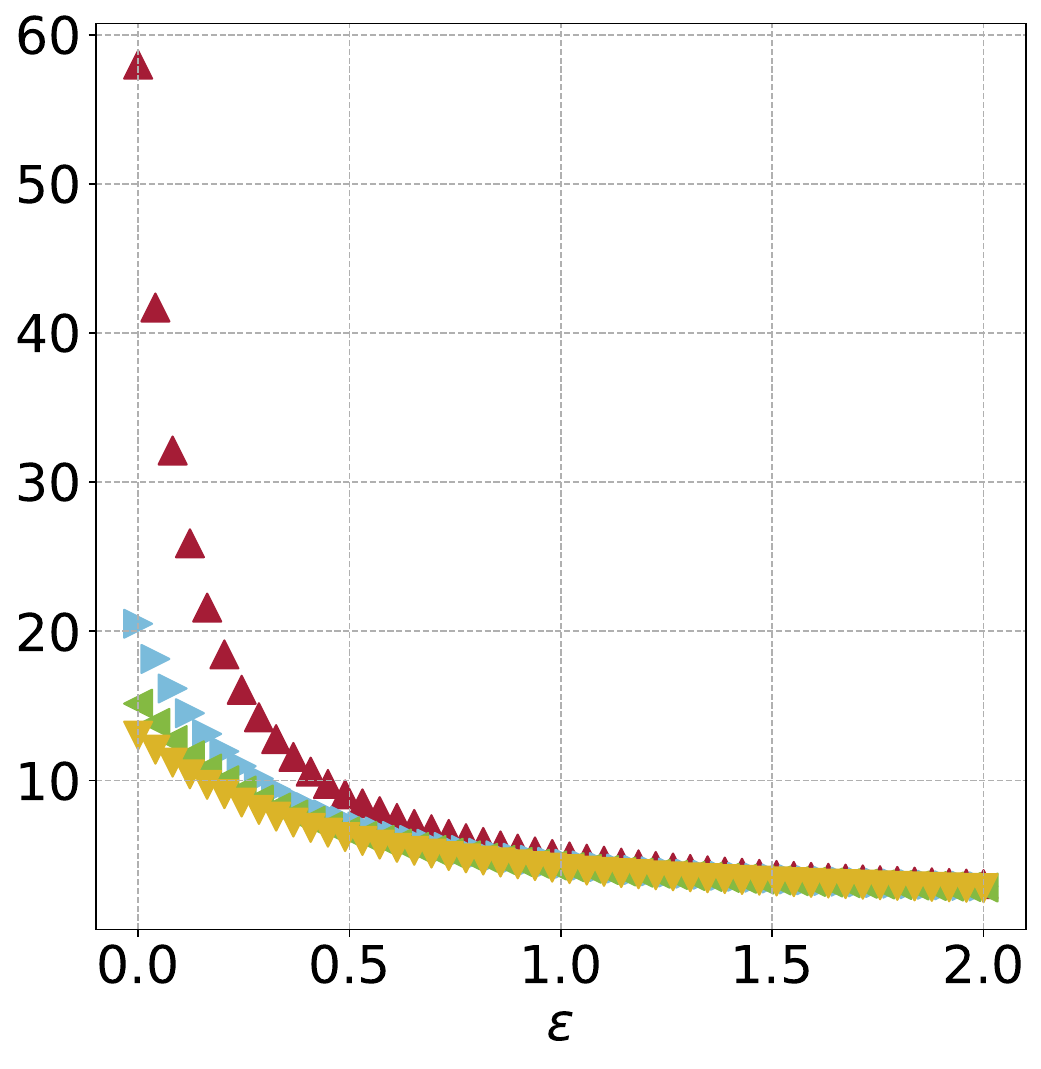}}
    \vspace{-3mm}

    \subfigure[$N=3$]{\includegraphics[width=0.5\linewidth]{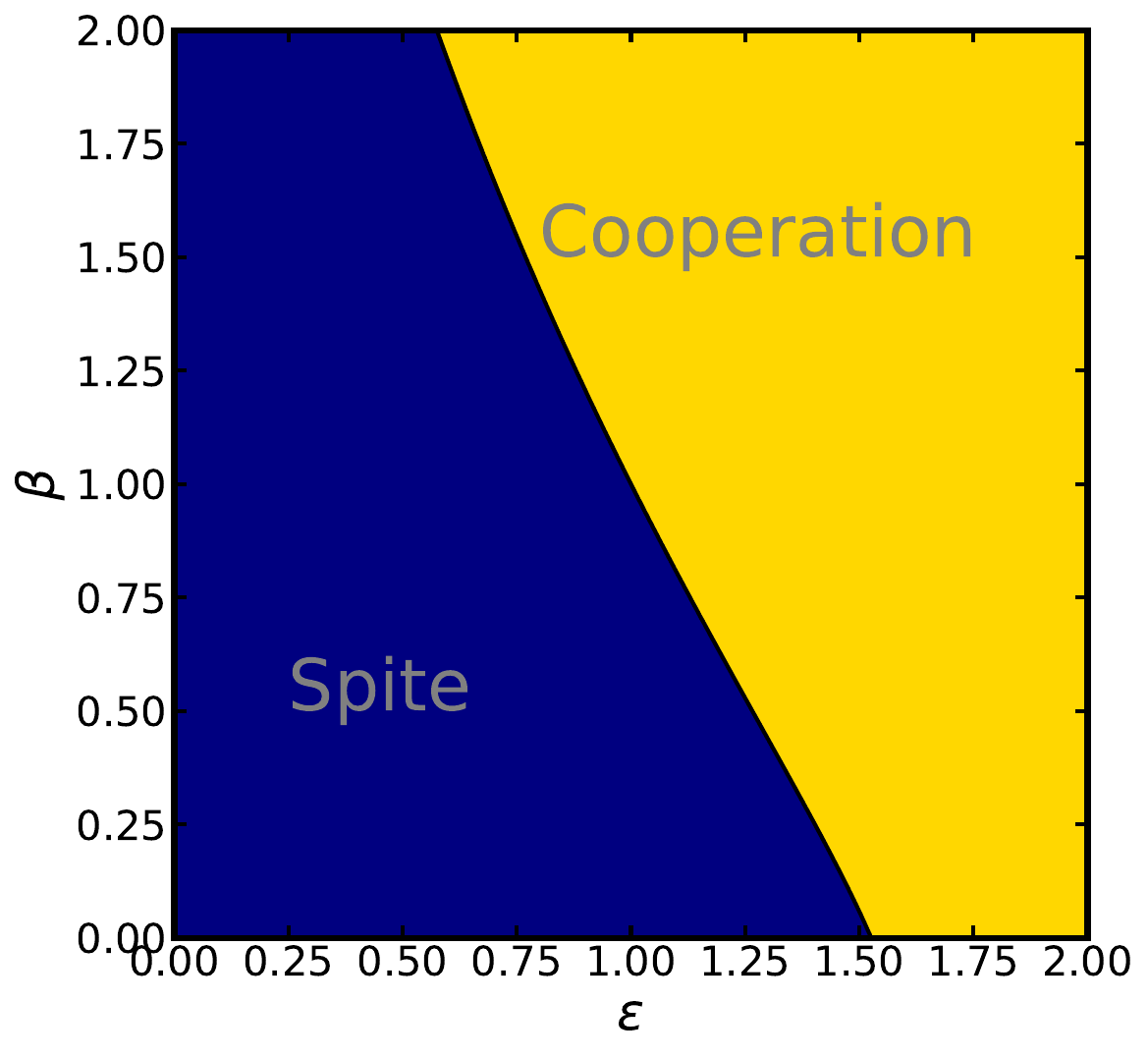}}
    \hspace{-3mm}
    \subfigure[$N=5$]{\includegraphics[width=0.5\linewidth]{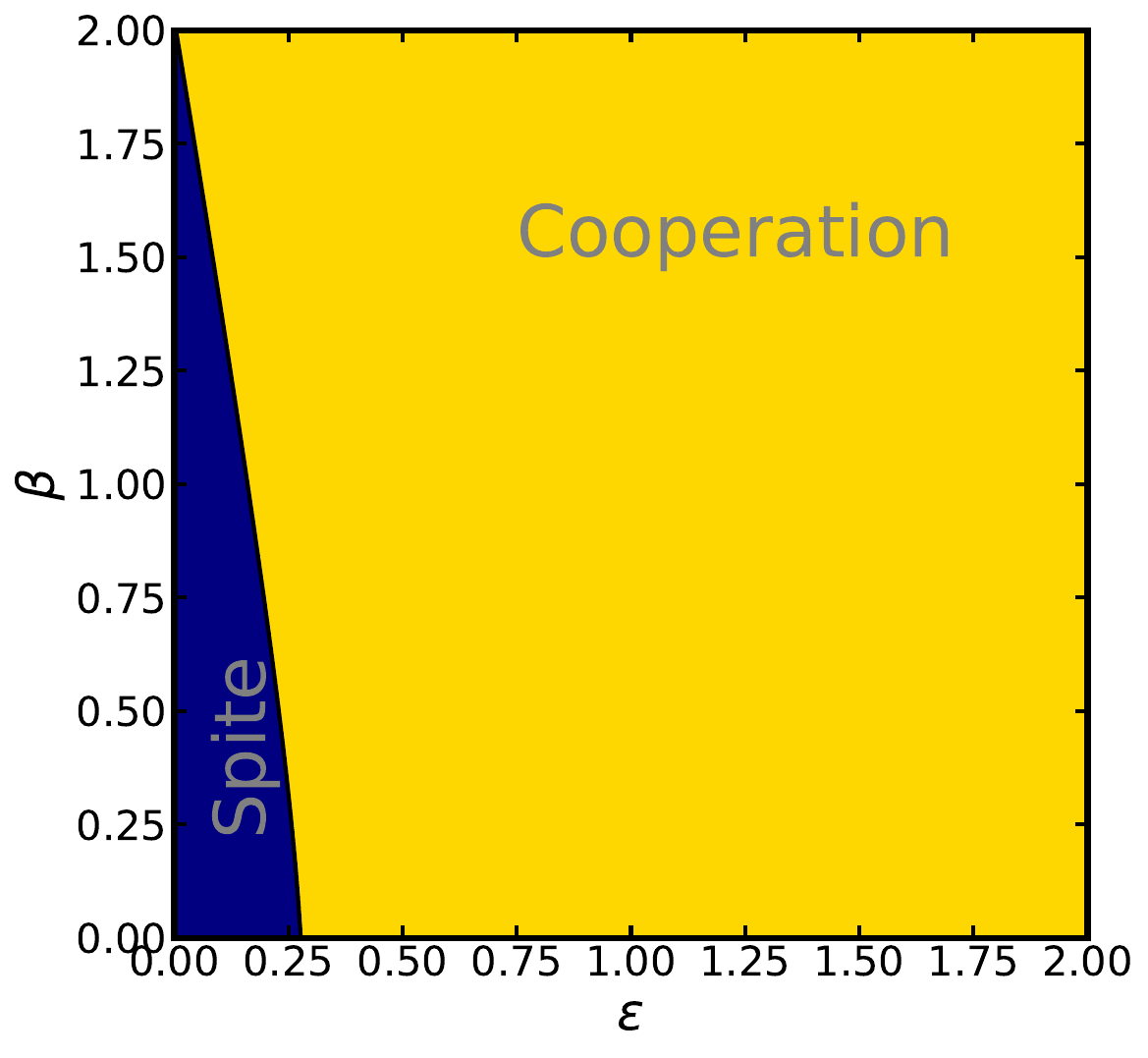}}
    \vspace{-3mm}
    \caption{\textbf{Evolution of cooperation and spite in ceiling fans with self-interaction.} $\epsilon$ and $\beta$ are the self-interaction strength for leaves and hubs, respectively. (a) The self-interaction only applies to the hub vertex, where $\epsilon=0$. (b) The self-interaction only applies to the leave vertices, where $\beta=0$. The selection always favors the evolution of cooperation if $N\geq7$. The system sizes are set as $N\in\left\{7,9,11,13\right\}$. (c) The exceptional case $N=3$. (d) The exceptional case $N=5$. In these two cases, the evolution of spite can be favored. The lines of borders in panel~(c) and (d) present the transition from spite (blue area) to cooperation (yellow area). There is minimal leaf self-interaction strength for the network to favor cooperation regardless of the system size $N$ and hub self-interaction strength $\beta$. (color online)}
    \label{fig: ceiling fan equation}
\end{figure}

A ceiling fan topology guarantees that every leaf can share information with another leaf. In Figs.~\ref{fig: ceiling fan equation}(a) and (b), we show the condition for cooperation with $N\geq7$ with $\epsilon=0$ and $\beta=0$ respectively. We can see that the cooperation condition is positive and cooperation can always be favored in both cases. Additionally, for $\epsilon=0$, when self-interaction only applies to the hub vertex, the cooperation condition is reduced by increasing self-interaction strength, and the same is valid for $\beta=0$. We have specified two exceptional cases where the selection may favor the evolution of spite, including $N=3$ and $N=5$. In Figs.~\ref{fig: ceiling fan equation}(c) and (d), we further discuss these two cases with special curves at $\beta=0$. These results show that $\epsilon$ and $\beta$ jointly decide the phase transition point from the evolution of spite to cooperation. However, as discussed previously, we can still find the critical point for a permanently positive cooperation condition. For the case $N=3$ in Fig.~\ref{fig: ceiling fan equation}(c), it shows that $\epsilon$ should be greater than the third root of $3\epsilon^3+13\epsilon^2-17\epsilon-15=0$ (approximately $1.53$) for cooperation regardless of $\beta$. For the case $N=5$, in Fig.~\ref{fig: ceiling fan equation}(d), the threshold of $\epsilon$ is at the third root of $9\epsilon^3+47\epsilon^2+8\epsilon-6=0$ (approximately $0.277$), providing more space for the favor of cooperation. If $\epsilon$ is greater than these critical points, the evolutionary dynamics can always favor cooperation, and there always exists a positive cooperation condition. 

In Figs.~\ref{fig: ceiling fan equation}(c)-(d), we further present the phase diagrams for the two cases $N=3$ and $N=5$. We find that the theoretical thresholds of $\epsilon$ for the favor of cooperation are consistent with the computational ones. $\epsilon>1.53$ and $\epsilon>0.277$ is the condition for the system to avoid the possibility of spite. Therefore, we can conclude that the favor of cooperation in a ceiling fan is easily achieved. 
\begin{figure}
    \centering
    \subfigure[The Ceiling Fan]{\centering\includegraphics[width=0.47\linewidth]{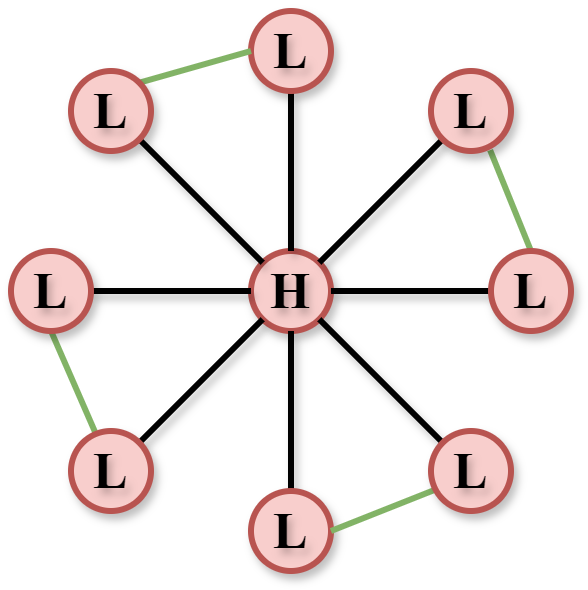}}
    \hspace{-3mm}
    \subfigure[Fixation Probability]{\includegraphics[width=0.53\linewidth]{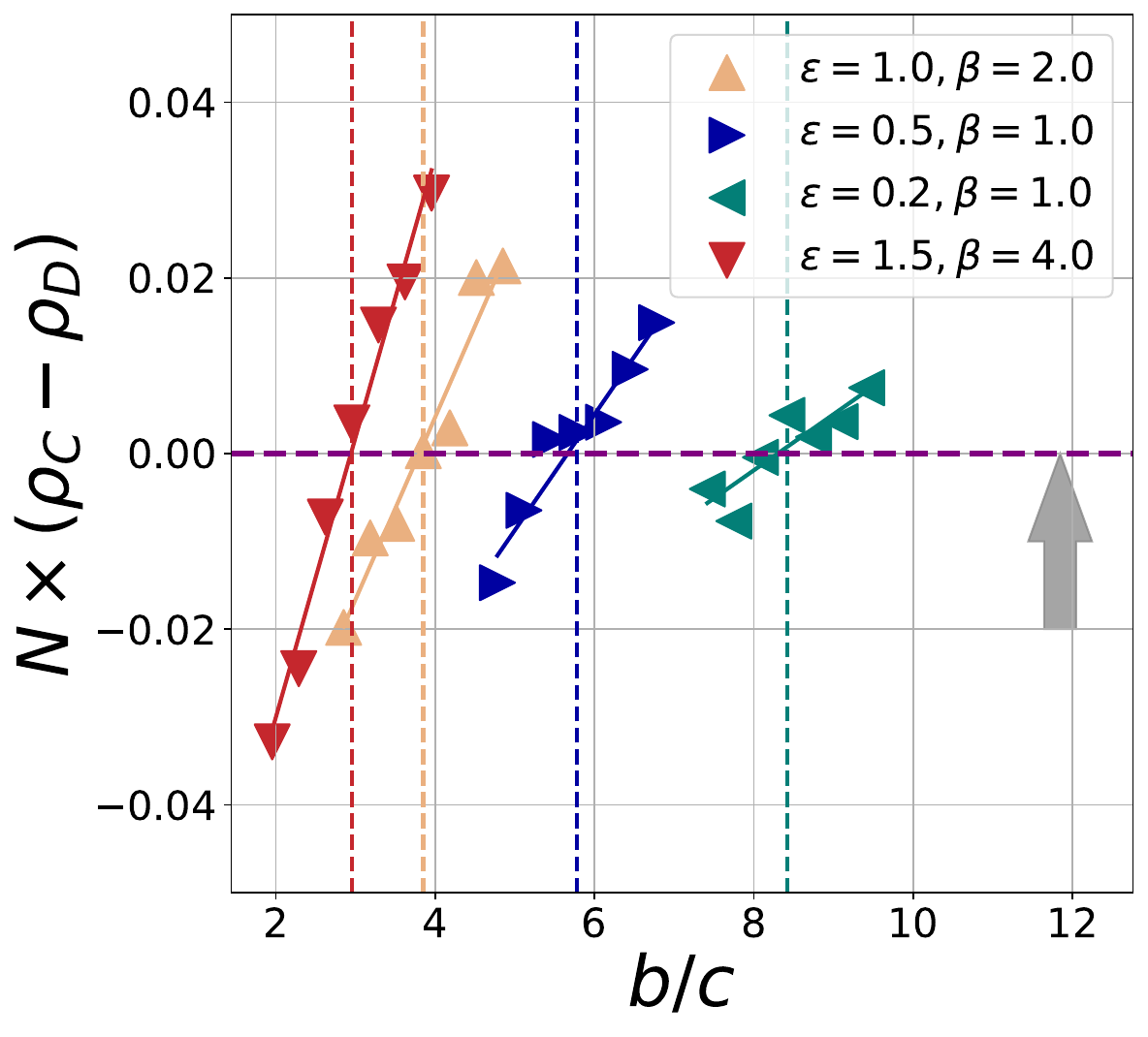}}
    \vspace{-3mm}
    \caption{\textbf{Fixation probabilities and conditions for cooperation in ceiling fans. }(a) An example of the ceiling fan. Each leaf is connected to only one other leaf, and the additional connections are shown in green lines. The system size should be odd. (b) The fixation probability for different self-interaction parameters $(\alpha,\gamma)\in\{(1.0,2.0),(0.5,1.0),(0.2,1.0), (1.5, 4.0)\}$. The system size is $N=15$. The vertical lines are the theoretical results in Eq.~\ref{eq: thm4 cf}. The grey arrow indicates the cooperation condition without self-interaction. Each data point is obtained by calculating the ratio of cooperation fixation in $2\times10^6$ independent realizations. (color online)} 
    \label{fig: ceiling fan fixation probability}
\end{figure}

In Fig.~\ref{fig: ceiling fan fixation probability}, we further show the fixation probabilities and the condition for cooperation with several self-interaction parameter groups. Fig.~\ref{fig: ceiling fan fixation probability}(a) shows an example of a ceiling fan. Each leaf vertex has two neighbors, including one hub and another leaf vertex. To fulfill this criterion the system size should only be odd number. In Fig.~\ref{fig: ceiling fan fixation probability}(b), we show the fixation probabilities with four groups of parameters in a ceiling fan with $N=15$ vertices. Compared to the condition without self-interaction, we find that a slight self-interaction strength can significantly reduce the condition for cooperation. With the increase of $\epsilon$ and $\beta$ the cooperation becomes easy to achieve. Therefore, the ceiling fan is an effective modified star topology for the selection to favor cooperation. 

\subsection{Random Networks}
\begin{figure*}[h!]
    \centering
    \subfigure[$e^{-k},b/c$]{\includegraphics[width=0.265\linewidth]{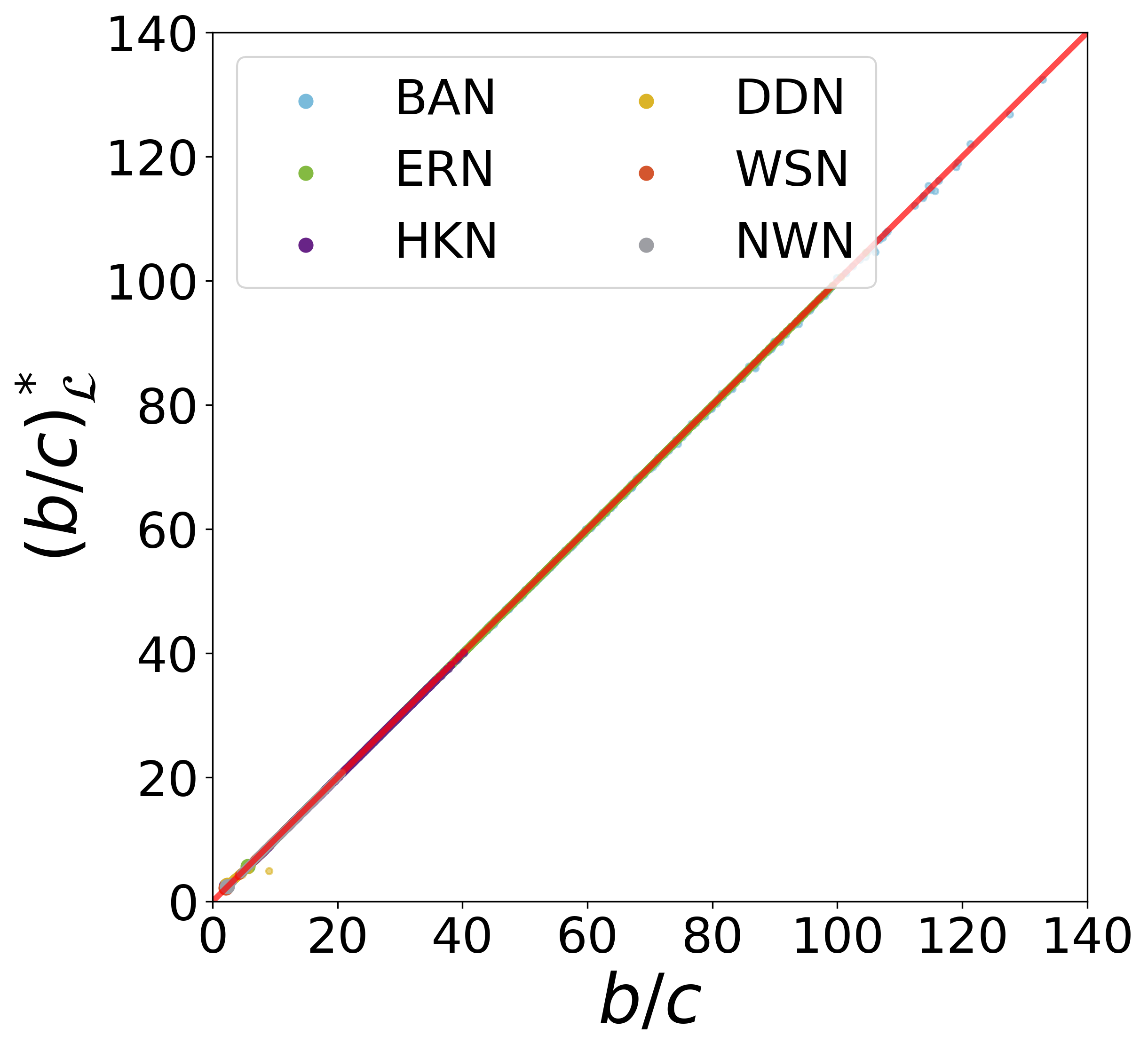}}
    \hspace{-3mm}
    \subfigure[$e^{-k},\langle k\rangle$]{\includegraphics[width=0.24\linewidth]{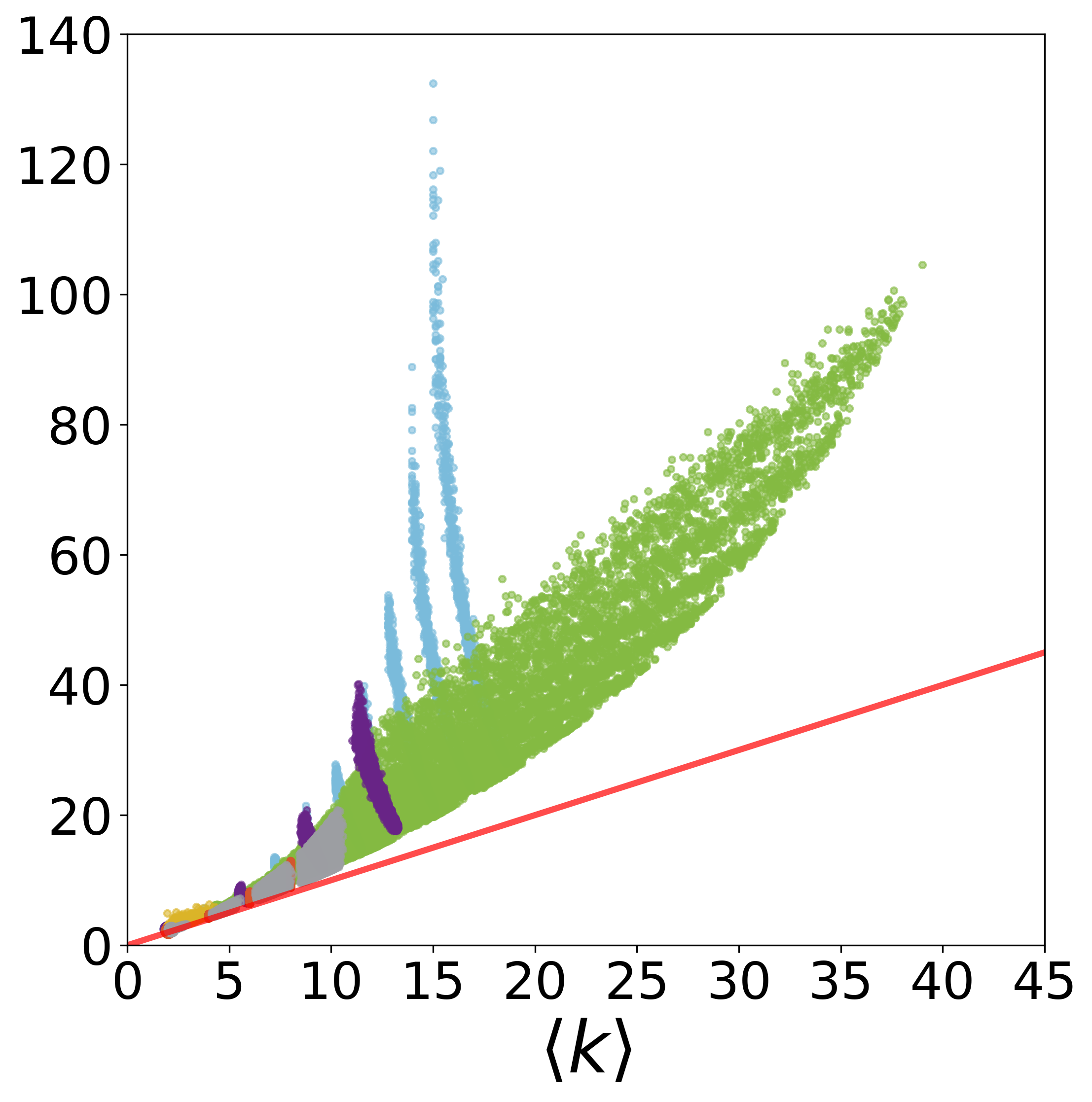}}
    \hspace{-3mm}
    \subfigure[$e^{-k},\langle k_{nn}\rangle$]{\includegraphics[width=0.24\linewidth]{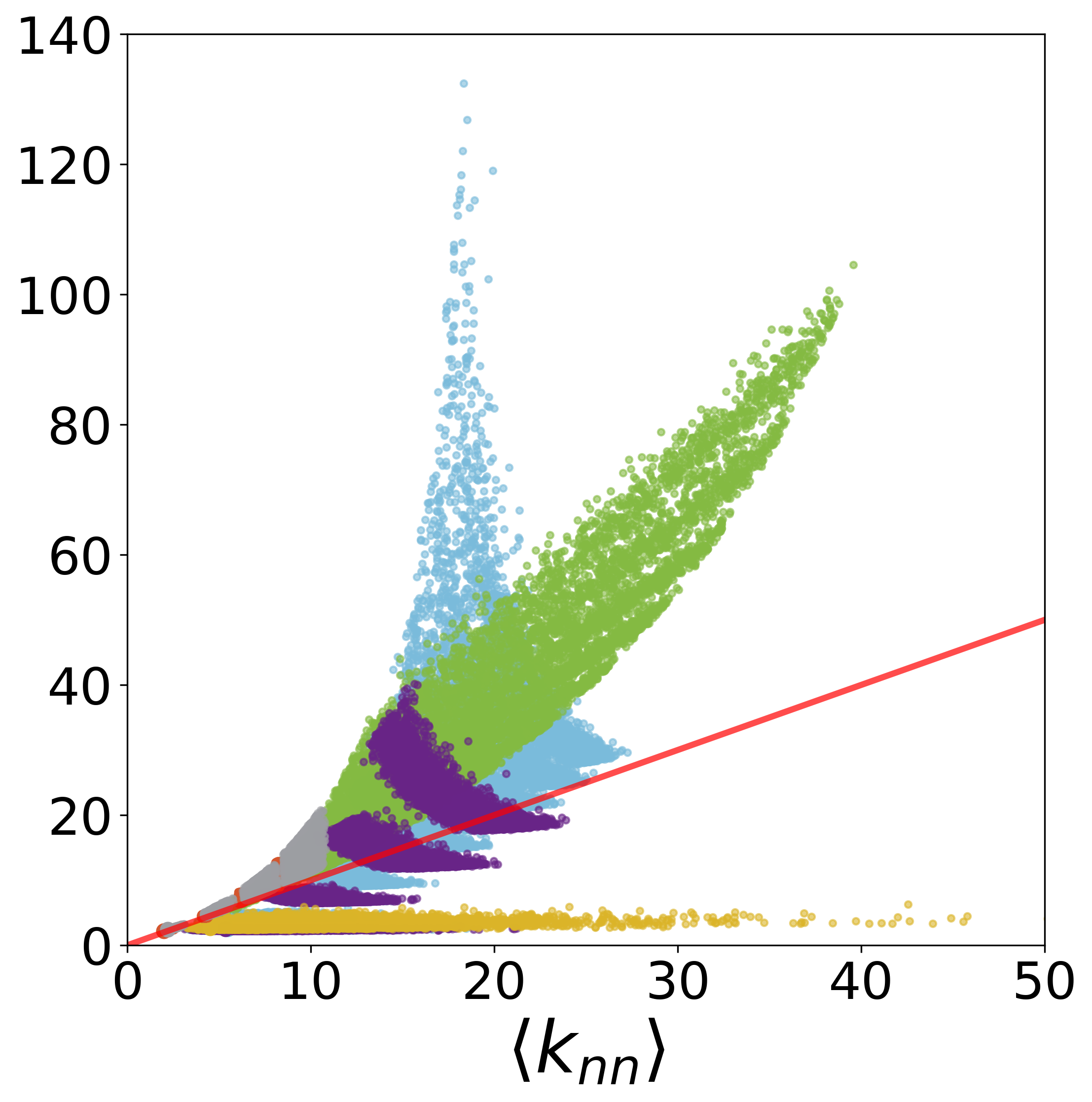}}
    \hspace{-3mm}
    \subfigure[$e^{-k}, \sigma$]{\includegraphics[width=0.255\linewidth]{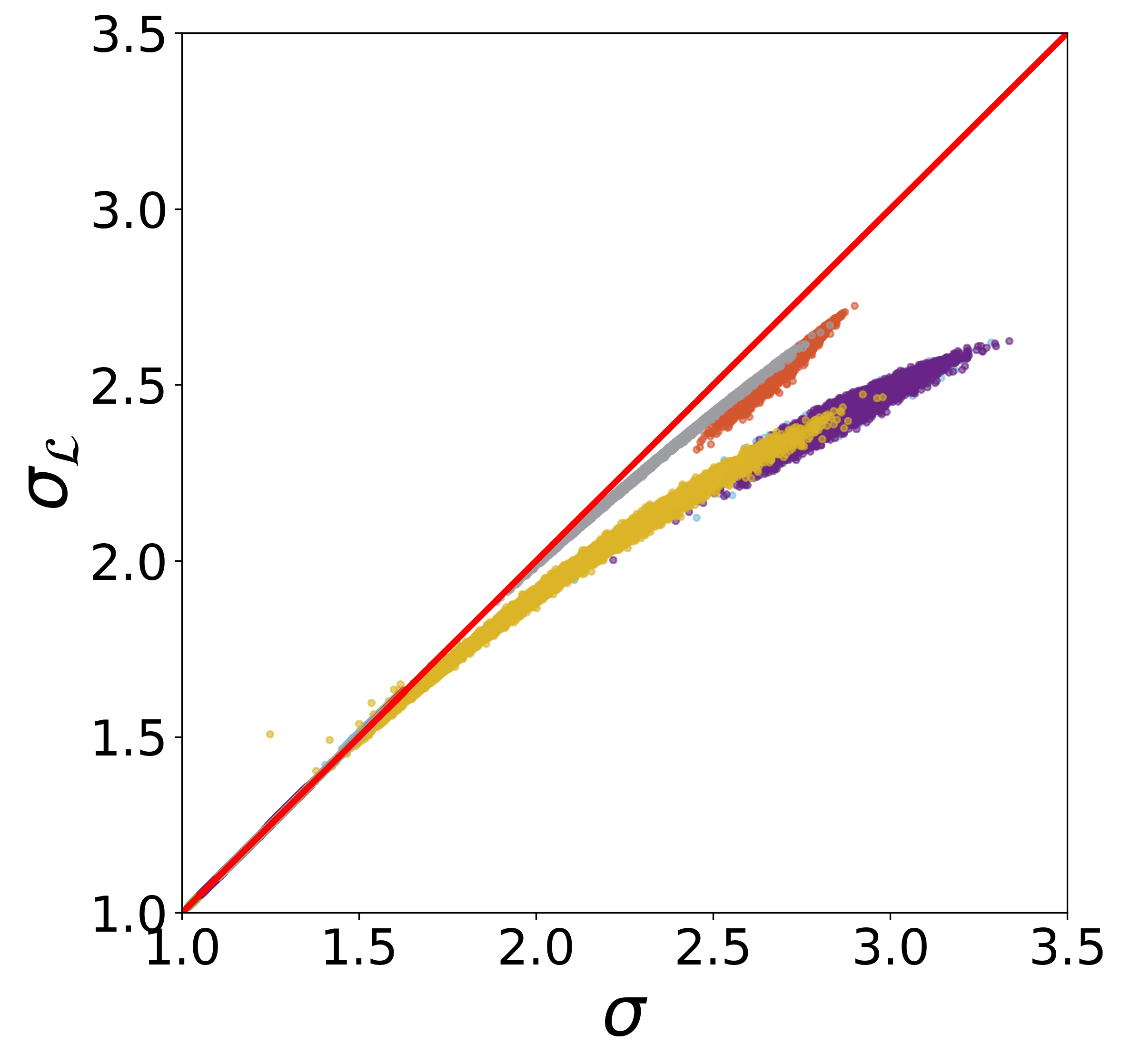}}
    \vspace{-3mm}

    \subfigure[$\ln{k},b/c$]{\includegraphics[width=0.265\linewidth]{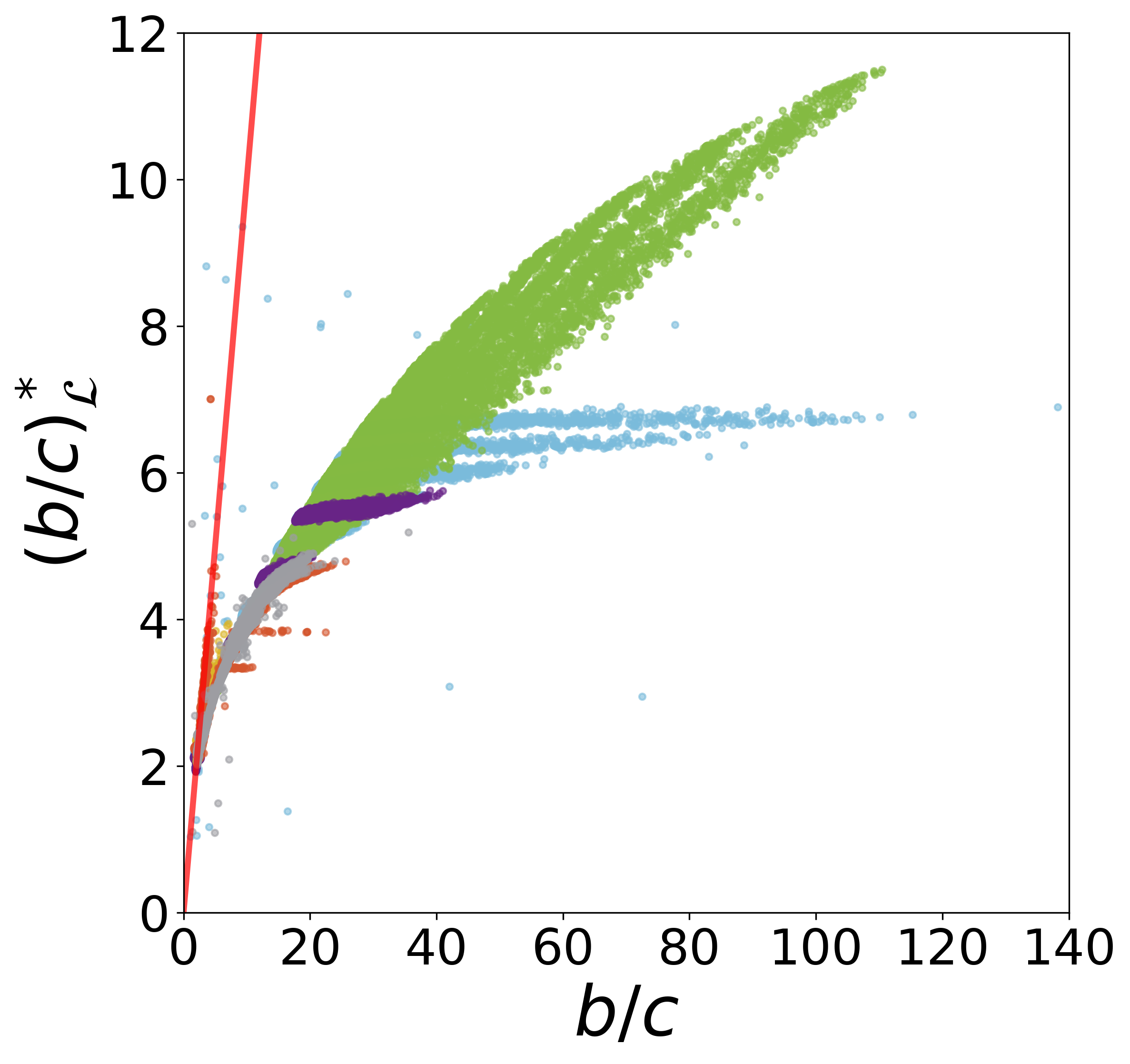}}
    \hspace{-3mm}
    \subfigure[$\ln{k},\langle k\rangle$]{\includegraphics[width=0.24\linewidth]{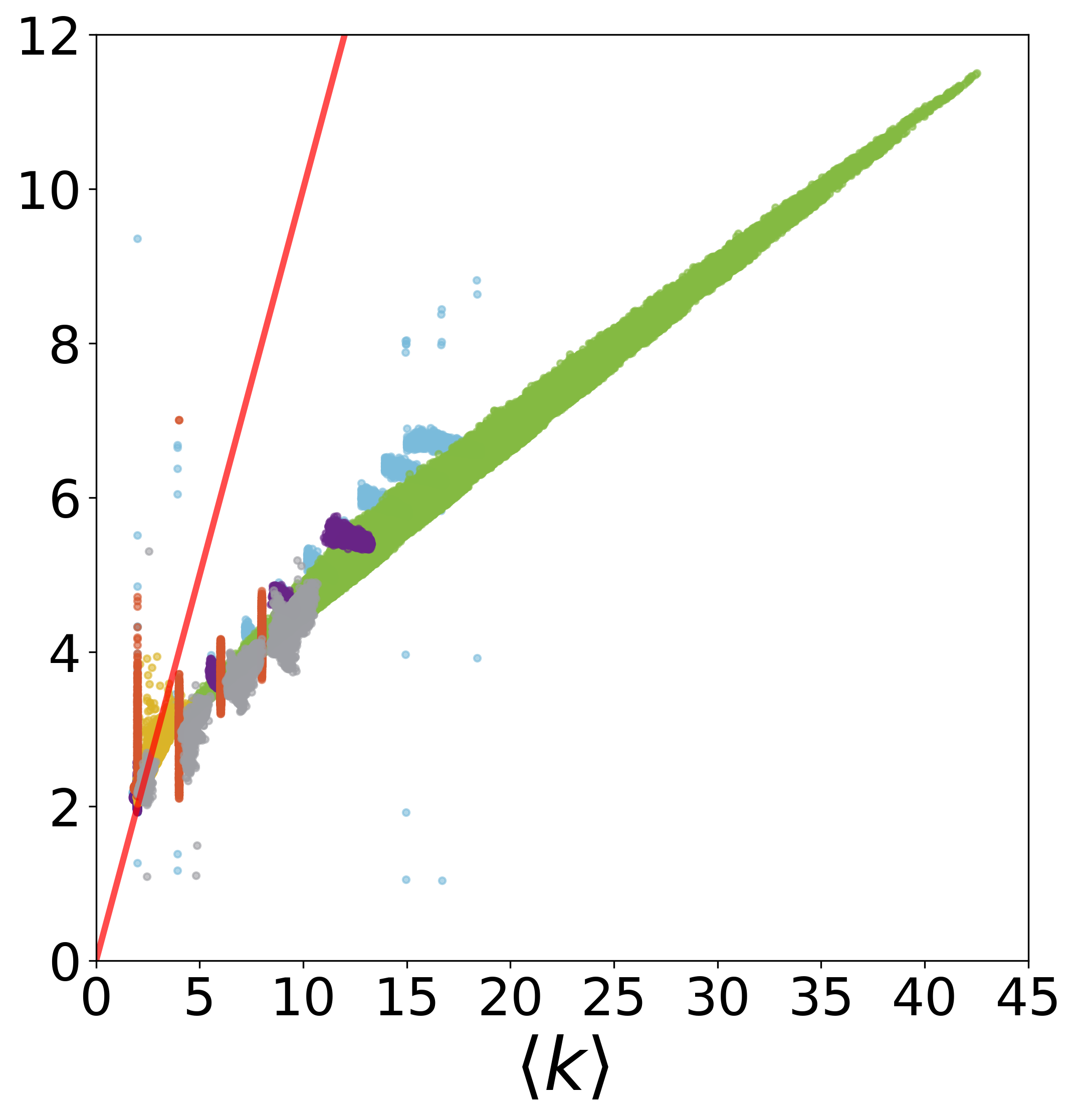}}
    \hspace{-3mm}
    \subfigure[$\ln{k},\langle k_{nn}\rangle$]{\includegraphics[width=0.24\linewidth]{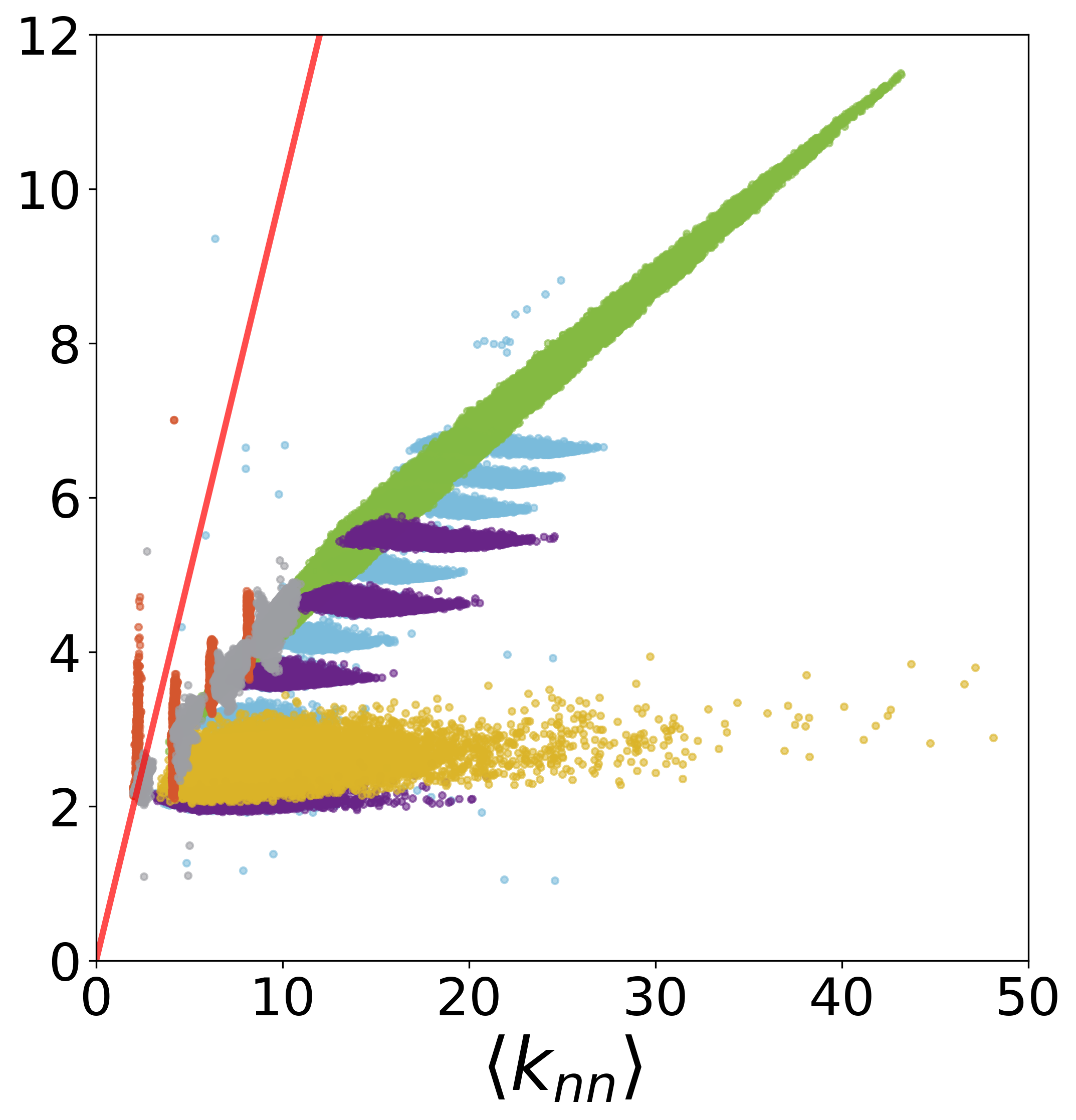}}
    \hspace{-3mm}
    \subfigure[$\ln{k}, \sigma$]{\includegraphics[width=0.25\linewidth]{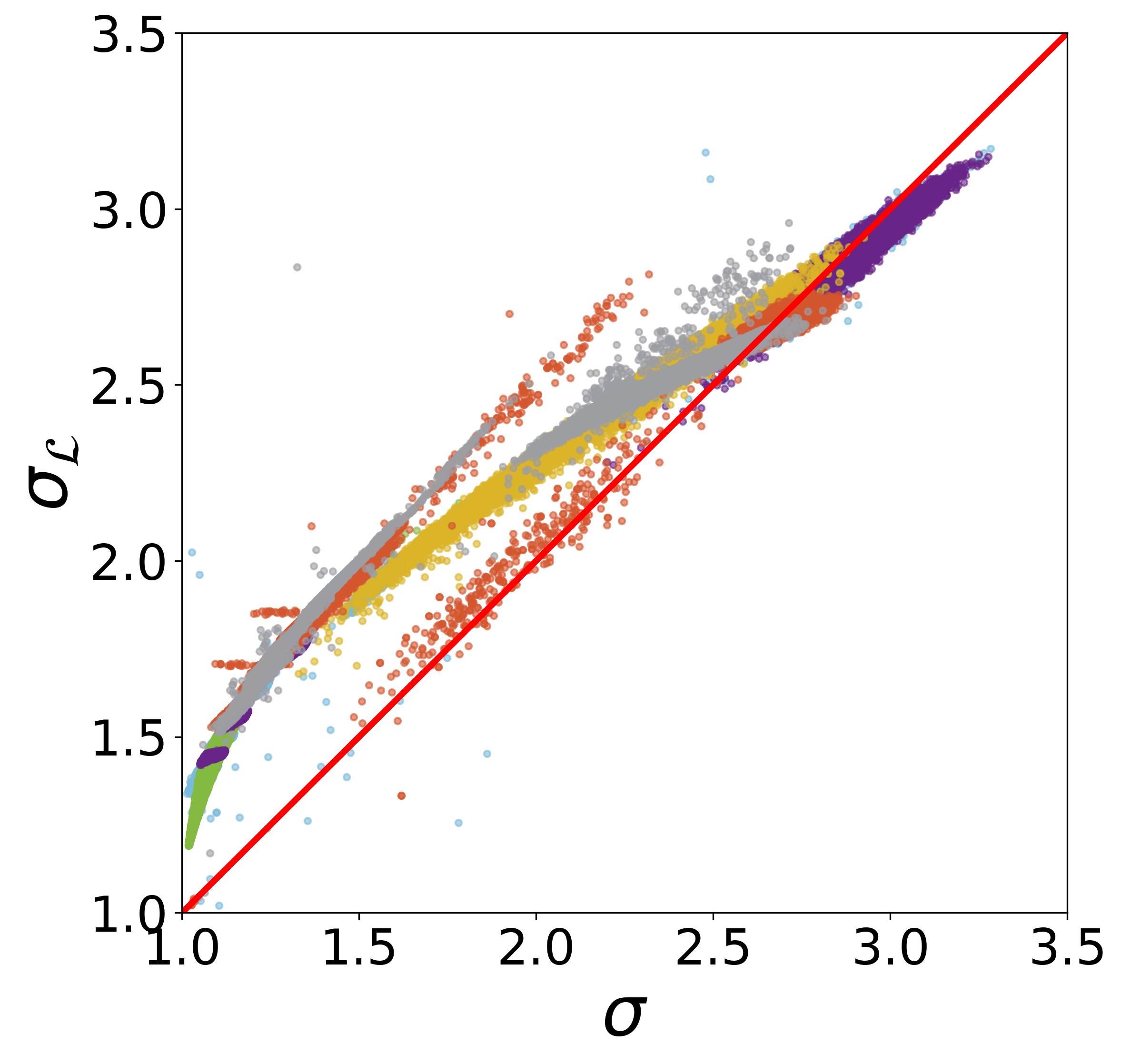}}
    \vspace{-3mm}

    \subfigure[$1-k^{-1},b/c$]{\includegraphics[width=0.265\linewidth]{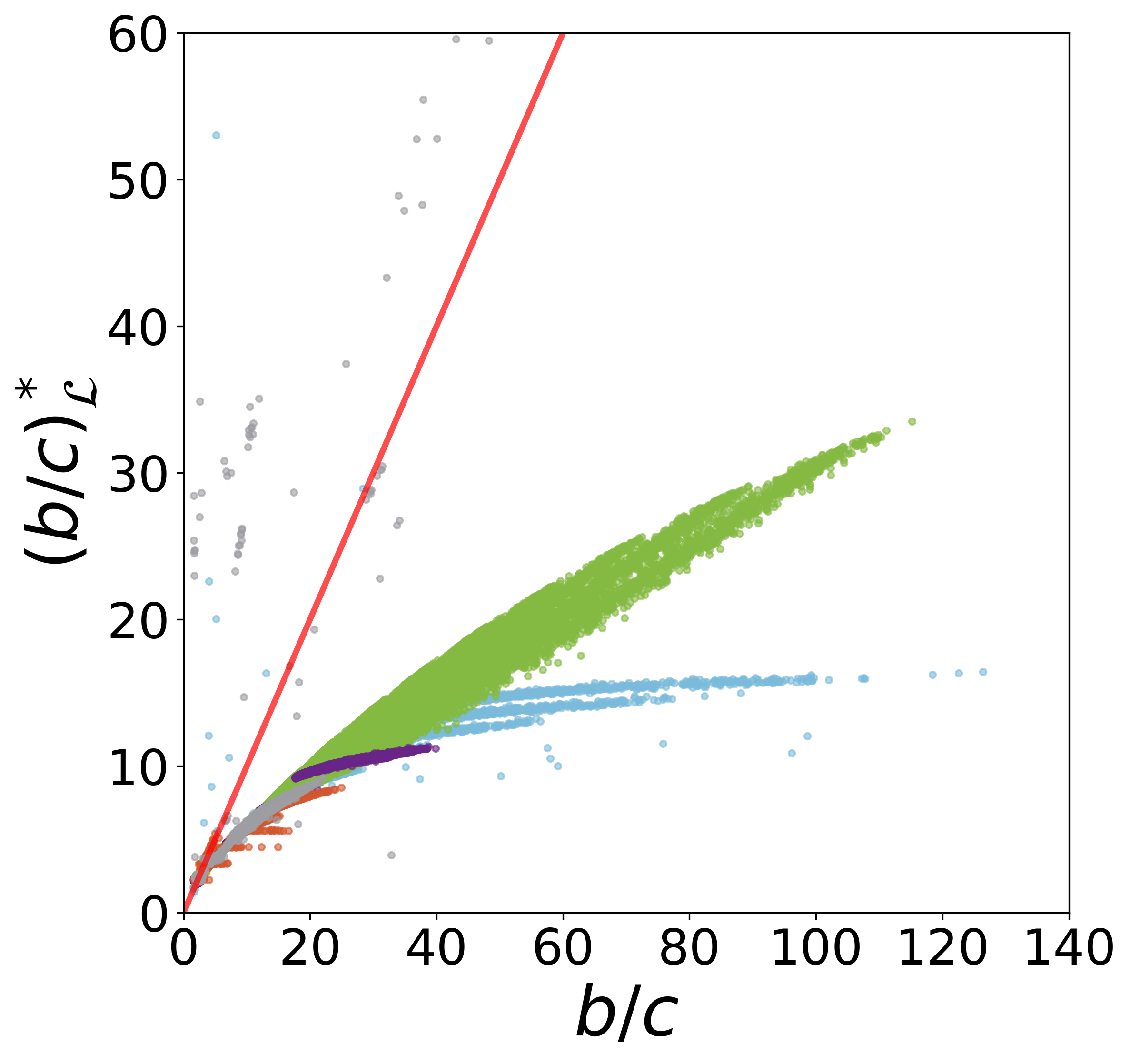}}
    \hspace{-3mm}
    \subfigure[$1-k^{-1},\langle k\rangle$]{\includegraphics[width=0.24\linewidth]{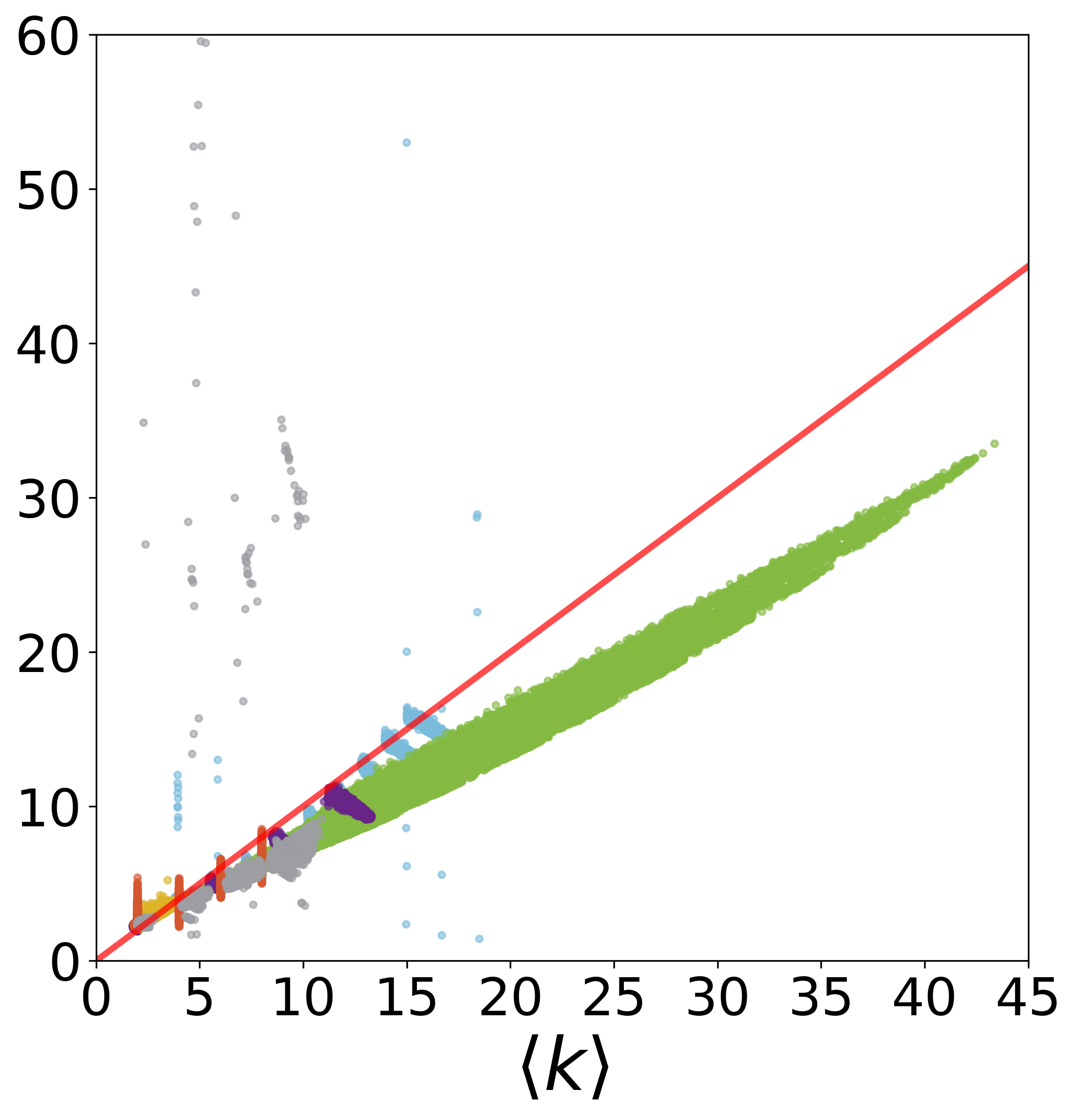}}
    \hspace{-3mm}
    \subfigure[$1-k^{-1},\langle k_{nn}\rangle$]{\includegraphics[width=0.24\linewidth]{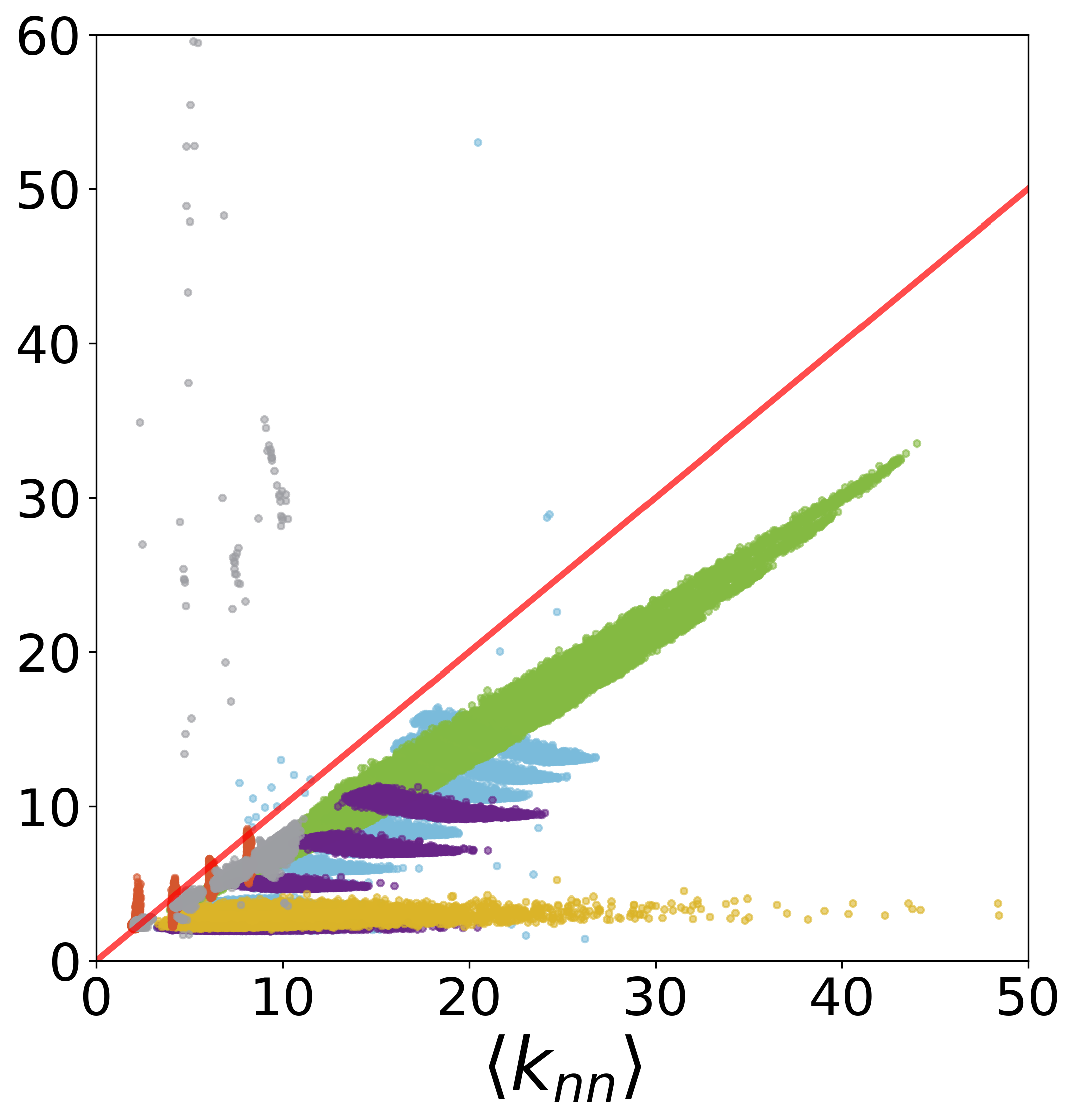}}
    \hspace{-3mm}
    \subfigure[$1-k^{-1}, \sigma$]{\includegraphics[width=0.25\linewidth]{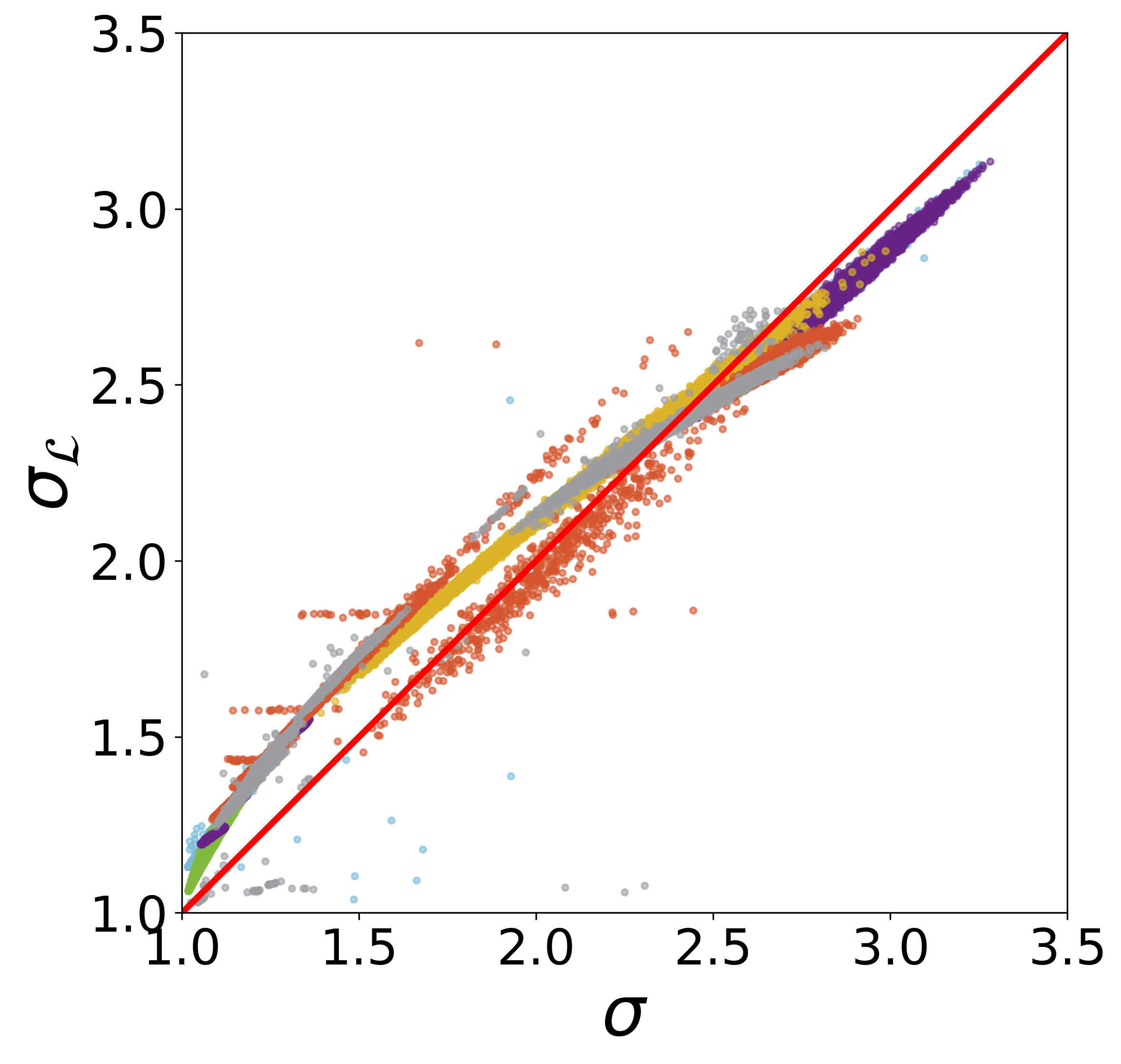}}
    \vspace{-3mm}

    \subfigure[$(k+1)^{-1},b/c$]{\includegraphics[width=0.265\linewidth]{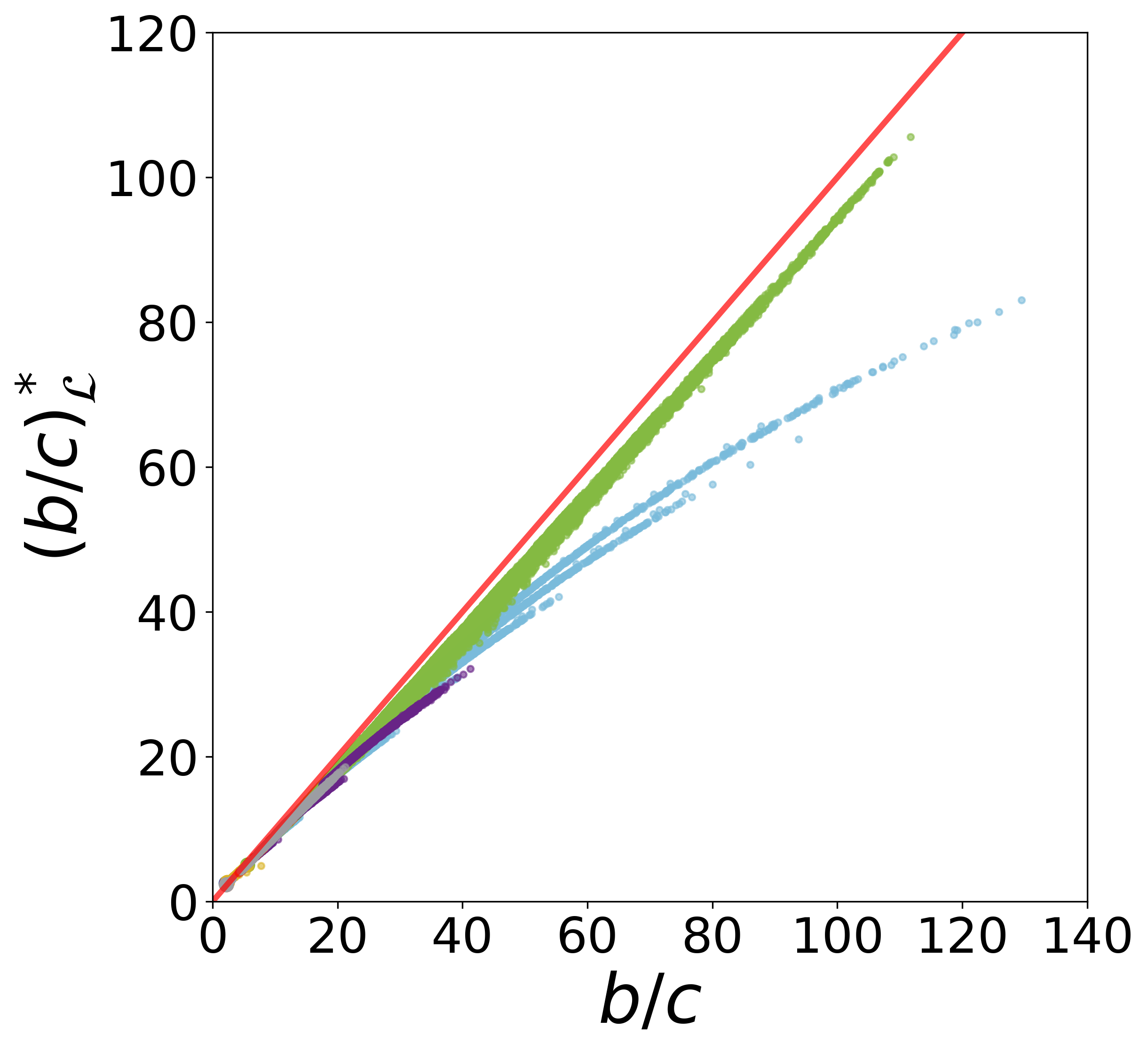}}
    \hspace{-3mm}
    \subfigure[$(k+1)^{-1},\langle k\rangle$]{\includegraphics[width=0.24\linewidth]{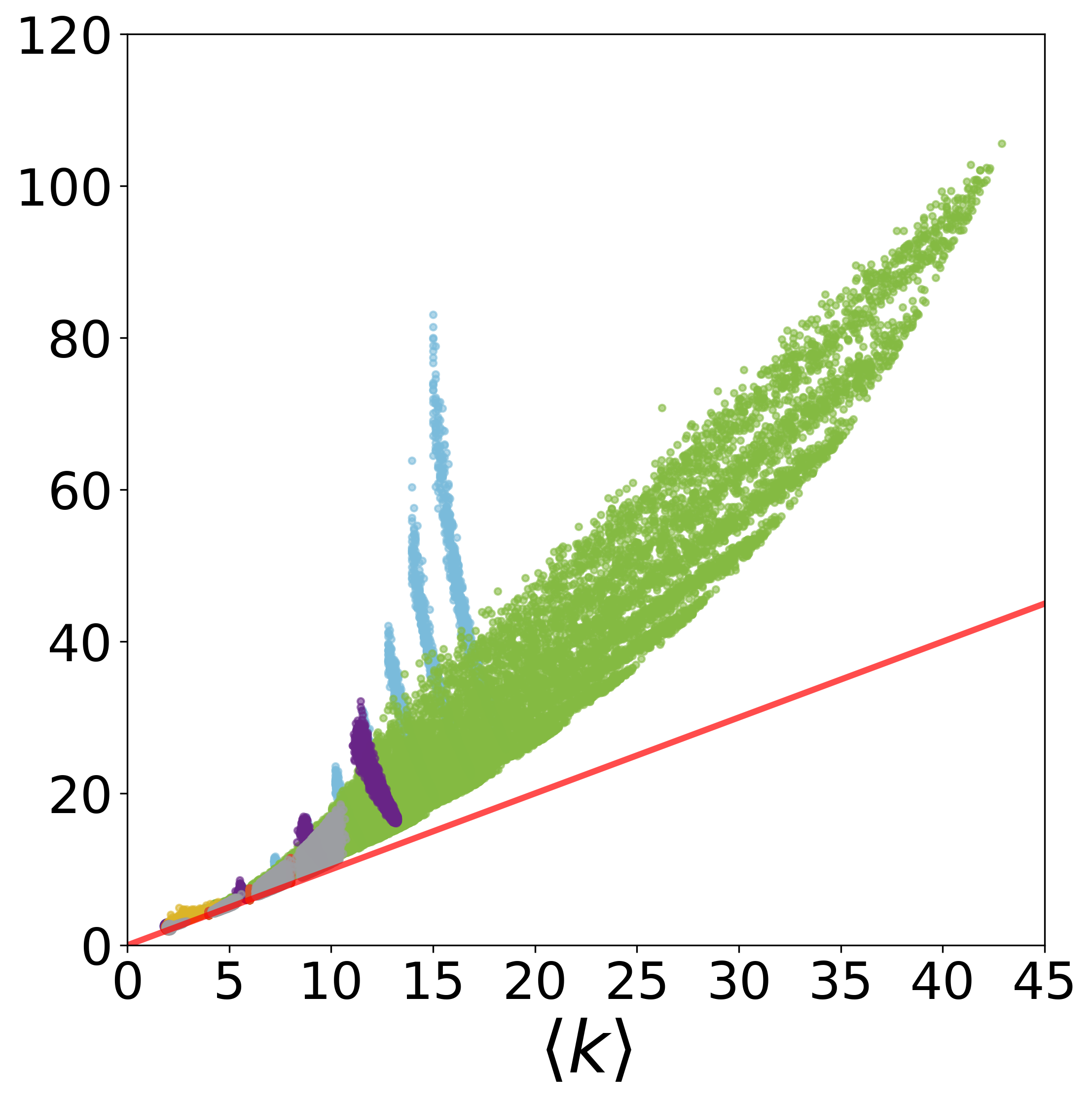}}
    \hspace{-3mm}
    \subfigure[$(k+1)^{-1},\langle k_{nn}\rangle$]{\includegraphics[width=0.24\linewidth]{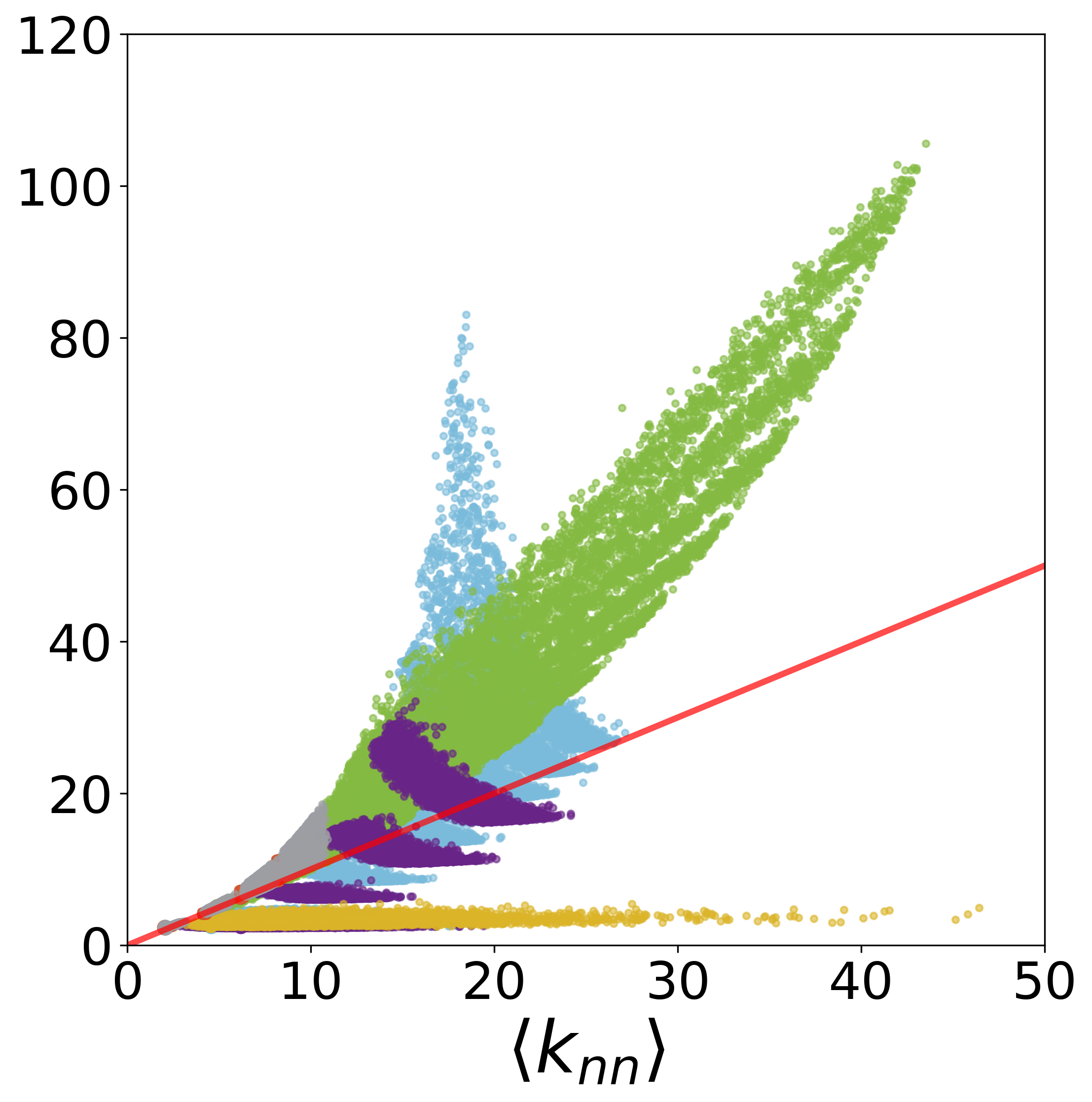}}
    \hspace{-3mm}
    \subfigure[$(k+1)^{-1}, \sigma$]{\includegraphics[width=0.25\linewidth]{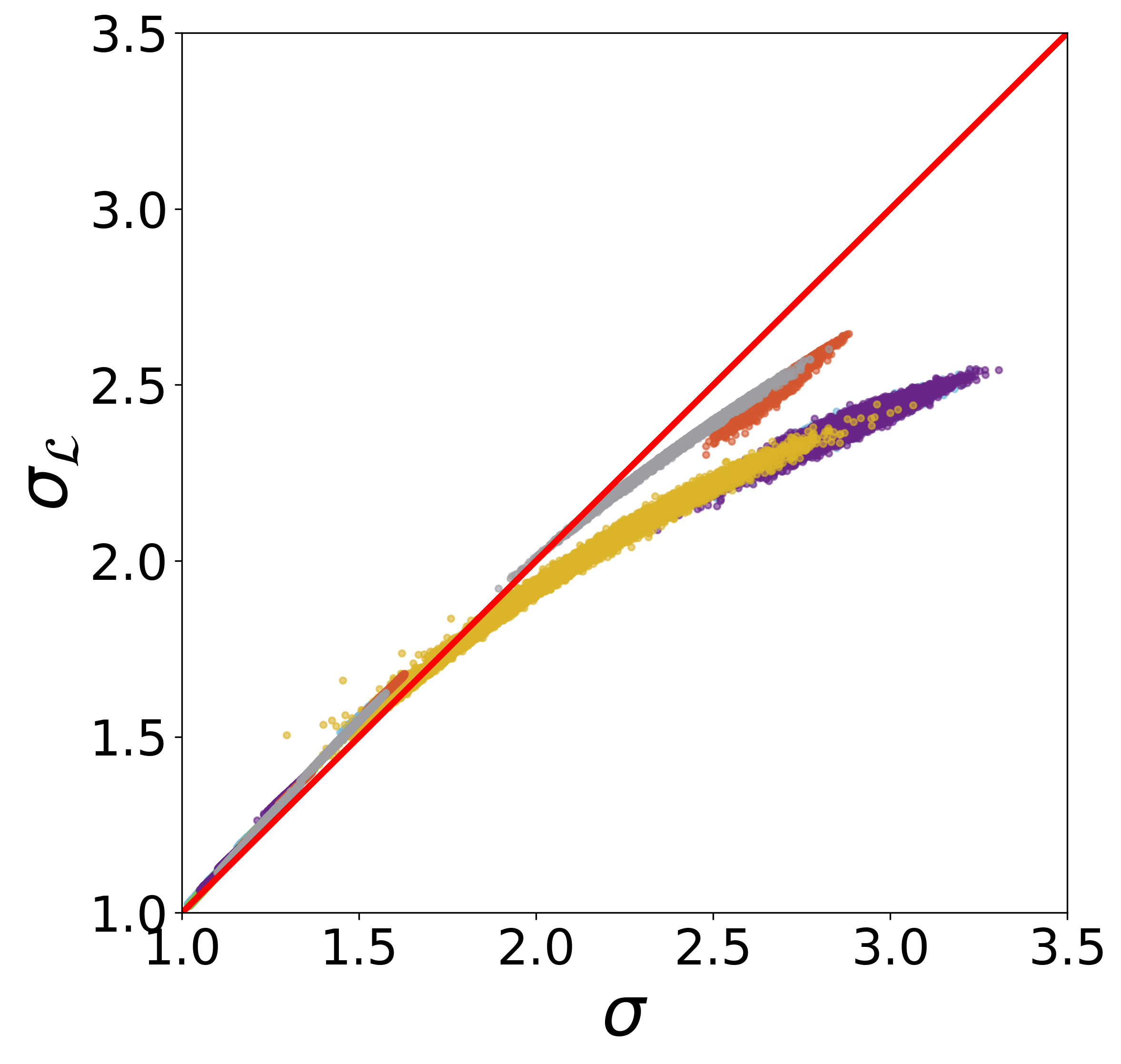}}
    \vspace{-3mm}
    \caption{\textbf{Conditions for cooperation with self-interaction in random networks compared to different network properties.} (a)-(d) $e^{-k}$. (e)-(h) $\ln{k}$. (i)-(l) $1-k^{-1}$. (m)-(p) $(1+k)^{-1}$. The considered networks are BAN, ERN, HKN, DDN, WSN, and NWN. We compare the cooperation conditions before and after the consideration of self-interaction in (a), (e), (i), and (m). In (b)(f)(j)(n) and (c)(g)(k)(o) we compare the cooperation condition with self-interaction with the mean degree $\langle k\rangle$ and the mean degree of neighbors $\langle k_{nn}\rangle$, respectively. In (d)(h)(l)(p) we compare the structure coefficient before and after self-interaction. Each panel has over $1.25\times10^5$ data points that present the critical cooperation conditions. The red solid line indicates when values of vertical and horizontal axes are equal. (color online)}
    \label{fig: random}
\end{figure*}

In this section we consider random networks that are generated by algorithms to capture some typical nature of empirical network topologies. Here, we compare the cooperation conditions before and after the consideration of self-interaction. We consider six random network models: (1) Barab\'{a}si-Albert network (BAN)~\cite{barabasi1999emergence}, each new agent connects to $m$ existing vertices with the probability proportional to degrees, (2) Erd{\H o}s-R\'{e}nyi network (ERN)~\cite{erdos1960evolution}, an agent has the probability $p$ to connect every other agent, (3) Holme-Kim network (HKN)~\cite{holme2002growing}, based on the BAN, an additional edge is added between the new vertex and a randomly chosen second-order neighbor of the focal new vertex, (4) Duplication-Divergence network (DDN)~\cite{ispolatov2005duplication}, a random agent is duplicated, and the replica connects to the neighbors of the protoplast with the retention probability $p$, (5) Watts-Strogatz network (WSN)~\cite{watts1998collective}, each vertex has the rewiring probability $p$ in the nearest neighbor coupled network with the degree $k$, (6) Newman-Watts network (NWN)~\cite{newman1999renormalization}, each edge is added between two randomly selected vertices in the nearest neighbor coupled network with the degree $k$. We stress that a WSN should be connected, otherwise, the condition for cooperation is $\infty$. Here we also consider the structure coefficient theory~\cite{tarnita2009strategy}. We denote the structure coefficients before and after self-interaction as $\sigma$ and $\sigma_{\mathbf{\mathcal{L}}}$, respectively. In a donation game, this theory suggests that cooperation is favored if $(b-c)\sigma-c>b$, i.e., $\sigma=[(\frac{b}{c})^*+1]/[(\frac{b}{c})^*-1]$ and $\sigma_{\mathbf{\mathcal{L}}} = [(\frac{b}{c})^*_{\mathbf{\mathcal{L}}}+1]/[(\frac{b}{c})^*_{\mathbf{\mathcal{L}}}-1]$. In Fig.~\ref{fig: random}(d), we show the comparison between $\sigma$ and $\sigma_{\mathbf{\mathcal{L}}}$. 

In Fig.~\ref{fig: random}, we show cooperation conditions in the mentioned six random network models and compare them before and after the consideration of self-interaction. In Figs.~\ref{fig: random}(a), we find that the usage of self-interaction function $e^{-k}$ makes difficult to promote cooperation. The critical conditions almost remain the same value obtained without self-interaction. This is because the self-interaction strength is extremely small in large-degree networks, e.g., for an agent with degree $k=10$, the self-interaction strength is only $e^{-10}\approx4.5\times10^{-5}$, which hardly changes the strategy selection process. Additionally, for a network with a small degree, the condition for cooperation is already small without self-interaction. Therefore, $e^{-k}$ is not an ideal self-interaction function for networked systems. Figs.~\ref{fig: random}(b)-(c) compare the cooperation condition to $\langle k\rangle$ and $\langle k_{nn}\rangle$, which show the necessary conditions in regular~\cite{ohtsuki2006simple} and heterogeneous graphs~\cite{konno2011condition} respectively. For $e^{-k}$, we still need $b/c>k$ condition to reach cooperation.

If the self-interaction function is $\ln{k}$, the cooperation condition can be significantly reduced. As shown in Fig.~\ref{fig: random}(e), $\ln{k}$ often helps the system to reduce the condition exponentially. agents only need to pay a quite small cost to achieve cooperation compared to the case without self-interaction. In Figs.~\ref{fig: random}(f) and (g), we also compare the cooperation condition with $\langle k\rangle$ and $\langle k_{nn}\rangle$. Our results suggest that self-interaction of $\ln{k}$ can considerably reduce the cooperation condition for networks with $k>5$. The conditions can be smaller than both $\langle k\rangle$ and $\langle k_{nn}\rangle$. Regarding $1-k^{-1}$ and $(k+1)^{-1}$ in Figs.~\ref{fig: random}(i)-(k) and (m)-(o), we find that the cooperation conditions are reduced as well, but not as significant as $\ln{k}$. The effect of self-interaction strength with $1-k^{-1}$ can also lead the system to a cooperation condition smaller than the mean degree, but the same is not true with $(k+1)^{-1}$. For $\ln{k}$ and $1-k^{-1}$, we can see that some data points are above the red lines, indicating small degree networks. In this case, the self-interaction may be counterproductive. 

In Figs.~\ref{fig: random}(d), (h), (l), and (p), we find that if a self-interaction strength can significantly reduce the cooperation conditions, the structure coefficients are mostly increased after self-interaction, e.g., $\ln{k}$ and $1-k^{-1}$. However, if the original $\sigma$ is large, the self-interaction is reduced after considering self-interaction, e.g., when $\sigma\approx3$, where the majority is HKN. 

\subsection{Real-world Networks}
\begin{figure*}
    \centering
    \subfigure[Dolphins]{\includegraphics[width=0.255\linewidth]{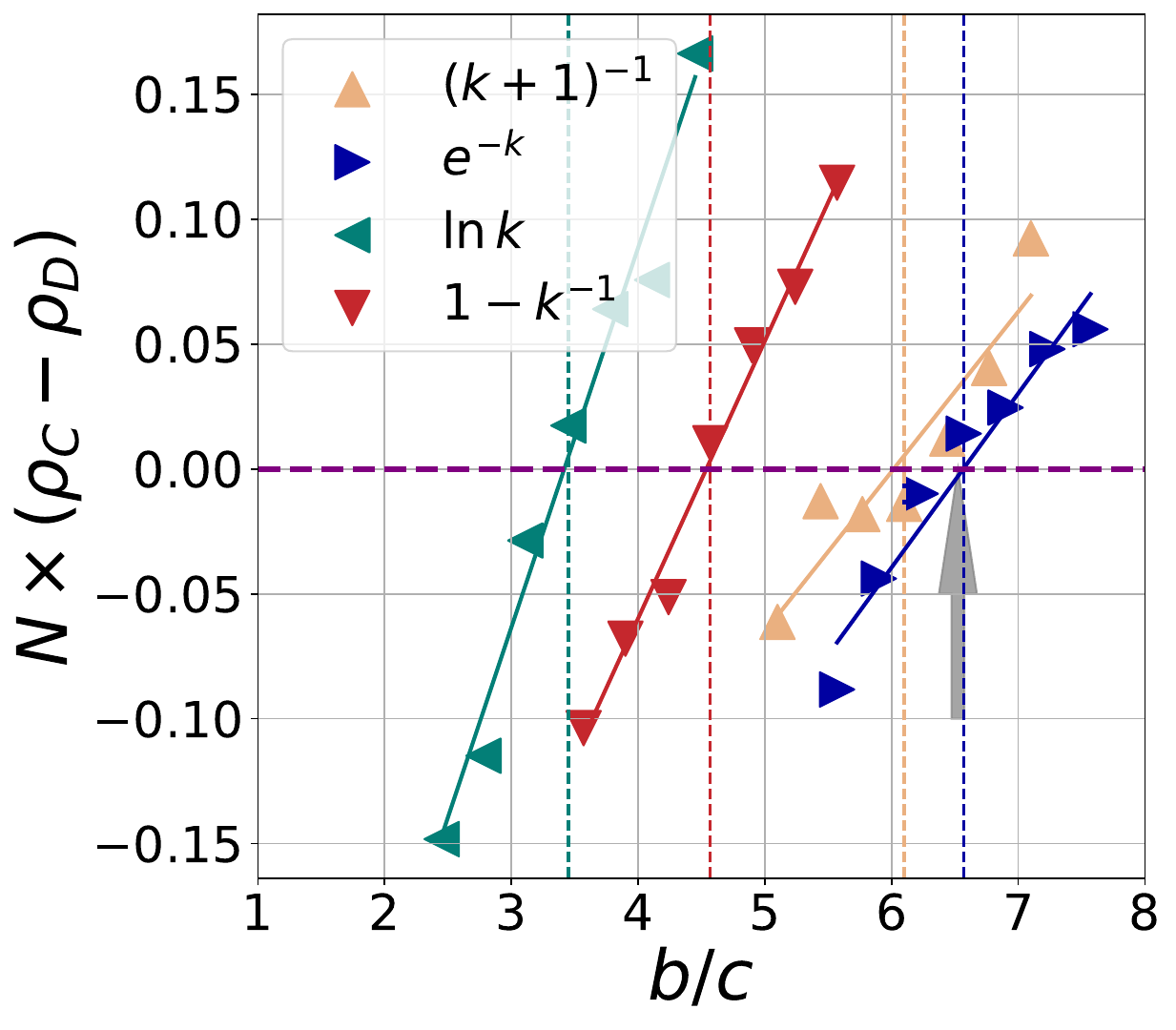}}
    \hspace{-3mm}
    \subfigure[American Football]{\includegraphics[width=0.24\linewidth]{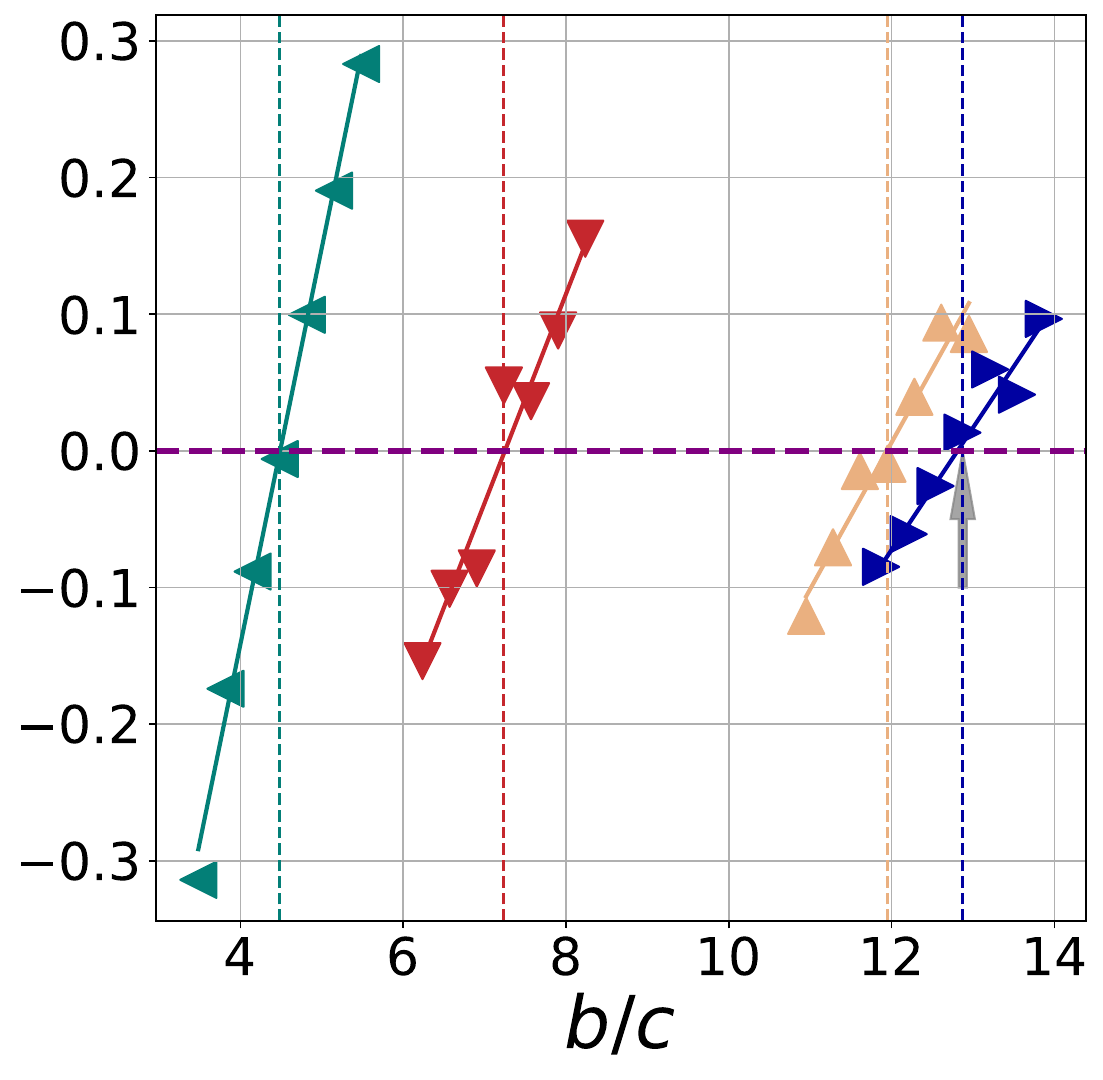}}
    \hspace{-3mm}
    \subfigure[USA Contiguous Border]{\includegraphics[width=0.24\linewidth]{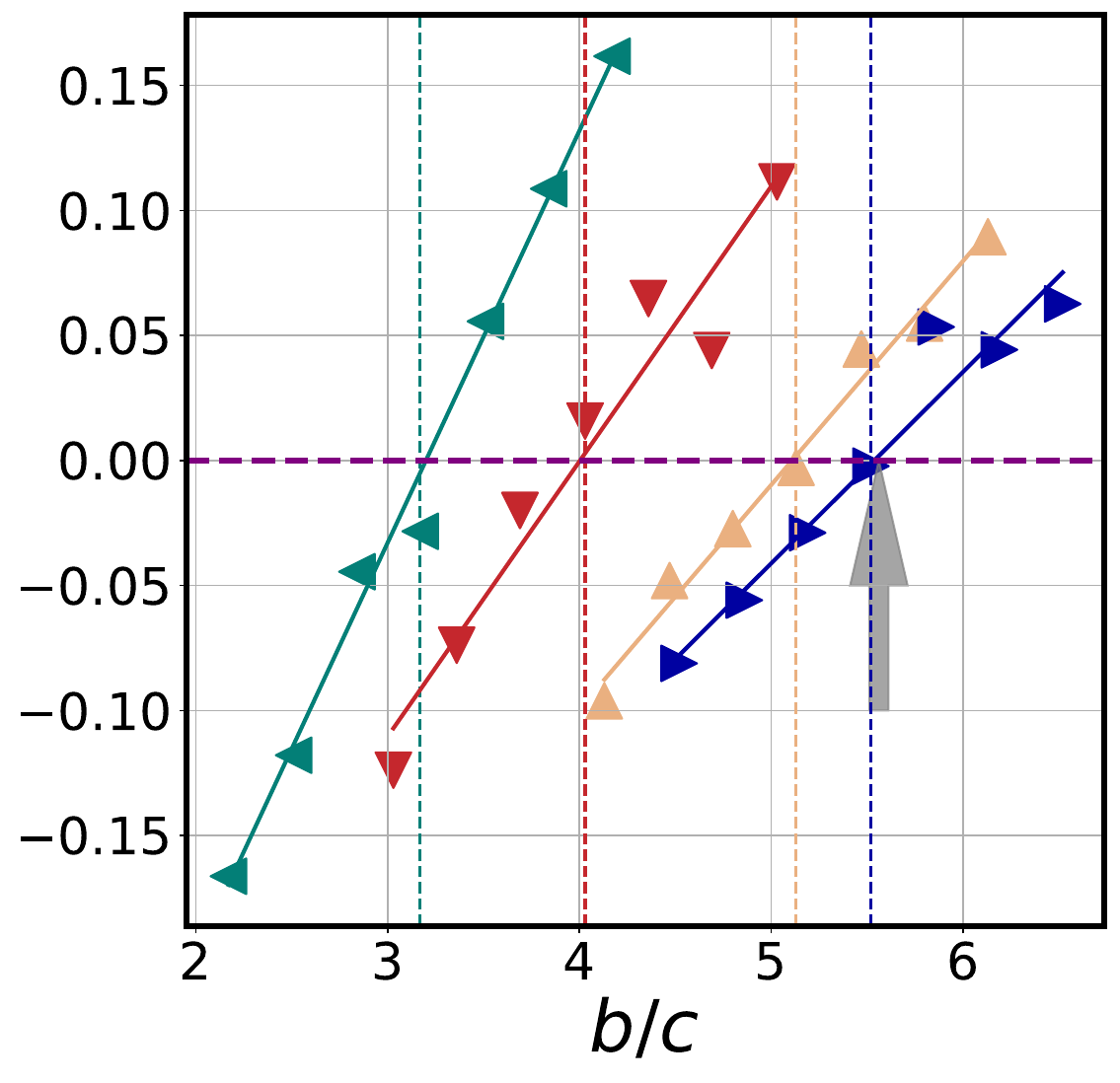}}
    \hspace{-3mm}
    \subfigure[Retweet]{\includegraphics[width=0.245\linewidth]{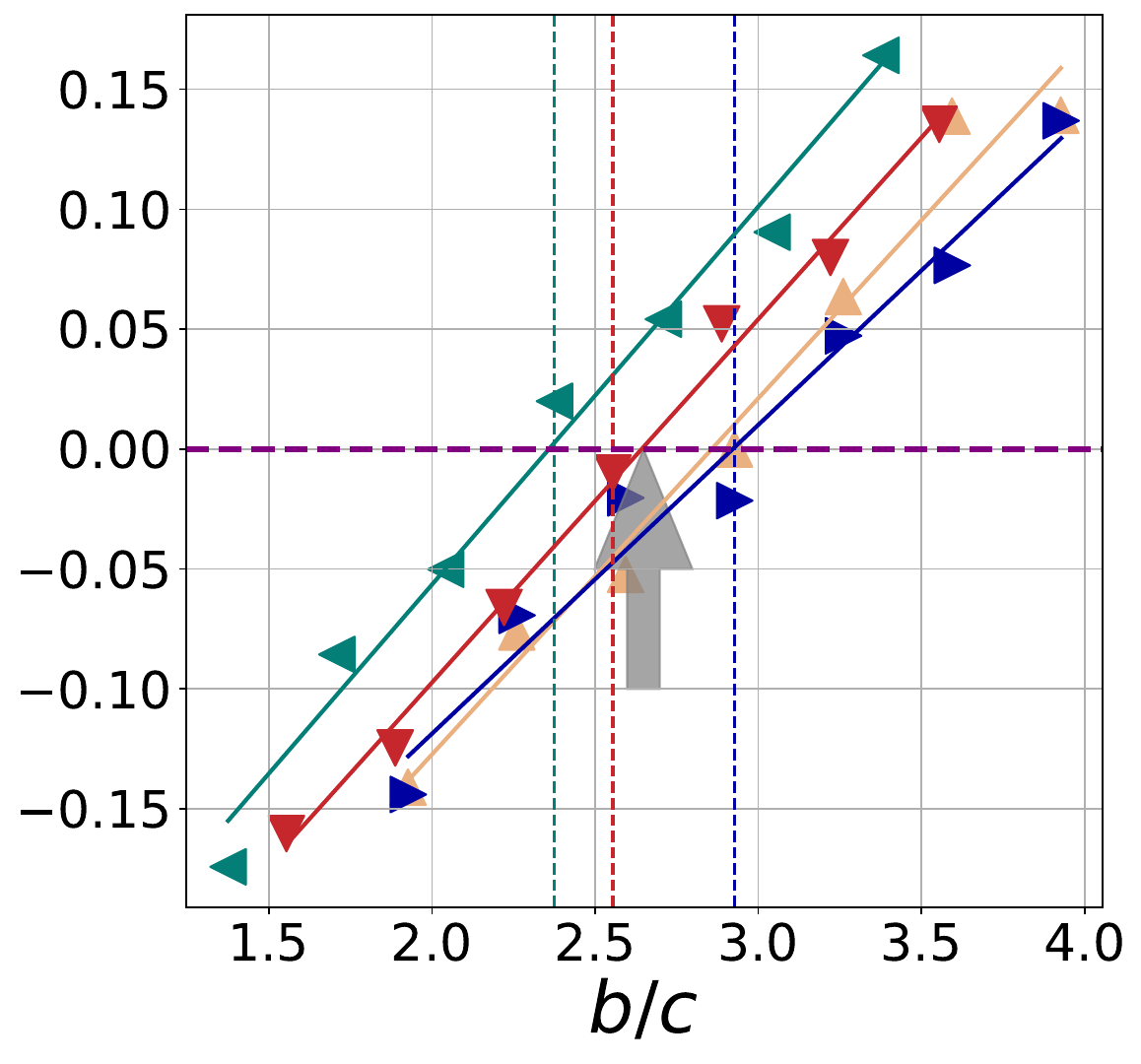}}
    \vspace{-3mm}

    \subfigure[Sandi Collaboration]{\includegraphics[width=0.25\linewidth]{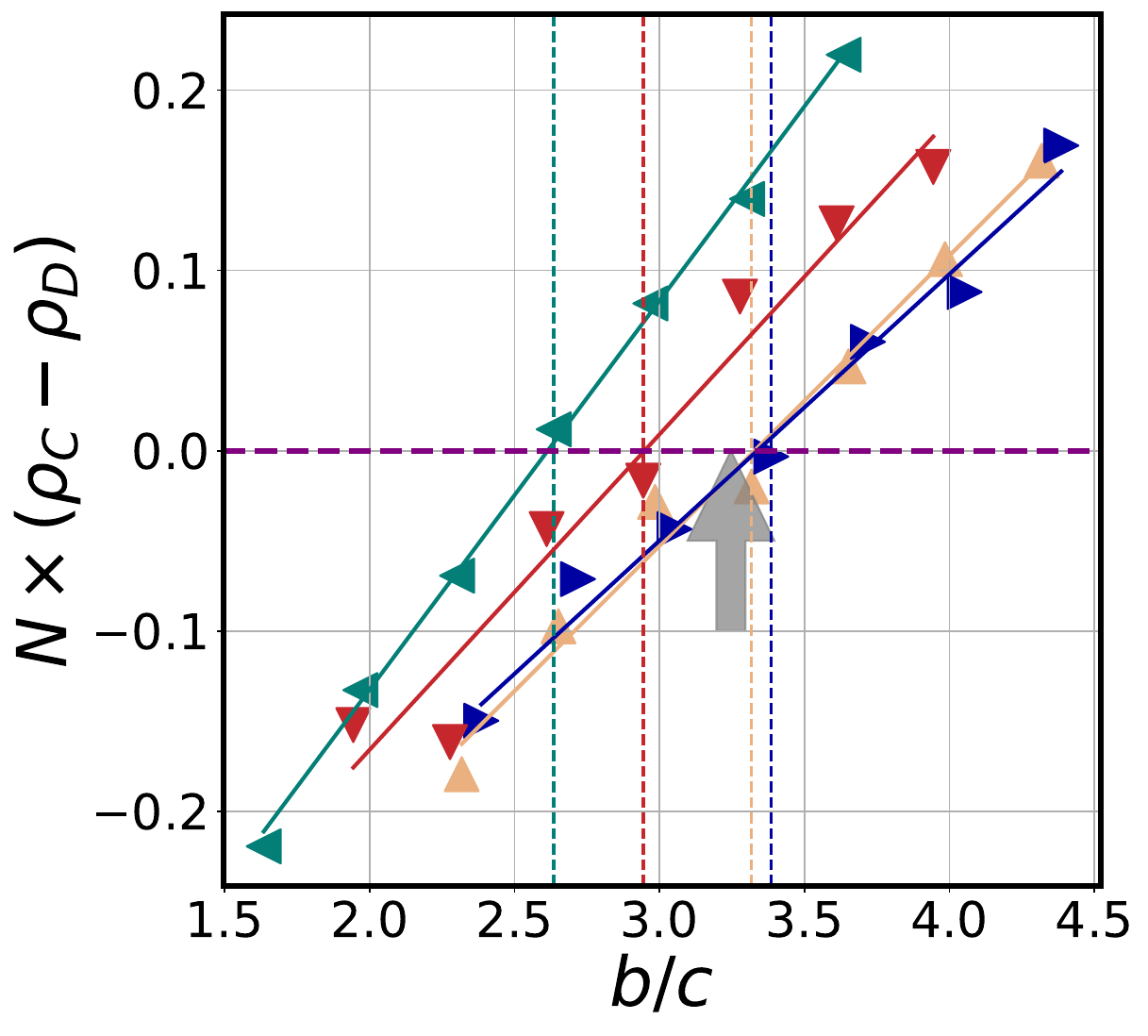}}
    \hspace{-3mm}
    \subfigure[Karate Club]{\includegraphics[width=0.24\linewidth]{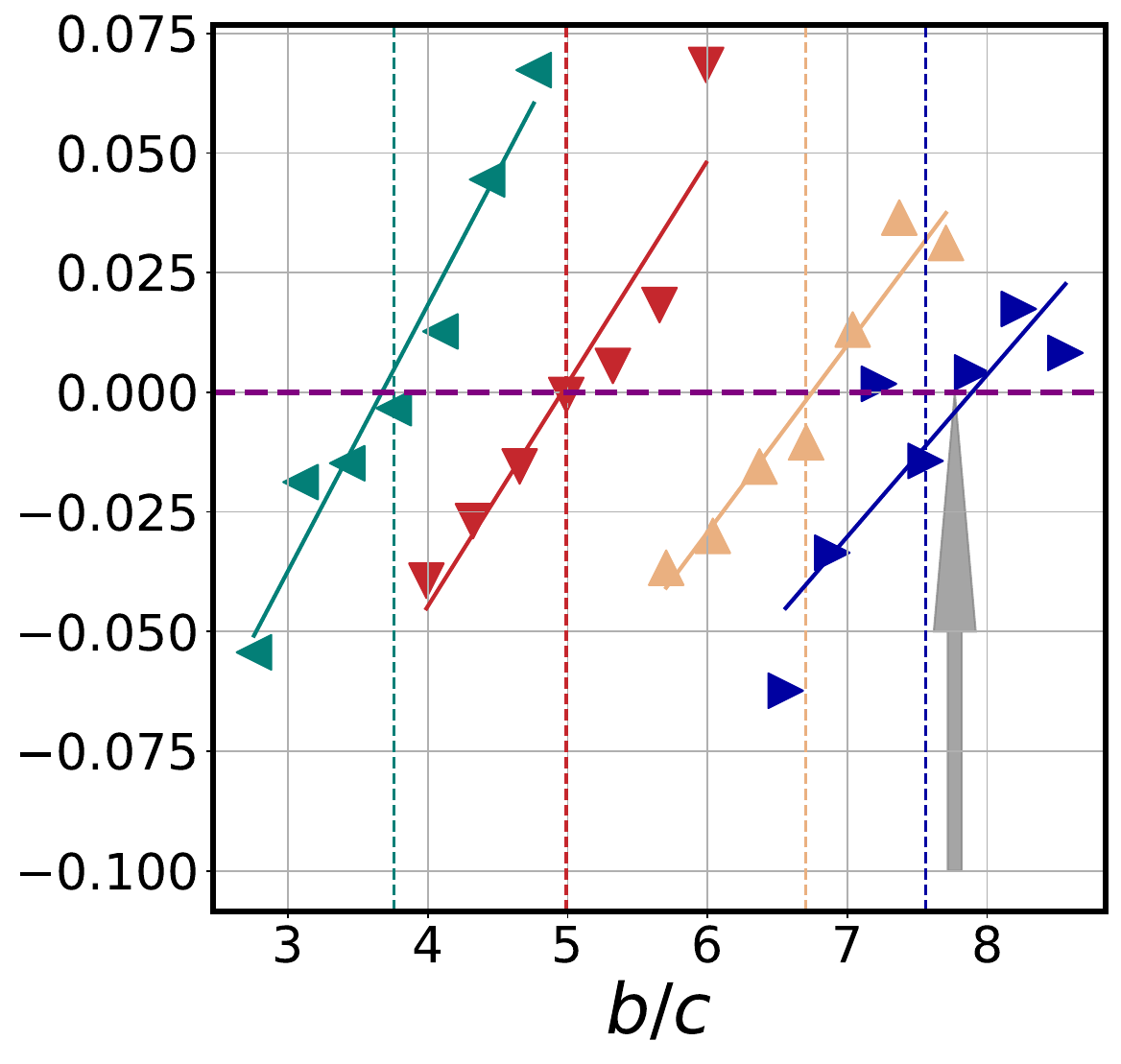}}
    \hspace{-3mm}
    \subfigure[Chesapeake Mesohaline Trophic]{\includegraphics[width=0.24\linewidth]{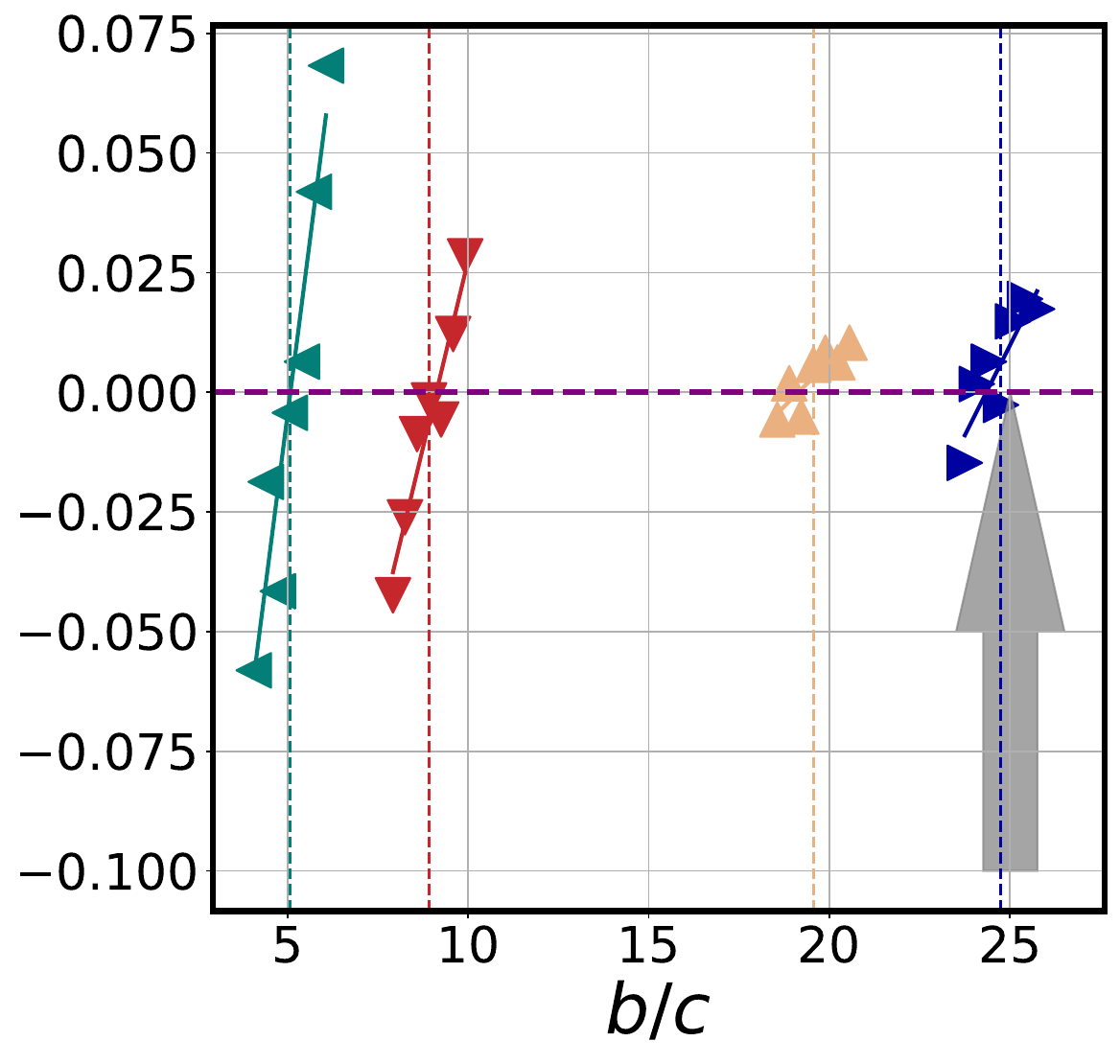}}
    \hspace{-3mm}
    \subfigure[Iceland]{\includegraphics[width=0.24\linewidth]{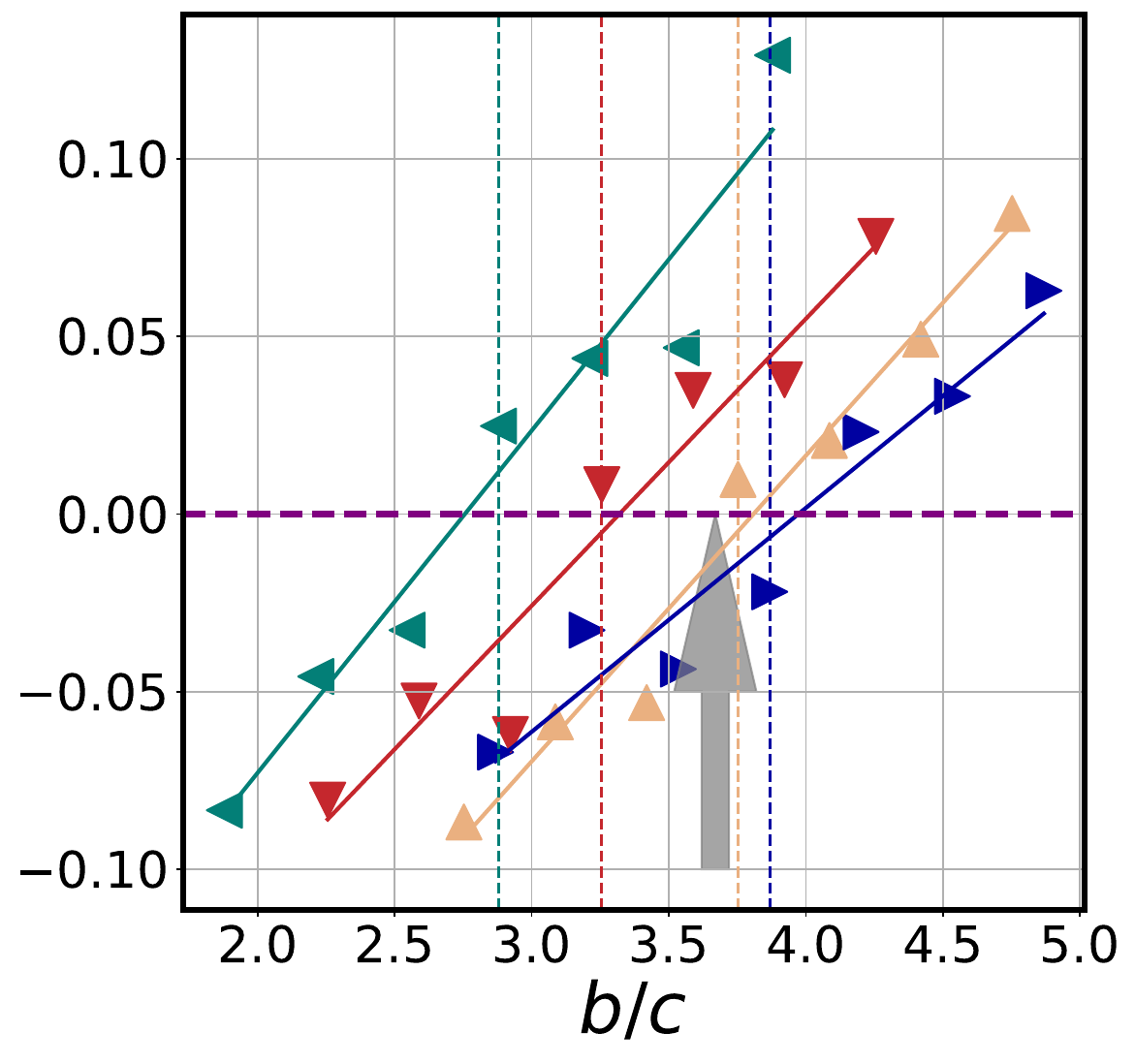}}
    \vspace{-3mm}
    \caption{\textbf{Fixation probabilities in real-world network data sets. }(a) Dolphin network with $N=62$. (b) American football network with $N=115$. (c) Contiguous border network of USA with $N=49$. (d) Retweet hashtag network with $N=96$. (e) Sandi collaboration network with $N=86$. (f) Social Karate club network with $N=34$. (g) Mesohaline trophic network of organisms carbon change in Chesapeake with $N=39$. (h) Iceland network with $N=75$. The vertical lines present the cooperation condition in specific self-interaction functions. The grey arrows denote the cooperation condition without self-interaction. Each data point is the ratio of cooperation fixation in $10^6$ independent runs. (color online)}
    \label{fig: real}
\end{figure*}
To check the robustness of our observations, we further explore the effect of self-interaction on the fixation of cooperation in four real-world network data sets. We first briefly introduce these networks: (a) the social association network of bottlenose dolphins with $N=62$ vertices and $\langle k\rangle=5$ in New Zealand~\cite{lusseau2003bottlenose}, (b) the network of American football games between Division IA colleges with $N=115$ vertices with $\langle k\rangle=10.66$~\cite{girvan2002community}, (c) the USA border connection network of every two states with $N=49$ vertices and $\langle k\rangle=4$~\cite{newman2015development}, and (d) the retweet network from various social and political hashtag with $N=96$ vertices and $\langle k\rangle=2$~\cite{rossi2014fast}. (e) the Sandi authors network with $N=86$ vertices and $\langle k\rangle=2$~\cite{rossi2015network}. (f) Zachary karate club network with $N=34$ vertices and $\langle k\rangle=4.59$~\cite{zachary1977information}. (g) the mesohaline trophic network of Chesapeake Bay with $N=39$ and $\langle k\rangle=8.72$~\cite{sanders2014benchmarking}. (h) the sexual contacts of male homosexuals in Iceland with $N=75$ and $\langle k\rangle=3.04$~\cite{liljeros2001web}. These network data sets can be found in~\cite{rossi2015network}. 

As shown in Fig.~\ref{fig: real}, we apply the mentioned four self-interaction strength functions. In a network with relatively high average connections, the self-interaction is weak and tends to zero if we consider $e^{-k}$. We can see that $e^{-k}$ hardly influences the cooperation condition in Figs.~\ref{fig: real}(a)-(c) and (e)-(g), but increases the condition in Fig.~\ref{fig: real}(d). Additionally, in Figs.~\ref{fig: real}(a)-(c), the cooperation conditions are significantly reduced for $\ln{k}$ and $1-k^{-1}$, and $(k+1)^{-1}$ is also slightly helpful in promoting the cooperation. However, for the retweet, Sandi collaboration network, and iceland networks, the self-interaction may not be effective in reducing the cooperation conditions. The reason may be the small average degree $\langle k\rangle=2$ and $3$. Our previous finding in Thm.~\ref{Theorem: 1}(b) for the regular graph family is very similar to this phenomenon. In a small degree case, the self-interaction does not always reduce the cooperation condition. 

\section{Conclusions and Outlooks}\label{sec: conclusion}
We study an evolutionary donation game model with self-interaction learning in networked systems. We defined a self-interaction landscape to describe the strength of the strategy self-replacement. We identify the condition for cooperation in several specific graph families, including regular graphs, stars, hub-hub joined stars, and ceiling fans. Additionally, we further present the condition of self-interaction strength that makes cooperation possible to overcome the evolution of spite. Our simulation results further show that with a proper self-interaction landscape, the condition for cooperation can be significantly reduced even with small weights of self-interaction, especially for the large-degree systems that are arduous to achieve cooperation. 

Although our results suggest that self-interaction shows fundamental advantages in promoting cooperation, there are still some vital issues to address. (i) The self-interaction can reduce the condition for cooperation and save the system from the evolution of spite. However, it also increases the average fixation time, because a mutant as well as its neighbor is very likely to maintain the original strategies, especially a large self-interaction strength. How to balance the strong cooperation and the long-range fixation time beyond Moran process~\cite{tkadlec2021fast} is a crucial problem. A mathematical method is needed to determine the appropriate marginal slowdown. (ii) Self-interaction is essentially adding self-loops to the network structure. Is there any strategy to modify the network structure in a more general way, so that the cooperation condition can be reduced, with minimal modification? (iii) What type of self-interaction can reduce the cooperation condition of an arbitrary but fixed networked system?

The problem discussed here can also be further studied from the following  perspectives. One can consider the self-interaction in different interaction and replacement graphs. The structure of the game interaction for acquiring payoff and fitness is different from the structure of strategy replacement or update. The effect of self-interaction in higher-order networks and temporal networks could also be a challenge. The effects of interaction among modular subnetworks on the evolution of cooperation are a promising direction to study evolutionary dynamics~\cite{dong2021optimal}. The question of how to achieve targeted cooperation control in a networked system is also challenging~\cite{liu2021efficient}. 
\appendices
\section{Proof for LEM.~\ref{Lemma: 1} and Some Useful Terms}\label{Appendix: A}
\small
We briefly review the proof of Lemma~\ref{Lemma: 1} because some technical terms will be used in the following analysis. We start by calculating each agent's payoff. For the mentioned donation game, the agent $i$'s payoff is $f_i(\mathbf{\mathcal{S}})=-cs_i+b\sum_j p_{ij}^{(1)}s_j$, where $p_{ij}^{(n)}$ is the probability of the $n$-step random walk from $i$ to $j$. If the selection strength is $0$, the evolutionary game process degenerates into the neutral drift model as a martingale~\cite{chen2013sharp}. In this case, the fixation probability of the cooperation is $1/N$. The non-zero selection strength then leads to the destabilization of such a martingale, which can be denoted by
\begin{small}
\begin{equation}\label{Eq: martingale}
\rho_C=\frac{1}{N}+\delta \langle D'\rangle+O(\delta^2), 
\end{equation}
\end{small}
where $\langle D'\rangle$ is the first derivative of the degree-weighted cooperation frequency change rate. Substituting each agent's fitness using Taylor's formula $F_i(\mathbf{\mathcal{S}})=\exp[\delta f_i(\mathbf{\mathcal{S}})]=1+\delta f_i(\mathbf{\mathcal{S}}) +O(\delta^2)$ into $\langle D'\rangle$, with the weak selection assumption, we have
\begin{small}
\begin{equation}
\langle D'\rangle=\langle\delta \sum_{i\in\mathcal{V}}\pi_i \left[-cs_i(s^{(0)}_i-s^{(2)}_i)+bs_i(s^{(1)}_i-s^{(3)}_i)\right]\rangle.
\end{equation}
\end{small}
The spatial assortment condition~\cite{allen2017evolutionary, mcavoy2021fixation} of two strategies suggests that $\langle\sum_{i\in\mathcal{V}}\pi_is_i(s^{(n_1)}_i-s^{(n_2)}_i)\rangle=(\eta^{(n_2)}-\eta^{(n_1)})/2N$, where $\eta^{(n)}=\sum_{i,j\in\mathcal{V}}\pi_i p_{ij}^{(n)}\eta_{ij}$, $\pi_i$ denotes the stationary distribution of a random walk in $\mathcal{G}$, and $\eta_{ij}$ is the expected coalescence time of $i$ and $j$ that can be calculated by the following recurrence
\begin{small}
\begin{equation}\label{Eq: Recurrence}
\eta_{ij}=
(1-\theta_{ij})[1+\frac{1}{2}\sum_{k\in\mathcal{V}}(p_{ik}\eta_{jk}+p_{jk}\eta_{ik})], 
\end{equation}
\end{small}
where $\theta$ is a Kronecker function with $\theta_{ij}=1$ if $i=j$ and $\theta_{ij}=0$ if $i\neq j$. Combining these equations directly induces Eq.~\ref{Eq: martingale}. 
$\hfill\blacksquare$

We can simplify some technical terms for our following analysis. We define $\eta_i=1+(\sum_{k\in\mathcal{V}}p_{ik}\eta_{ik})/2$, via the recurrence condition based on Eq.~\ref{Eq: Recurrence}, we have $\eta^{(n+1)}=\sum_{i\in\mathcal{V}}\pi_i p_{ii}^{(n)}\eta_i+\eta^{(n)}-1$. Letting $\eta^{(n)}\rightarrow\infty$, we have $\sum_{i\in\mathcal{V}}\pi_i^2\eta_i=1$. The main notations of this paper are shown in Tab. \ref{tab: notations}. 
\begin{table}[h]
\centering
\caption{Notations}\label{tab: notations}
\begin{tabular}{cc}
   \toprule
   Symbol & Definition\\
   \midrule
   $\eta^{(n)}$ & The expected coalescence time of all $n$-step random walks\\
   $\rho_C$ ($\rho_D$) & The fixation probability of cooperation (defection)\\
   $k$ & The degree of a regular network\\
   $\ell(k)$&The self-loop strength of the vertex with a degree $k$\\
   $\alpha$ & The self-interaction weight of the vertex with degree $1$\\
   $\beta$ & The self-interaction weight of the vertex with degree $N-1$ \\
   $\gamma$ & The self-interaction weight of the vertex with degree $N$ \\
   $\epsilon$ & The self-interaction weight of the vertex with degree $2$ \\
   $p_{ij}^{(n)}$ & The probability that a random walk takes $n$ steps from $i$ to $j$ \\
   $\eta_{ij}$ & The expected coalescence time of a random walk from $i$ to $j$  \\
   $\pi_i$ & The stationary probability of a random walk\\
   $\mathcal{S}$ & The strategy vector of the network\\
   $b$ & The benefit of a donation game\\
   $c$ & The cost to donate in a donation game\\
   $f_i(\mathcal{S})$ & The payoff of the vertex $i$ given the strategy vector $\mathcal{S}$\\
   $F_i(\mathcal{S})$ & The fitness of the vertex $i$ given the strategy vector $\mathcal{S}$\\
   $\delta$ & The selection strength\\
   \bottomrule
\end{tabular}
\end{table}
\section{Proof for THM.~\ref{Theorem: 1}: Regular Graphs}\label{Appendix: B}
\small
We first prove $(a)$. In regular graphs, we assume that all vertices have the same degree and self-interaction strength. Therefore, the corresponding stationary distribution of a random walk is $1/N$ for all vertices. Accordingly, our previous discussions suggest that $\sum_{i\in\mathcal{V}}\pi_{i}\eta_{i}=N$. Based on the recurrence equation, we have $\eta^{(1)}=\sum_{i\in\mathcal{V}}\pi_{i} \eta_{i}-1$, $\eta^{(2)}=\sum_{i\in\mathcal{V}}\left[1+(\ell(k))/(k+\ell(k))\right]-2$, and $\eta^{(3)}=\sum_{i\in\mathcal{V}}\pi_{i} \eta_i\left[1+(\ell(k))/(k+\ell(k))+(\ell^2(k)+1)/(k+\ell(k))^2\right]-3$. Substituting $\eta^{(1)}$, $\eta^{(2)}$, and $\eta^{(3)}$ into Lem. \ref{Lemma: 1}, we can directly obtain the condition for cooperation to be favored as Eq.~\ref{Eq: bcr regular}. 

Next we prove (b). To explore if self-interaction helps to form cooperation or not, we can compare Eq.~\ref{Eq: bcr regular} to $(N-2)/(N/k-2)$, which is the critical condition for cooperation without self-interaction. If the regular graph is sparse enough, we assume that the self-interaction increases the critical condition. If $N>2k$, this assumption can be indicated and reduced as
\begin{small}
\begin{equation}
\begin{aligned}
&\frac{N\left[k^2+3k\ell(k)+2\ell^2(k)\right]-2\left[k+\ell(k)\right]^2}{N\left[k\ell(k)+k+2\ell^2(k)\right]-2\left[k+\ell(k)\right]^2}-\frac{N-2}{N/k-2}>0\iff\\
&N\{\left[k+\ell(k)\right]^2-\ell(k)k\left[k+\ell(k)\right]-k\left[k+\ell^2(k)\right]+\ell(k)\left[k+\ell(k)\right]\}\\
&+2\{k\left[k+\ell^2(k)\right]-\left[k+\ell(k)\right]^2\}>0\iff\\
&N\left[2\ell(k)+3k-2k\ell(k)-k^2\right]+2\left[k\ell(k)-\ell(k)-2k\right]>0\iff\\
&\ell(k)(2N-2kN+2k-2)+3kN-k^2N-4k>0. 
\end{aligned}
\end{equation}
\end{small}
This condition can be finally reduced to 
\begin{small}
\begin{equation}
\ell(k)<\frac{k\left[N(k-3)+4\right]}{2(k-1)(1-N)}. 
\end{equation} 
\end{small}
Apparently, the right-hand side of inequality is negative if $N>2$ and $k>2$, inducing $\ell(k)<0$, which is against Def.~\ref{def: 1}. Therefore, with our mentioned condition, we have $(\frac{b}{c})^*_{\mathbf{\mathcal{L}}}<(N-2)/(N/k-2)$. For a graph that is sparse enough, we have $k+2\ell(k)-1>0$ and $(\frac{b}{c})^*_{\mathbf{\mathcal{L}}}>1$, resulting that the evolution of spite will never be favored. The self-interaction in the networked system relaxes instead of increasing the critical condition for cooperation. This ends our proof. 

Last we prove (c). For a dense graph with $N<2k$, $\mathbf{\mathcal{L}}$ can enhance cooperation within a certain interval of $\ell(k)$. The numerator of Eq.~\ref{Eq: bcr regular} is greater than $0$. If the denominator is greater than $0$ as well, there will be a chance for the system to favor cooperation, which is
\begin{small}
\begin{equation}\label{eq: reg2}
N\left[k\ell(k)+k+2\ell^2(k)\right]-2\left[k+\ell(k)\right]^2>0. 
\end{equation}
\end{small}
Reduce Eq.~\ref{eq: reg2} directly leads to Thm.~\ref{Theorem: 1}(c). 
$\hfill\blacksquare$

\section{Proof for THM.~\ref{Theorem: 2}: Star Graphs}\label{Appendix: C}
\small
We start from the critical condition for cooperation in (a). The probabilities for a one-step random walk from a leaf to itself or the hub are $p_{LL}=\alpha/(1+\alpha)$ and $p_{LH}=1/(1+\alpha)$ respectively, and from a hub to itself or a leaf are $p_{HH}=\beta/(N-1+\beta)$ and $p_{HL}=1/(N-1+\beta)$ respectively. According to the recurrence condition Eq.~\ref{Eq: Recurrence}, we have the expected coalescence time between the hub and a leaf as 
\begin{equation}
\eta_{HL}=1+\frac{1}{2}[p_{HH}\eta_{HL}+(N-2)p_{HL}\eta_{LL}+p_{LL}\eta_{HL}],
\end{equation}
and between a leaf and another leaf as $\eta_{LL}=1+p_{LL}\eta_{LL}+p_{LH}\eta_{HL}$. Here we note that $\eta_{LL}$ denotes the quantity from a focal leaf to another arbitrary leaf because the expected coalescence time is $0$ if two ends of the coalescence process are identical. Solving the system of equations, we find that
\begin{small}
\begin{equation}
\eta_{HL}=\frac{(1+\alpha)(N\alpha-2\alpha+2\beta+3N-4)}{N+\alpha+\beta},
\end{equation}
\end{small}
and
\begin{small}
\begin{equation}
\eta_{LL}=\frac{N\alpha^2+5N\alpha-\alpha^2+3\alpha\beta+4N-5\alpha+3\beta-4}{N+\alpha+\beta}. 
\end{equation}
\end{small}
Therefore, we have
\begin{small}
\begin{equation}
\begin{aligned}
&\eta_H=1+(N-1)p_{HL}\eta_{HL}\\
&=1+\frac{(N-1)(1+\alpha)(N\alpha-2\alpha+2\beta+3N-4)}{(N-1+\beta)(N+\alpha+\beta)},\\
\end{aligned}
\end{equation}
\end{small}
and
\begin{small}
\begin{equation}
\eta_L=1+p_{LH}\eta_{LH}=\frac{N\alpha+4N-\alpha+3\beta-4}{N+\alpha+\beta}. 
\end{equation}
\end{small}
The state space of a random walk in a star topology can be divided into two categories, including the leaves and the hub, where each leaf has an equivalent stationary distribution. The stationary probability for a random walk in the hub is $\pi_H=(N-1+\beta)/(2N+N\alpha-2-\alpha+\beta)$, and in a leaf is $\pi_L=(1+\alpha)/(2N+N\alpha-2-\alpha+\beta)$. There are $N-1$ leaves and $1$ hub in the star, therefore the critical condition for cooperation is
\begin{small}
\begin{equation}\label{eq: star condition}
(\frac{b}{c})^*_{\mathbf{\mathcal{L}}}=\frac{(N-1)\pi_L\eta_L(1+p_{LL})+\pi_H\eta_H(1+p_{HH})-2}{(N-1)\pi_L\eta_L(p_{LL}+p_{LL}^{(2)})+\pi_H\eta_H(p_{HH}+p_{HH}^{(2)})-2}, 
\end{equation}
\end{small}
where $p_{LL}^{(2)}=p_{LL}^2+p_{LH}p_{HL}$ and $p_{HH}^{(2)}=p_{HH}^2+(N-1)p_{HL}p_{LH}$. We can reduce Eq.~\ref{eq: star condition} to directly obtain Eq.~\ref{eq: thm2 star}. 

Now we consider the effect of the hub self-interaction and the leaf self-interaction, denoted by $\alpha=0$ or $\beta=0$ respectively. We first prove (b) with $\alpha=0$ and only the hub vertex undergoes the self-interaction. In this case, the condition for cooperation is
\begin{small}
\begin{equation}\label{eq: star alpha0}
(\frac{b}{c})^*_{\mathbf{\mathcal{L}}\vert\alpha=0}=\dfrac{\parbox{6cm}{$\dfrac{3}{2}(N-1+\beta)[-\dfrac{8}{3}+\dfrac{4}{3}N^3+\dfrac{14}{3}\beta-\dfrac{4}{3}\beta^2+N^2(3\beta-\dfrac{16}{3})+N(\dfrac{4}{3}\beta^2-\dfrac{23}{3}\beta+\dfrac{20}{3})]$}}{\beta(-N^3+N^2(-\dfrac{1}{2}\beta+2)+N(-\dfrac{1}{2}\beta-1)+\beta)}.
\end{equation}
\end{small}
Apparently, Eq.~\ref{eq: star alpha0} is always negative if $N>2$, $\alpha>0$, and $\beta>0$. Therefore, if the self-interaction only works for the hub vertex, the system always favors the evolution of spite instead of cooperation. The agents are likely to pay a cost to impair others' interests. If $N\rightarrow+\infty$, the limitation of Eq.~\ref{eq: star alpha0} is $-\infty/\beta$. Due to our previous demonstration that the self-interaction strength is finite, this critical condition is $-\infty$. 

Then, we focus on the case that only leaves undergo self-interaction. With $\beta=0$, the critical condition is
\begin{small}
\begin{equation}\label{eq: star beta0}
(\frac{b}{c})^*_{\mathbf{\mathcal{L}}\vert\beta=0}=\frac{\parbox{6cm}{$\dfrac{3}{2}(N-1)(1+\alpha)[N^3(\alpha^2+\dfrac{11}{3}\alpha+\dfrac{4}{3})+N^2(-4\alpha^2-\dfrac{40}{3}\alpha-\dfrac{16}{3})+N(5\alpha^2+\dfrac{47}{3}\alpha+\dfrac{20}{3})-2\alpha^2-6\alpha-\dfrac{8}{3}]$}}{\parbox{6cm}{$\alpha^3(N^4-5N^3+9N^2-7N+2)+\alpha^2(4N^4-20N^3+36N^2-28N+8)+\alpha(N^4-7N^3+15N^2-13N+4)$}}. 
\end{equation}
\end{small}
We can find that Eq.~\ref{eq: star beta0} is always positive if $N>3$, meaning that the selection favors cooperation instead of spite. Seeking the limit of Eq.~\ref{eq: star beta0} with $N\rightarrow+\infty$ directly leads to Eq.~\ref{eq: star limit beta0}. In fact, this limitation holds for both $\alpha>0$ and $\beta>0$. 
$\hfill\blacksquare$

\section{Proof for THM.~\ref{Theorem: 3}: Hub-Hub Joined Star Graphs}\label{Appendix: D}
\small
In a graph where two stars are joined by hubs, the edge $HH$ is the bridge between two star panels. We denote $p_{HH}^{\sim}=1/(N+\gamma)$ as the probability that a random walk from a hub agent passes the bridge to another hub agent. If the walker starts from a leaf agent, it steps into itself with the probability $p_{LL}=\alpha/(1+\alpha)$, and the hub agent of its located panel with the probability $p_{LH}=1/(1+\alpha)$. If the walker starts from a hub agent, it steps into itself with the probability $p_{HH}=\gamma/(N+\gamma)$, and a leaf agent with the probability $p_{HL}=1/(N+\gamma)$. There are two types of expected coalescence time quantity in the same panel ($\eta_{HL}$ and $\eta_{LL}$) and three types in the different panels ($\eta_{HL}^{\sim}$, $\eta_{LL}^{\sim}$, and $\eta_{HH}^{\sim}$), where $\sim$ marks that the bridge from $H$ to another $H$ separates two vertices. These five types of expected coalescence time can be presented in the following system of equations
\begin{small}
\begin{equation}
\begin{cases}
\eta_{HL}=1+\dfrac{1}{2}\left[(N-2)p_{HL}\eta_{LL}+(p_{LL}+p_{HH})\eta_{HL}+p_{HH}^{\sim}\eta_{HL}^{\sim}\right]\\
\eta_{LL}=1+p_{LH}\eta_{LH}+p_{LL}\eta_{LL}\\
\begin{aligned}
\eta_{HL}^{\sim}=1+\dfrac{1}{2}\left[p_{HH}^{\sim}\eta_{HL}+\right.&(p_{HH}+p_{LL})\eta_{HL}^{\sim}\\
&\left.+(N-1)p_{HL}\eta_{LL}^{\sim}+p_{LH}\eta_{HH}^{\sim}\right]\\
\end{aligned}
\\
\eta_{LL}^{\sim}=1+p_{LL}\eta_{LL}^{\sim}+p_{LH}\eta_{HL}^{\sim}\\
\eta_{HH}^{\sim}=1+p_{HH}\eta_{HH}^{\sim}+(N-1)p_{HL}\eta_{LH}^{\sim}
\end{cases}
.
\end{equation}
\end{small}
By solving these equations, we find that
\begin{small}
\begin{equation}
\eta_{HL}=\dfrac{\parbox{7cm}{$(1 + \alpha)\{ N^2 (10 + 9\alpha + 2\alpha^2) + \gamma (-2 - 2\alpha + 3\gamma)+N[-5-3\alpha^2+11\gamma + \alpha (-8 + 5\gamma)]\}$}}
{N^2 (2 + \alpha) + N (3 + \alpha)(1 + \alpha + \gamma) + \gamma (2 + 2\alpha + \gamma)}, 
\end{equation}
\end{small}
\begin{small}
\begin{equation}
\eta_{LL}=\dfrac{\parbox{7cm}{$2 (1 + \alpha) \{ N^2 (6 + 5\alpha + \alpha^2) - N [1 + \alpha^2 + \alpha(2 - 3\gamma) - 7\gamma] + 2\gamma^2 \}$}}
{N^2 (2 + \alpha) + N (3 + \alpha)(1 + \alpha + \gamma) + \gamma(2 + 2\alpha + \gamma)},
\end{equation}
\end{small}
\begin{small}
\begin{equation}
\eta_{HL}^{\sim}=\frac{\parbox{7cm}{$N^3 (2 + \alpha)^2 + \gamma^2 (2 + 2 \alpha + \gamma) + 
    N^2 (2 + \alpha) \left[5 + 3\alpha^2 + 4\gamma + \alpha (8 + \gamma)\right] + 
    N[-4 - 4\alpha^3 + 9\gamma + 5\gamma^2 + \alpha^2 (-12 + 5\gamma) + 
       2\alpha (-6 + 7\gamma + \gamma^2)]$}}
{N^2 (2 + \alpha) + N (3 + \alpha) (1 + \alpha + \gamma) + 
    \gamma (2 + 2\alpha + \gamma)},
\end{equation}
\end{small}
\begin{small}
\begin{equation}
\eta_{LL}^{\sim}=\frac{\parbox{7cm}{$N^3 (2 + \alpha)^2 + N^2 (2 + \alpha) [6 + 3\alpha^2 + 4\gamma + \alpha (9 + \gamma)] + 
    \gamma \left(2 + 2\alpha^2 + 3\gamma + \gamma^2 + \alpha (4 + 3\gamma)\right) + 
    N [-1 - 3\alpha^3 + 12\gamma + 5\gamma^2 + \alpha^2 (-7 + 6\gamma) + 
       \alpha (-5 + 18\gamma + 2\gamma^2)]$}}
{N^2 (2 + \alpha) + N (3 + \alpha) (1 + \alpha + \gamma) + 
    \gamma (2 + 2\alpha + \gamma)},
\end{equation}
\end{small}
\begin{small}
\begin{equation}
\eta_{HH}^{\sim}=\frac{\parbox{7cm}{$4 + 4 \alpha^3 + N^3 (2 + \alpha)^2 - 4 \alpha^2 (-3 + \gamma) - 4 \gamma + \gamma^2 + \gamma^3 + 
    \alpha (12 - 8 \gamma + \gamma^2) + N^2 (2 + \alpha)[3 \alpha^2 + 4 (1 + \gamma) + \alpha (7 + \gamma)] + 
    N [-11 - 7 \alpha^3 + 6 \gamma + 5 \gamma^2 + \alpha^2 (-25 + 4 \gamma) + 
       \alpha (-29 + 10 \gamma + 2 \gamma^2)]$}}
{N^2 (2 + \alpha) + N (3 + \alpha) (1 + \alpha + \gamma) + 
    \gamma (2 + 2 \alpha + \gamma)}.
\end{equation}
\end{small}
Therefore, we have
\begin{small}
\begin{equation}\label{eq: star hh h}
\begin{aligned}
&\eta_H=1+p_{HH}^{\sim}+(N-1)p_{HL}\eta_{HL}^{\sim}\\
&=\frac{\parbox{7cm}{$2 \{2 + 2 \alpha^3 + N^3 (2 + \alpha)^3 - 
      N^2 (2 + \alpha) [1 + \alpha^2 + \alpha (2 - 3 \gamma) - 6 \gamma] - \alpha^2 (-6 + \gamma) - 
      2 \alpha (-3 + \gamma) - \gamma + \gamma^3 - 
      N [3 + 2 \alpha^3 + \gamma - 6 \gamma^2 + \alpha^2 (7 + \gamma) + 
         \alpha (8 + 2 \gamma - 3 \gamma^2)]\}$}}
{(N + \gamma) [N^2 (2 + \alpha) + 
      N (3 + \alpha) (1 + \alpha + \gamma) + \gamma (2 + 2 \alpha + \gamma)]},
\\
\end{aligned}
\end{equation}
\end{small}
\begin{small}
\begin{equation}\label{eq: star hh l}
\begin{aligned}
&\eta_L=1+p_{LH}\eta_{LH}\\
&=\frac{2 N^2 (6 + 5 \alpha + \alpha^2) - 2 N [1 + \alpha^2 + \alpha (2 - 3 \gamma) - 7 \gamma] + 4 \gamma^2}
{N^2 (2 + \alpha) + N (3 + \alpha) (1 + \alpha + \gamma) + \gamma (2 + 2 \alpha + \gamma)}. 
\end{aligned}
\end{equation}
\end{small}
A random walker can be in a leaf or a hub. The state space of a random walk can be divided into two categories. The stationary distribution of a random walker in a leaf or a hub is $\pi_H=(N+\gamma)/(4N+2\gamma+2N\alpha-2\alpha-2)$ and $\pi_L=(1+\alpha)/(4N+2\gamma+2N\alpha-2\alpha-2)$ respectively. There are $2(N-1)$ leaf vertices and $2$ hub vertices, then we have
\begin{small}
\begin{equation}\label{eq: star hh condition}
(\frac{b}{c})^*_{\mathbf{\mathcal{L}}}=\frac{(N-1)\pi_L\eta_L(1+p_{LL})+\pi_H\eta_H(1+p_{HH})-1}{(N-1)\pi_L\eta_L(p_{LL}+p_{LL}^{(2)})+\pi_H\eta_H(p_{HH}+p_{HH}^{(2)})-1}. 
\end{equation}
\end{small}
Here, because two hub vertices are connected, we have $p_{LL}^{(2)}=p_{LL}^2+p_{LH}p_{HL}$ and $p_{HH}^{(2)}=p_{HH}^{(2)}+p_{HL}[(N-1)p_{LH}+p_{HH}^{\sim}]$. Substituting Eqs. \ref{eq: star hh h}, \ref{eq: star hh l}, and \ref{eq: star hh condition}, we can obtain Eq.~\ref{eq: thm3 starhh} as presented in (a). 

Through the range analysis, it follows immediately that $Num_{StarHH}>0$ and $Den_{StarHH}>0$ if $N>2$. Therefore, we have $(\frac{b}{c})^*_{\mathbf{\mathcal{L}}}>0$, and there exists a positive threshold for the system to favor cooperation as shown in (b). Accordingly, the evolution of spite is never favored. Letting $N\rightarrow+\infty$ in Eq.~\ref{eq: thm3 starhh} directly leads to (c). 
$\hfill\blacksquare$

\section{Proof for THM.~\ref{Theorem: 4}: Ceiling Fans}\label{Appendix: E}
\small
We start by proving (a). We denote $\circ$ as a mark of the random walk quantity from a leaf to its adjacent leaf. If a random walker starts from a leaf, it steps into itself with the probability $p_{LL}=\epsilon/(2+\epsilon)$, into the hub with the probability $p_{LH}=1/(2+\epsilon)$, and into its neighboring leaf with the probability $p_{LL}^{\circ}=1/(2+\epsilon)$. If a walker starts from a hub, it steps into itself with the probability $p_{HH}=\beta/(N-1+\beta)$ and into a leaf with the probability $p_{HL}=1/(N-1+\beta)$. We denote $\eta_{HL}$, $\eta_{LL}$, $\eta_{LL}^{\circ}$ as the expected coalescence time between the hub and a leaf, two nonadjacent leaves, and two adjacent leaves, respectively. These three quantities satisfy the following system of equations
\begin{small}
\begin{equation}
\begin{cases}
\begin{aligned}
\eta_{HL}=1+&\dfrac{1}{2}\left[p_{HH}\eta_{HL}+(N-3)p_{HL}\eta_{LL}+p_{LL}\eta_{HL}\right.\\
&\left.+p_{LL}^{\circ}\eta_{HL}+p_{HL}\eta_{LL}^{\circ}\right]\\
\end{aligned}\\
\eta_{LL}=1+p_{LH}\eta_{HL}+p_{LL}\eta_{LL}+p_{LL}^{\circ}\eta_{LL}\\
\eta_{LL}^{\circ}=1+p_{LH}\eta_{HL}+p_{LL}\eta_{LL}^{\circ}\\
\end{cases}.
\end{equation}
\end{small}
Therefore, we find that
\begin{small}
\begin{equation}
\eta_{HL}=\frac{(2 + \epsilon) \left[-14 - 5 \epsilon + 2N(4 + \epsilon) + 4\beta\right]}{4 + 2N + 3\epsilon + 2\beta},
\end{equation}
\end{small}
\begin{small}
\begin{equation}
\eta_{LL}=\frac{2 (2 + \epsilon) \left[-5 - \epsilon + N(5 + \epsilon) + 3\beta\right]}{4 + 2N + 3\epsilon + 2\beta}, 
\end{equation}
\end{small}
\begin{small}
\begin{equation}
\eta_{LL}^{\circ}\frac{(2 + \epsilon) \left[-5 - \epsilon + N(5 + \epsilon) + 3\beta\right]}{4 + 2N + 3\epsilon + 2\beta}. 
\end{equation}
\end{small}
Accordingly, we obtain
\begin{small}
\begin{equation}
\begin{aligned}
&\eta_{H}=1+(N-1)p_{HL}\eta_{HL}\\
&=\frac{\parbox{7cm}{$24 + 21\epsilon + 5\epsilon^2 + 2N^2(3 + \epsilon)^2 - 6\beta - \epsilon\beta + 2\beta^2 + N(-42 - 33\epsilon - 7\epsilon^2 + 12\beta + 4\epsilon\beta)$}}{(-1 + N + \beta)(4 + 2N + 3\epsilon + 2\beta)},\\
\end{aligned}
\end{equation}
\end{small}
\begin{small}
\begin{equation}
\begin{aligned}
&\eta_{L}=1+p_{LH}\eta_{HL}+p_{LL}^\circ\eta_{LL}^\circ=\frac{3 (-5 - \epsilon + N (5 + \epsilon) + 3 \beta)}{4 + 2N + 3\epsilon + 2\beta}.\\
\end{aligned}
\end{equation}
\end{small}
There are $N-1$ leaves and $1$ hub in the system, then we have
\begin{small}
\begin{equation}\label{eq: ceiling fan condition}
(\frac{b}{c})^*_{\mathbf{\mathcal{L}}}=\frac{(N-1)\pi_L\eta_L(1+p_{LL})+\pi_H\eta_H(1+p_{HH})-2}{(N-1)\pi_L\eta_L(p_{LL}+p_{LL}^{(2)})+\pi_H\eta_H(p_{HH}+p_{HH}^{(2)})-2}. 
\end{equation}
\end{small}
Here, $p_{LL}^{(2)}=p_{LL}^2+{{p}^{\circ}_{LL}}^2+p_{LH}p_{HL}$ and $p_{HH}^{(2)}=p_{HH}^2+(N-1)p_{HL}p_{LH}$. The stationary distribution of a random walk in a ceiling fan with our given self-interaction strength is $\pi_H=(N-1+\beta)/\left[3(N-1)+\beta+N\epsilon-\epsilon\right]$ and $\pi_L=(2+\epsilon)/\left[3(N-1)+\beta+N\epsilon-\epsilon\right]$. Then, we can directly obtain the condition for cooperation as presented in Eq.~\ref{eq: thm4 cf}. 

By analyzing the value range, we can find that the evolutionary dynamics can favor the evolution of spite if $N=3$ or $N=5$. Additionally, if $N\geq7$, Eq.~\ref{eq: thm4 cf} is always positive, which results in our conclusion in (b). Letting $N\rightarrow+\infty$ directly induces the conclusion in (c). 
$\hfill\blacksquare$

\ifCLASSOPTIONcaptionsoff
  \newpage
\fi
\bibliographystyle{IEEEtran}
\small
\bibliography{citations}
\newpage
\begin{IEEEbiography}[{\includegraphics[width=1in,height=1.25in,clip,keepaspectratio]{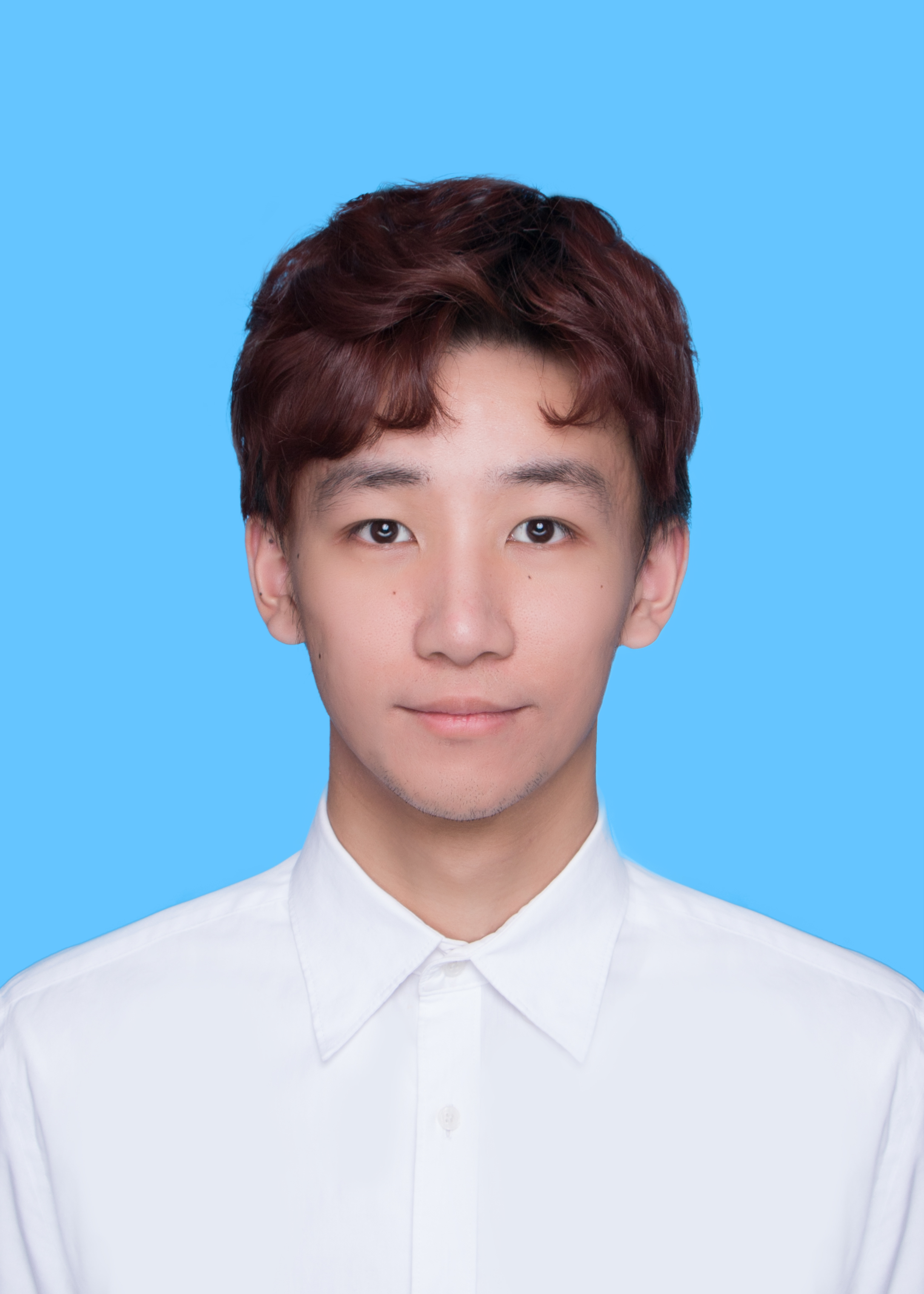}}]{Ziyan Zeng}
received the bachelor’s degree from the College of Artificial Intelligence, Southwest University, Chongqing, China. He is currently pursuing the master’s degree in computer science. His research interests include complex networks, evolutionary games, stochastic processes, and nonlinear science. 
\end{IEEEbiography}
\begin{IEEEbiography}
[{\includegraphics[width=1in,height=1.25in,clip,keepaspectratio]{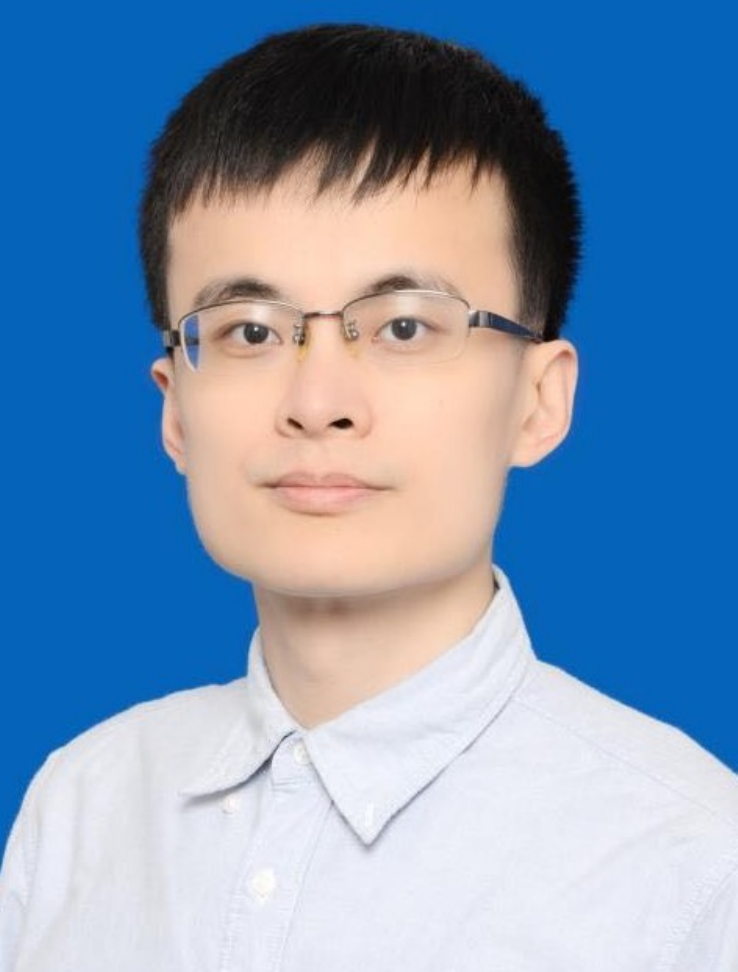}}]{Minyu Feng}
received his Ph.D. degree in Computer Science from a joint program between University of Electronic Science and Technology of China, Chengdu, China, and Humboldt University of Berlin, Berlin, Germany, in 2018. Since 2019, he has been an associate professor at the College of Artificial Intelligence, Southwest University, Chongqing, China. 

Dr. Feng is a Senior Member of IEEE, China Computer Federation (CCF), and Chinese Association of Automation (CAA). Currently, he serves as a Subject Editor for \textit{Applied Mathematical Modelling} and an Editorial Board Member for \textit{Humanities \& Social Sciences Communications, Chaos, and Scientific Reports}. Besides, he is a Reviewer for Mathematical Reviews of \textit{the American Mathematical Society}.

Dr. Feng's research interests include Complex Systems, Evolutionary Game Theory, Computational Social Science, and Mathematical Epidemiology.
\end{IEEEbiography}
\begin{IEEEbiography}
[{\includegraphics[width=1in,height=1.25in,clip,keepaspectratio]{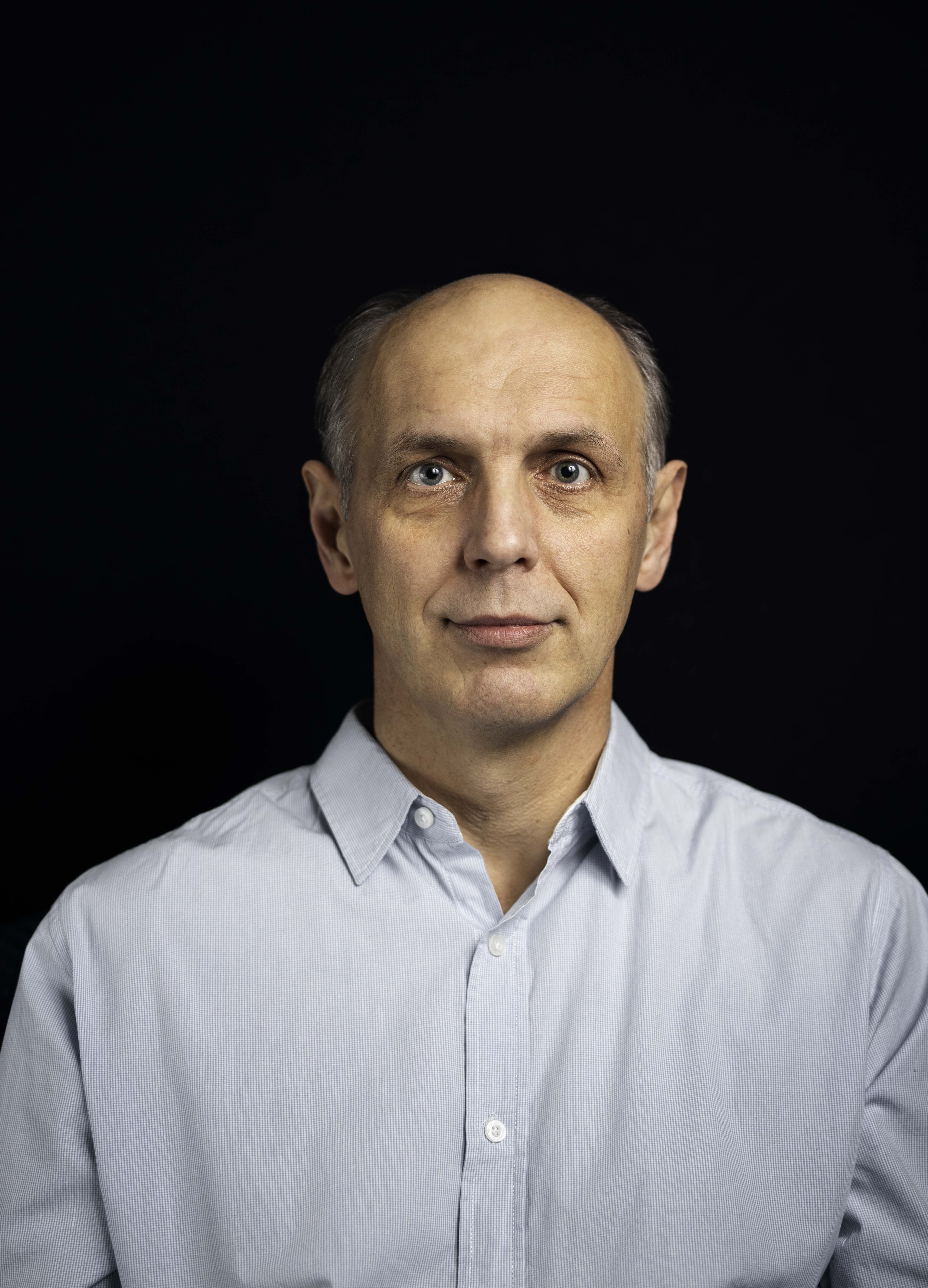}}]{Attila Szolnoki}
is currently a Research Advisor with the Centre for Energy Research, Budapest, Hungary. He has authored around 200 original research articles with over 25,000 citations and an H-index of 80. His current research interests include evolutionary game theory, phase transitions, statistical physics, and their applications.

He is an outstanding or a distinguished referee of several internationally recognized journals. Besides, he serves or had served as an Editor for scientific journals including \textit{PNAS Nexus, Journal of Royal Society Interface, EPL, Physica A, Scientific Reports, Applied Mathematics and Computation, Frontiers in Physics, PLoS ONE, Entropy, and Indian Journal of Physics}. He is among the top 1\% most cited physicists according to Thomson Reuters Highly Cited Researchers. 
\end{IEEEbiography}
\vfill
\end{document}